\DeclareUnicodeCharacter{2212}{-}
\documentclass[epj]{svjour}
\usepackage{amsmath}
\usepackage{braket}

\usepackage{enumitem}   
\usepackage{hyperref}

\usepackage[doi=false,sorting=none,isbn=false,url=false, eprint=false,style=numeric-comp]{biblatex}

\usepackage{graphics}
\usepackage{float}
\usepackage[demo]{graphicx}
\usepackage{subcaption}

\begin{document}
\title{Dissipative dynamics of quantum correlation quantifiers under decoherence channels}
\author{Nitish Kumar Chandra\inst{1}, Sarang S. Bhosale\inst{1} and Prasanta K. Panigrahi\inst{1}
\thanks{\emph{email:}
\href{mailto:nitishkrchandra@gmail.com}{nitishkrchandra@gmail.com}, \href{mailto:sarangbhosale11@iiserkol.ac.in}{sarangbhosale11@iiserkol.ac.in}, \href{mailto:pprasanta@iiserkol.ac.in}{pprasanta@iiserkol.ac.in} (Corresponding author)}%
}                     
%
%
\institute{$^{1}$ Department of Physical Sciences, Indian Institute of Science Education and Research Kolkata, India }
\date{Received: date / Revised version: date}

\abstract{ In this work, we investigate the dynamics of quantum correlations captured by entropic and geometric measures of discord under the influence of dissipative channels for widely used two qubit $X$ state with maximally mixed marginals. Identifying the phenomena of the preservation, sudden change and revival of quantum correlation quantifiers, we determine their hierarchy of robustness under memory-less decoherence channels. We find the analytical expressions of decoherence probabilities at which sudden change occur when the first qubit is subjected to the decoherence channels for multiple times. We deduce the exact analytical expression for the preservation time, the duration for which quantum correlations remain unaffected by noise. We also find the time duration between the two sudden change phenomena of trace distance discord, and show its inverse proportionality to the number of times a channel operates on the state. The constraint relations of decoherence probabilities corresponding to the sudden change region and preservation phenomena are obtained when both the qubits are subjected to locally independent quantum channels for multiple times. Our investigation provides physical insights and possible practical implementation of the discord measures in noisy environments.
\PACS{
      {PACS-key}{discribing text of that key}   \and
      {PACS-key}{discribing text of that key}
     } 
} 

\maketitle
\section{Introduction}
The presence of non-local correlations is one of the distinguishing features of quantum systems that demarcates them from the classical ones \cite{Bell1964,1935,Blaylock}. The non-local properties arise as a consequence of the superposition principle and the tensorial product structure of the Hilbert space, describing the composite quantum system. It is well established that quantum correlations between two subsystems of a global pure state are completely captured by entanglement measures \cite{calabrese2004entanglement,calabrese2012entanglement,castro2018entanglement,Horodecki,plenio2005logarithmic}. However, there exist more general quantum correlations beyond entanglement that manifest even in certain separable mixed states \cite{Usen}. Among the various quantum correlation measures, the entropic quantum discord proposed by Oliver and Zurek \cite{Ollivier} and independently by Henderson and Vedral \cite{henderson} has received tremendous attention due to its application 
in quantum computation with both pure and mixed states. It has found applications in quantum speedup using deterministic quantum computation with one quantum bit (DQC1) \cite{Animesh,knill}. Apart from the computational advantages, quantum discord has found applications in quantum communication protocols \cite{Animesh2}, remote state preparation \cite{dakic}, and various other quantum tasks \cite{Pirandola,Madhok}.

The concept of quantum discord extends the preceding distinction between the entangled and separable quantum states. However, its explicit expression is known for a few special classes of two-qubit $X$ state \cite{Luo, PhysRevA.81.042105}. The calculation for other forms is highly nontrivial due to the involvement of the optimization procedure. This has led to the development of geometric measures of quantum discord for bipartite systems as they are easily computable and have found operational interpretation in teleportation \cite{PhysRevA.86.062313} and state distinguishability \cite{spehner}. These measures are defined through a metric induced on the space of quantum states of the bipartite system. In case of geometric measures, distance to the nearest classical-quantum state is a measure of quantumness in the system. One such geometric quantum discord (GQD)  is defined using the Hilbert-Schmidt distance, whose analytical expression is calculable for a general two-qubit system \cite{PhysRevLett.105.190502}. It has been shown to increase under local operations performed on the unmeasured party, and thus, it is a physically unreliable measure of quantum correlation \cite{PhysRevA.86.034101}. As a consequence, other GQD's were defined based on distance measures like Bures distance, trace distance and Hellinger distance \cite{spehner,PhysRevA.87.042115,Siciu}. These measures satisfy all the axioms, to be considered as bonafide measure of quantum correlations \cite{Roga_2016}. 

In practical situations, the physical systems are invariably subjected to noise which leads to degradation of classical and quantum correlations \cite{SCHLOSSHAUER20191,Kenfack2017}. The phenomenon of loss of correlations under the influence of the external environment is known as decoherence. It is thus crucial to understand the decoherence dynamics of correlations for an open quantum system that has implications on quantum information processing tasks \cite{10.5555/1972505}. Based on the type of interaction, the dynamics can be Markovian or non-Markovian depending on whether the coupling strength of the system with the environment is weak or strong. It has been shown that entanglement between the subsystems under a noisy environment shows phenomena like sudden death in a finite time \cite{PhysRevLett.93.140404}. However, quantum discord does not suffer from the phenomenon of sudden death and decays asymptotically under Markovian dynamics \cite{PhysRevA.80.024103}, and disappears only at certain points under non-Markovian dynamics \cite{PhysRevA.81.014101}. It was shown that Bures distance discord is more robust than entanglement under decoherence channels and revives under depolarizing channel, whereas entanglement damps to sudden death without being restored \cite{SHI2016843}. A comparative study of different correlation measures under depolarising channel and the hybrid channel showed that quantum discord is less robust against noise in comparison to other correlation measures \cite{Shi2016}. It has been shown that an initial set of Bell-diagonal states under decoherence channels undergo a sudden change phenomenon i.e., discontinuity in the decay rate of quantum discord \cite{PhysRevA.80.024103}. This phenomenon was experimentally observed in Ref. \cite{Xu2010}.  Similar investigations for discord measure based on trace norm revealed that it can exhibit double sudden change phenomenon, whereas the entropic measure and the Hilbert-Schmidt measure do not eventuate this behavior \cite{PhysRevA.87.042115}. The sudden change of correlation measures can be associated with several physical scenarios like classical and quantum phase transitions \cite{Luigi,Werlang,Pal,Pal2012}. A deeper physical understanding of this phenomenon remains an open question \cite{Céleri2017}.


For a specific class of states belonging to Bell diagonal states, the quantum discord remains unaffected after interaction with Markovian channels. This behavior, known as the freezing phenomenon, was interpreted in terms of sudden transition when the classical correlations of the system stop decaying and the quantum correlations start deteriorating \cite{Mazzola}. Moreover, the freezing phenomenon was also observed for other geometric measures under a non-dissipative decohering environment, and shown to be independent of the choice of the bonafide geometric discord measures and, thus, recognized as a universal phenomenon \cite{Cianciaruso2015}. The freezing behavior under a noisy environment is also observed for quantum correlations captured by coherence measures. For a particular class of states with maximally mixed marginals under independent non-dissipative channels, the exact conditions for freezing of the distance-based measures of quantum coherence were investigated in Ref. \cite{Bromley}. Based on this study, the freezing phenomenon of coherence monotones was analyzed when a two qubit system was subjected to various decoherence channels for multiple times \cite{Zhao,Wang}. In Ref. \cite{Zhao}, the conditions for observing freezing coherence measures under the action of amplitude damping channel was obtained. In Ref. \cite{Wang}, the dynamics of coherence measures under different Markovian channels were investigated and it was shown that the phenomenon of frozen coherence depends on the choice of coherence measure. Motivated from these studies involving the dynamics of coherence, we do a detailed investigation of entropic and geometric measures of quantum discord under various Markovian channels operated multiple times on the first sub-system and then on both the sub-systems.
We derive the exact expressions for sudden change phenomena of discord measures and preservation time for trace distance discord when the first qubit is subjected to various Markovian channels. Further, we deduce the constraints on decoherence probabilities corresponding to the regions of sudden change in discord measures for the case when both the qubits are subjected to bi-Markovian channels of same and different types.    


The paper is organized as follows. In Sect. $2$, we briefly review the definitions of entropic and geometric measures of discord and describe their analytical expressions for Bell diagonal states. In Sect. $3$, we describe the quantum channels and investigate the dynamics of discord measures when the first subsystem experiences various Markovian channels multiple times. In Sect. $4$, we investigate the dynamics of discord measures under bi-sided Markovian channels of the same and different types when applied multiple times on both the subsystems. Finally, we present the conclusions of our work in Sect. $5$.

\section{Quantum Correlation Measures}
To begin with, we will review the definitions of some of the quantum correlations based on entropy as well as geometry-based measures, which can be analytically computed for two-qubit Bell diagonal states. The discord measures capture quantum correlations beyond the entanglement measures. They capture the minimal disturbance on the bi-partite state $\rho^{AB}$ after performing a local projective measurement on either of the subsystems. The distance-based measures quantify the deviation of the given state from the classical state. A good quantum correlation quantifier (asymmetric measure) defined for a bipartite quantum state $\rho^{AB}$ with subsystems $A$ and $B$ satisfies the following axioms \cite{Cianciaruso2015},

\begin{enumerate}[label=(\roman*)]
    \item It vanishes for classical-quantum states, i.e., states for which there exists local measurement on $A$, such that the total state remains unperturbed. Such states in general can be written  in the form $\rho_{CQ} = \sum_{i}p_{i}|i^{A}\rangle\langle i^{A}| \otimes \rho_{i}^B$, where, $p_{i}\geq 0 $ with $\sum_{i}p_{i} = 1$, $\{|i^{A}\rangle\}$ an orthonormal basis for $A$, and $\rho^{B}_{i}$ is the density matrix for subsystem $B$.
    \item It remains invariant under local unitary transformations i.e., $\rho^{AB}\rightarrow (U_{A} \otimes U_{B}) \rho^{AB} (U_{A}^{\dagger} \otimes U_{B}^{\dagger})$ where $U_{A}$ and $U_{B}$ are local unitary operations applied on subsystems A and B respectively.
    \item It should not increase under the action of Completely Positive Trace Preserving (CPTP) linear maps, when acted on the subsystem B.
    \item It should reduce to an entanglement monotone for pure states.
\end{enumerate}

The entropic quantum discord and distance based measures like Schatten 1-norm and squared Bures distance are known to satisfy all the properties required for defining quantum correlation measures. The distance used in defining geometric quantum correlation measures has to satisfy the mathematical requirements of contractivity under CPTP maps, invariance under transposition 
and joint convexity to be considered a bonafide distance \cite{Cianciaruso2015}. 

In this work, we investigate the states belonging to the special class of two qubit $X$ states known as Bell diagonal states. In the computational basis, it takes the form,
\begin{equation}
    \rho^{AB} = \frac{1}{4}(I^{A}\otimes I^{B} + \sum_{i = 1}^{3}d_{i}\sigma_{i}^{A}\otimes\sigma_{i}^{B})
\end{equation}
 
where $d_{i} = tr(\rho^{AB}(\sigma_{i}^{A}\otimes\sigma_{i}^{B}))$, $I$ is the two dimensional identity matrix, $\sigma_{i}$ are Pauli matrices, $d_{i}$ are real numbers satisfying $0\leq |d_{i}| \leq 1$. We can use the vector $d = \{d_{1},d_{2},d_{3}\}$ to represent the Bell-diagonal state. Its eigenstates (as the name suggests) are four Bell states,

\begin{equation}
    \left|\psi_{a b}\right\rangle \equiv\left(|0, b\rangle+(-1)^{a}|1,1 \oplus b\rangle\right) / \sqrt{2}
\end{equation}

whose eigenvalues are,
\begin{equation}
\lambda_{a b}=\frac{1}{4}\left[1+(-1)^{a} d_{1}-(-1)^{a+b} d_{2}+(-1)^{b} d_{3}\right]
\end{equation}
 where a, b $ \in \{0,1\}$. In matrix form, it is given by
 \begin{equation}
     \rho^{AB}=\frac{1}{4}\left(\begin{array}{cccc}
1+d_{3} & 0 & 0 & d_{1}-d_{2} \\
0 & 1-d_{3} & d_{1}+d_{2} & 0 \\
0 & d_{1}+d_{2} & 1-d_{3} & 0 \\
d_{1}-d_{2} & 0 & 0 & 1+d_{3}
\end{array}\right)
 \end{equation}

 The condition on eigenvalues $(\lambda_{ab})\geq0$ restricts the value of $\Vec{d} = (d_{1},d_{2},d_{3})$, so the physically realizable Bell diagonal states can be represented as points lying inside the tetrahedron having vertices $(1,1,-1),(1,-1,1),(-1,-1,-1)$ and $(-1,1,1)$. This geometric description has been widely used in literature to explain the evolution of states retaining this form under different dynamical scenarios \cite{SHI2016843,Cianciaruso2015}. 
\subsection{Quantum Discord}

The density operator $\rho$ of the system is a hybrid object that encompasses both classical and quantum correlations. The task of distinguishing classical and quantum correlations is of pivotal importance in quantum information theory. This issue was addressed by defining the classically correlated states that satisfy the generalized Bell-inequalities \cite{Werner}. However, certain phenomena like quantum speed-up with separable states \cite{PhysRevLett.85.2014,PhysRevLett.83.1054} and quantum non-locality without entanglement \cite{PhysRevA.59.1070} required the description of quantum correlations by a more fundamental and generalized correlation measure \cite{Ollivier,henderson}. 

For a bipartite quantum system, the quantum mutual information quantifies the total correlations present in the system,
\begin{equation}
    I(\rho^{AB})=S(\rho^{A})+S(\rho^{B})-S(\rho^{AB})
\end{equation}
 where, $\rho^{AB}$ is the density matrix of the bipartite system, $\rho^{A(B)}$ is the reduced density matrix of subsystem $A(B)$, and  $S(\rho) = -\operatorname{Tr}(\rho\operatorname{log}\rho)$ is the von- Neumann entropy.

The quantum mutual information includes the total correlations of the system, including classical and quantum correlations, if the classical component of the correlations is subtracted, the remaining quantity is the quantifier of the  quantum correlations, known as quantum discord,
 
 \begin{equation}
     Q(\rho^{AB})=I(\rho^{AB})-C(\rho^{AB})
 \end{equation}
where the quantity $C(\rho^{AB})$ is the classical correlation and is defined as,
\begin{equation}
    C(\rho^{AB})= max_{\{\Pi^{B}_{k}\}}[S(\rho^{A})-S_{\{{\Pi^{B}_{k}}\}}(\rho^{A|B})]
\end{equation}
$S_{\{\Pi^{B}_{k}\}}(\rho^{A|B})=\sum_{k}p_{k}S(\rho^{A}_{k})$ is the average conditional von Neumann entropy of the state after measurement where $p_{k}={Tr_{AB}(\Pi^{B}_{k}\rho^{AB})}$ and $\Pi_{k}^{B}=|k\rangle\langle k|$ are the positive operator-valued measure (POVM) on the second sub-system (B) for the standard basis states $(k=0,1)$.
The post measurement state is given by,
\begin{equation}
    \rho^{A}_{k}= \frac{Tr_{B}(\Pi^{B}_{k} \rho^{AB}\Pi^{B}_{k})}{Tr_{AB}(\Pi^{B}_{k}\rho^{AB})}
\end{equation}
Hence, the quantum discord is given by the expression,
 \begin{equation}
     Q(\rho^{AB})=S(\rho^{B})-S(\rho^{AB})+min_{\{\Pi^{B}_{k}\}}S(\rho^{A|B})
 \end{equation}

It is a non-negative quantity and is zero for only classically correlated states, i.e., states that can be completely determined using local operations and classical communication (LOCC), without disturbing it \cite{Modi}. Hence, it can be non-zero even for certain separable states, unlike the entanglement, which is always zero for the separable states. Therefore, quantum discord is a non-local correlation that is more general than entanglement and even encompasses correlations that are not necessarily captured by entanglement measures. The analytical expression of discord is difficult to calculate due to the difficulty in minimization of all POVM's. However, for a special class of two-qubit $X$ states, quantum discord has been analytically obtained \cite{Luo}. For the Bell-diagonal form, the closed-form expression of quantum discord is given by,

\begin{equation}
\mathcal{D}\left(\rho^{AB}\right)=-\mathcal{H}\left(\lambda_{a b}\right)-\sum_{j=1}^{2} \frac{\left(1+(-1)^{j} d\right)}{2} \log _{2} \frac{\left(1+(-1)^{j} d\right)}{4}
\end{equation}
where $d=\max \left\{\left|d_{1}\right|,\left|d_{2}\right|,\left|d_{3}\right|\right\}$ and $\mathcal{H}\left(\lambda_{a b}\right)=-\sum_{a, b} \lambda_{a b} \log_{2} \lambda_{a b}$ is the Shannon entropy.

\subsection{Bures Distance Discord}

Geometric discord measures have been defined, which are akin to the well-known entropic-based discord measures. Apart from easier computability, they offer operational advantages in terms of state distinguishability. One such geometric quantum correlation measure defined using a contractive distance measure is true Bures distance \cite{Bures}. The distance defined using the Bures metric is monotonous as well as Riemannian. This metric matches with the Quantum Fisher Information \cite{helstrom1969quantum}, which has applications in interferometry. It satisfies all the criteria for a good correlation quantifier. It reduces to entanglement monotone for pure states, and in the case of mixed states, it corresponds to the maximal success probability in an ambiguous quantum state discrimination task \cite{spehner}. 
 The Bures distance (squared) between two states $\rho$ and $\sigma$ is given by the expression,
 \begin{equation}
d_{\mathrm{B}}(\rho, \sigma)=2[1-\sqrt{F(\rho, \sigma)}]
\end{equation}
\\
where the expression $F(\rho, \sigma)=\left[\operatorname{tr}(\sqrt{\rho} \sigma \sqrt{\rho})^{1/2}\right]^{2}$ is the Uhlmann fidelity.\\

 The Bures distance discord is defined similar to the Bures measure of entanglement and is given by \cite{Shi1},
 
\begin{equation}
D_{\mathrm{B}}(\rho)=\left\{(2+\sqrt{2})\left[1-\sqrt{F_{\max }(\rho)}\right]\right\}^{1 / 2}
\end{equation}

where, $F_{\max }(\rho)=\max _{\sigma \in \rho_{CQ}} F(\rho, \sigma)$ and $\rho_{CQ}$ is the set of classical quantum states. 

We can obtain the closed analytical form of $F_{\max }(\rho)$ explicitly for the case of Bell diagonal states,
\begin{equation}
F_{\max }(\rho)=\frac{1}{2}+\frac{1}{4} \max _{\langle i j k\rangle}\left[\sqrt{\left(1+d_{i}\right)^{2}-\left(d_{j}-d_{k}\right)^{2}}+\sqrt{\left(1-d_{i}\right)^{2}-\left(d_{j}+d_{k}\right)^{2}}\right]
\end{equation}
where the maximum value is taken over the cyclic permutations of $\{1,2,3\}$.


\subsection{Trace Distance Discord}

Trace distance discord is one of the quantum correlations quantifier defined using the Schatten $1$- norm \cite{Paula}. For a bipartite system $\rho^{AB}$ involving qubits A and B, it is defined as the minimal trace distance between $\rho^{A B}$ and $\rho_{C Q}$ which is the set of classical quantum states having zero quantum discord with respect to measured party,
$$
D_{T}(\rho)=\min _{\sigma \in \rho_{C Q}}\left\|\rho^{A B}-\sigma\right\|_{1}
$$
where $\|M\|_{1}=\operatorname{Tr} \sqrt{M^{\dagger} M}$ is the trace norm, and $\sigma$ takes the form,
\begin{equation}
\sigma = \sum_{i = 1}^{2} p_{i} |\phi_{i}^{A}\rangle \langle \phi_{i}^{A}| \otimes \rho_{i}^{B}
\end{equation}
 
 The set of states $\{|\phi_{i}^{A}\rangle\}$ forms an orthonormal basis for the subsystem A and $\rho_{i}^{B}$ is the general reduced matrix for the subsystem B. It is worth noticing that trace distance discord defined using the Schatten 1-norm is the only $p$-norm geometric measure that does not increase under local operations on the unmeasured party \cite{PianiM}. 
  The property of contractivity under trace-preserving quantum channels and invariance under local unitary transformations are two of the most significant advantages of employing this distance measure.
 It is monotonically related to entanglement for pure states. It has been shown that Negativity, an entanglement monotone is the entanglement counterpart of the trace distance discord \cite{Piani}. 
 
 The analytical expression of TDD for the two qubit $X$ states has been computed in Ref. \cite{Ciccarello_2014}. For the special case of Bell diagonal states characterized by the vector $\Vec{d} = (d_{1},d_{2},d_{3})$, the analytic expression is given by,
\begin{equation}
D_{T}(\rho)=\operatorname{\textit{int}}\left\{\left|d_{1}\right|,\left|d_{2}\right|,\left|d_{3}\right|\right\}
\end{equation}

where ``$int$" is the intermediate value among the absolute values of the correlation functions $d_{1}, d_{2}$ and $d_{3}$.

\section{Quantum Channels}
The decoherence phenomenon is an inevitable and irrevocable characteristic of the environment. There is always a finite possibility of two-way exchange of energy and information when a system is exposed to the environment.  Such interactions have the potential to entangle the quantum system with its surroundings. With the help of certain maps between the spaces of operators known as quantum channels, these phenomena are comprehensively investigated in the dynamics of open quantum systems.  A quantum channel is a linear map that takes density operators to density operators with or without respecting its unitarity character.  
Quantum channels, which incorporate the evolution of pure quantum states to mixed quantum states due to decoherence, are used to describe the evolution of quantum states in terms of density operators. Unitary transformation is taken into account in the coherent evolution of quantum states.
However, when the evolution of quantum states is not necessarily coherent, a differential equation known as the master equation can be used \cite{nielsen2002quantum}.

One can invert the action of the channel to obtain the original input state if the channel operators are unitary. Although, there exist channels showing non-invertible and non-unitary characteristics \cite{preskill}. A thorough investigation of the effect of channels on quantum systems can give insights into the methods to preserve and manipulate the fragile quantum correlations, a major ingredient of modern quantum technologies.
Mathematically, the description of the quantum channel using the Kraus operator formalism \cite{kraus1983states} is,
\begin{equation}
    \rho_{out}=\varepsilon (\rho_{in})
    = \sum_{a} K_{a}\rho K^{\dagger}_{a} 
    \quad \text{and} \quad
     \sum_{a} K_{a} K^{\dagger}_{a}=I
\end{equation}
where, $\varepsilon(\rho_{in})$ is the evolved state under some local decoherence channel, $\rho_{in}$ is the initial density operator of the system and $\rho_{out}$ is density operator after the evolution. The channel operators $K_{a}$ and $K^{\dagger}_{a}$ incorporate the decoherence parameter.
A quantum channel must satisfy the following properties,
\begin{enumerate}[label=(\alph*)]
\item Linearity:
 i.e., $\varepsilon(\alpha \rho_{1}+\beta \rho_{2})= \alpha \varepsilon (\rho_{1})+\beta \varepsilon (\rho_{2})$
  \item Hermiticity 
   i.e., $\varepsilon(\rho) =\varepsilon(\rho^{\dagger})$ 
   \item  Positivity 
  i.e., $\varepsilon(\rho)\geq0$
  \item Trace preserving 
    i.e., $tr(\varepsilon(\rho))=tr(\rho)$
\end{enumerate}
    
The premise that the evolution of a system at each future time is determined solely by its current state rather than its history is a crucial assumption towards understanding a particular class of dynamical systems. The process modeled under this assumption is known as a ``Markov process" \cite{bharucha1997elements}. There are two sets of channels, depending on whether the channel meets the aforementioned assumption. The channels with no memory effects or information backflow from the environment to the system fall under the Markovian channels. The channels with memory effect on the system or allow the backflow of the information from the environment to the system are known as Non-Markovian channels \cite{PhysRevLett.103.210401}. The exact description of quantum Markovianity and quantum Non-Markovianity are context-dependent and are still a matter of ongoing investigation \cite{Vacchini_2011,Glick2020,LI20181}. There have been a variety of notions that can be used to characterize quantum Markovianity in a specific setting \cite{LI20181}.


  Mathematically, a quantum channel $\varepsilon(\rho)$ is a completely positive and trace-preserving map (CPTP) which maps operators to operators. They are characterized by a set of Kraus operators $\{K_{i}\}$ which contain the information about the decoherence process of the single qubit,

\begin{equation}
    \varepsilon(\rho)=\sum_{i, j}\left(K_{i} \otimes K_{j}\right) \rho\left(K_{i} \otimes K_{j}\right)^{\dagger}
\end{equation}


If the first subsystem of a bi-partite state ($\rho_{AB}$) with sub-systems $A$ and $B$ passes through a quantum channel described by Kraus operators ($K_{0}$, $K_{1}$, $K_{2}$ and $K_{3}$) then the output state $\rho_{AB}^{(1)}$ is given by,
\begin{equation}
    \rho_{AB}^{(1)}=(K_{0} \otimes I) \rho_{AB} (K_{0}^{\dagger} \otimes I) + (K_{1} \otimes I) \rho_{AB} (K_{1}^{\dagger} \otimes I) + (K_{2} \otimes I) \rho_{AB} (K_{2}^{\dagger} \otimes I) + (K_{3} \otimes I) \rho_{AB} (K_{3}^{\dagger} \otimes I)
\end{equation}
After the first subsystem undergoes through a quantum channel $n$ times, the output state $\rho_{AB}^{(n)}$ is given by,
\begin{equation}
    \rho_{AB}^{(n)}=\sum_{i_{1}, i_{2}, \ldots, i_{n}= 0,1,2,3} (K_{i_{1} i_{2} \cdots i_{n}} \otimes I) \rho_{AB} (K_{i_{1} i_{2} \cdots i_{n}}^{\dagger} \otimes I)
\end{equation}
where $K_{i_{1} i_{2} \cdots i_{n}}= K_{i_{1}} K_{i_{2}} \cdots K_{i_{n}}$.

Similarly, when both the subsystems undergo through two independent quantum channels for $n$ times, the output state $\rho_{AB}^{(n)}$ is given by,

\begin{equation}
    \rho_{AB}^{(n)}=\sum_{i_{1}, i_{2}, \cdots, i_{n}, j_{1}, j_{2}, \cdots, j_{n}=0,1,2,3} (K_{i_{1} i_{2} \cdots i_{n}} \otimes K_{j_{1} j_{2} \cdots j_{n}}) \rho_{AB} (K_{i_{1} i_{2} \cdots i_{n}}^{\dagger} \otimes K_{j_{1} j_{2} \cdots j_{n}}^{\dagger})
\end{equation}



\subsection{Bit Flip Quantum Channel}

The bit flip channel explains the physical process in which the system is either impacted by a Pauli X operation with probability $(1-p)$ or left unimpacted with probability $p$ \cite{nielsen2002quantum}. For example,
 \begin{equation}
     \ket{0}\rightarrow\ket{1} \quad \text{and}\quad \ket{1}\rightarrow\ket{0}
 \end{equation}

The Kraus operators for bit flip quantum channel are defined as,
\begin{equation}
K_{0}=\sqrt{1-p/2}\begin{pmatrix}
1 & 0 \\
0 & 1
\end{pmatrix} \  \text{and} \quad K_{1}=\sqrt{p/2} \begin{pmatrix}
0 & 1 \\
1 & 0
\end{pmatrix}
\end{equation}
where, $p$ is the decoherence probability and it is given as, $p=1-e^{-\gamma t}$. The parameter $\gamma$ is the decoherence rate.\\
If the first subsystem of a bi-partite Bell-diagonal state ($\rho_{AB}$) with sub-systems $A$ and $B$ passes through a quantum channel described by Kraus operators ($K_{0}$, $K_{1}$) then the output state $\rho_{AB}^{(1)}$ is given by,
\begin{equation}
     \rho^{(1)}_{AB}=\frac{1}{4}\left(\begin{array}{cccc}
1+d_{3}(1-p) & 0 & 0 & d_{1}-d_{2}(1-p) \\
0 & 1-d_{3}(1-p) & d_{1}+d_{2}(1-p) & 0 \\
0 & d_{1}+d_{2}(1-p) & 1-d_{3}(1-p) & 0 \\
d_{1}-d_{2}(1-p) & 0 & 0 & 1+d_{3}(1-p)
\end{array}\right)
 \end{equation}

\begin{figure}[hbt!]
\begin{subfigure}{.5\textwidth}
  \centering
  \includegraphics[width=0.8\linewidth]{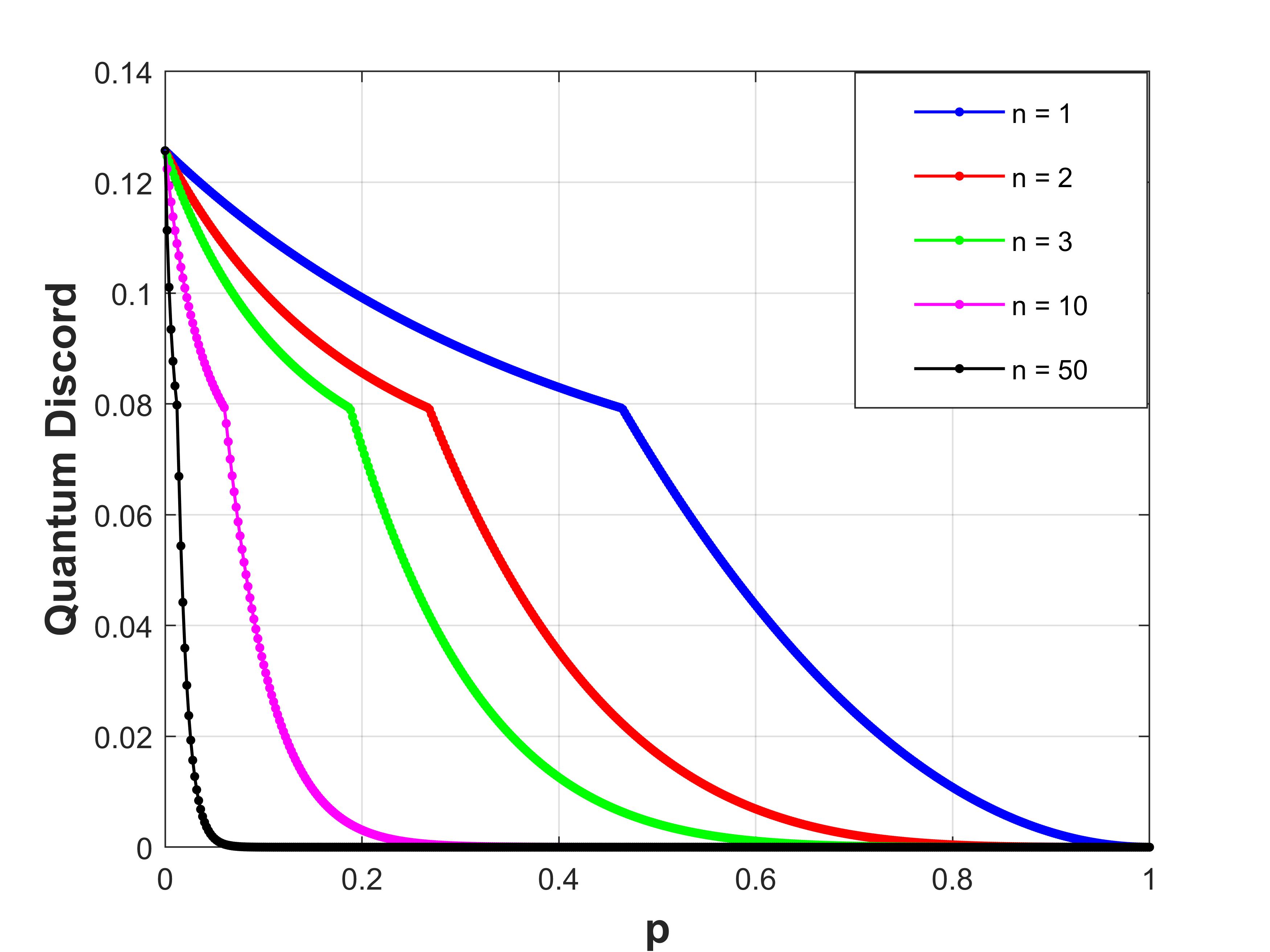}
  \label{fig001}
\end{subfigure}%
\begin{subfigure}{.5\textwidth}
  \centering
  \includegraphics[width=.8\linewidth]{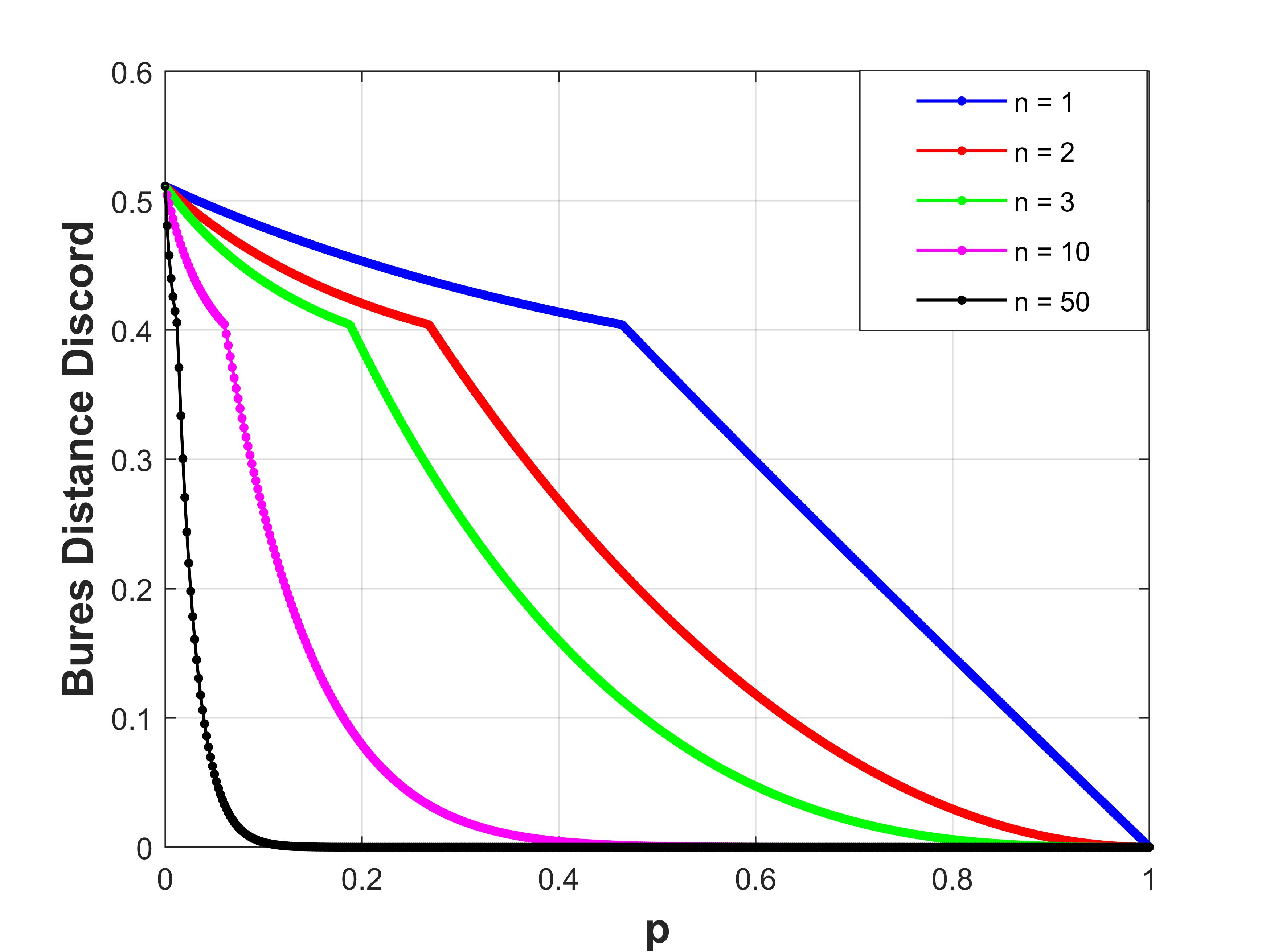}
  \label{fig002}
\end{subfigure}
  \centering
\begin{subfigure}{.5\textwidth}
  \includegraphics[width=.8\linewidth]{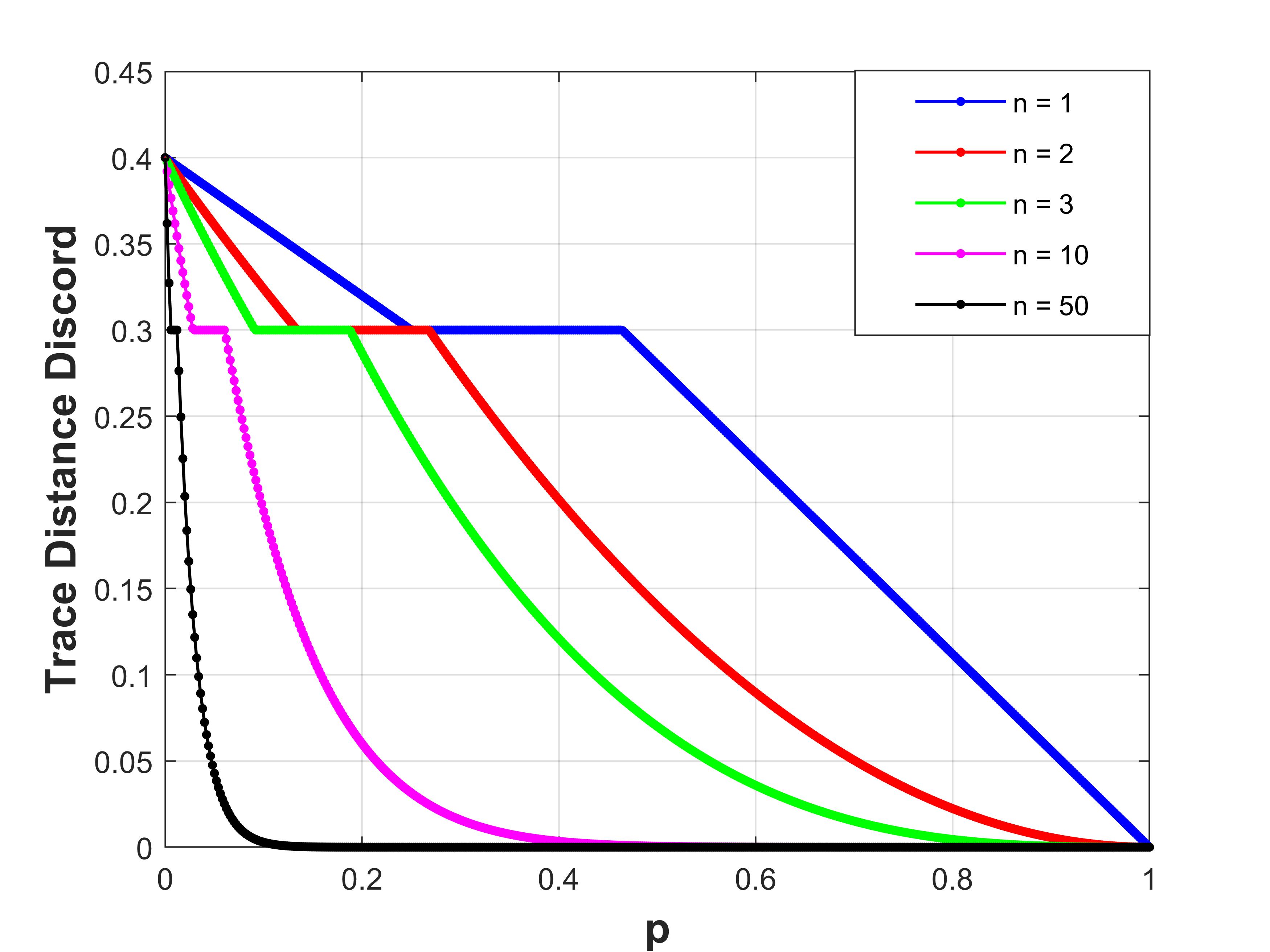}
  \label{fig003}
\end{subfigure}
\caption{The first subsystem undergoes through the bit-flip quantum channel $n$ times for Bell diagonal state ($d_{1} = 0.3$, $d_{2} = −0.4$ and $d_{3} = 0.56$). The evolution of (a) Quantum Discord, and (b) Bures Distance Discord show the sudden change phenomenon at $p_{sc}^{1} = 0.46$ (Blue, $n=1$), $p_{sc}^{2} = 0.26$ (Red, $n=2$), $p_{sc}^{3} = 0.18$ (Green, $n=3$), $p_{sc}^{10} = 0.06$ (Magenta, $n=10$) and $p_{sc}^{50} = 0.01$ (Black, $n=50$) (c) Trace Distance Discord shows two SCPs, the first SCP is at $p_{sc_{1}}^{1} = 0.24$, $p_{sc_{1}}^{2} = 0.13$, $p_{sc_{1}}^{3} = 0.09$, $p_{sc_{1}}^{10} = 0.02$ and $p_{sc_{1}}^{50} = 0.006$ and the second SCP matches with that of QD and BDD.}
\label{fig30}
\end{figure}

 After the first subsystem undergoes through the bit-flip channel $n$ times, the coefficients of the evolved Bell-diagonal state are given by, 
\begin{equation}
    d_{1}^{\prime} = d_1 \quad d_2^{\prime} = d_{2}(1-p)^{n} \quad d_{3}^{\prime} = d_{3}(1-p)^{n} 
\end{equation}
The effect of bit-flip channel on entropic measure (QD)  and geometric measures (BDD and TDD)  of discord are shown in the Fig. \ref{fig30}. We observe the phenomenon of sudden change in the dynamical evolution of discord measures under bit flip channel. The existence of such  phenomena occur as a result of optimization step involved in the definitions of such correlation quantifiers. 

We can explain the phenomenon of sudden change from the maximization step involved in the evaluation of discord measures. The calculation of quantum discord involves the step of evaluating, $d = \text{ max } (|d^{\prime}_{1}|,|d^{\prime}_{2}|,|d^{\prime}_{3}|)$. In our work, we have considered the initial set of conditions such that $|d_{1}|<|d_{2}|<|d_{3}|$. For this set of initial conditions, the correlation function $|d_{1}^{\prime}|$ remains invariant under the evolution of this channel and the other correlation functions $|d_{2}^{\prime}|$ and $|d_{3}^{\prime}|$ decay with the same factor implying that $|d_{3}^{\prime}|$ is always greater than $|d_{2}^{\prime}|$. The moment $|d_{3}^{\prime}|$ becomes less than $|d_{1}^{\prime}|$, the quantity $d$ switches the value from $|d_{3}^{\prime}|$ to $|d_{1}^{\prime}|$ and we observe a sudden change at $|d_{3}^{\prime}| = |d_{1}^{\prime}|$. Explicitly when,
\begin{equation}
    |d_{1}| = |d_{3}|(1-p)^{n}
\end{equation}
we observe a sudden change at, 
\begin{equation}
    p_{sc}^{Q} = 1 - \left(\frac{|d_{1}|}{|d_{3}|} \right )^{\frac{1}{n}}
\end{equation}

The calculation of BDD involves the step of cyclic maximization of the Uhlmann fidelity for Bell diagonal state  $F_{max} (\rho_{BD})$ i.e., it involves the step of finding the maximum from the cyclic combination of the term,
\begin{equation}
    A_{ijk} = \left[\sqrt{\left(1+d_{i}^{\prime}\right)^{2}-\left(d_{j}^{\prime}-d_{k}^{\prime}\right)^{2}}+\sqrt{\left(1-d_{i}^{\prime}\right)^{2}-\left(d_{j}^{\prime}+d_{k}^{\prime}\right)^{2}}\right]
\end{equation}
At any time, we need to find the maximum among the possible permutations ($A_{123},A_{231}$   and $A_{312}$). The point of sudden change phenomenon (SCP) will be one of the solution from the set of equations: $A_{123}=A_{231}$ $\implies$ $|d_{1}^{\prime}|=|d_{2}^{\prime}|$, $A_{231}=A_{312}$  $\implies$ $|d_{2}^{\prime}|=|d_{3}^{\prime}|$, $A_{123} = A_{312}$ $\implies$ $|d_{1}^{\prime}|=|d_{3}^{\prime}|$. 

However, the only solution which leads to the SCP is when the maximum value term (say $A_{ijk}$) during evolution becomes less than the term (say $A_{i^{\prime}j^{\prime}k^{\prime}}$) and the term ($A_{i^{\prime}j^{\prime}k^{\prime}}$) replaces its order from intermediate value to maximum value, thus, we observe a SCP at $A_{ijk} = A_{i^{\prime}j^{\prime}k^{\prime}}$. In the case of BDD for the Bit Flip channel, we observe the SCP when $A_{123}$ becomes equal to $A_{312}$. Interestingly, the solution of equating these two quantities yield the equivalent condition to that of the quantum discord i.e., $|d_{1}^{\prime}| = |d_{3}^{\prime}|$.



The evaluation of TDD involves the step of finding intermediate value among the absolute value of correlation functions $|d_{1}^{\prime}|, |d_{2}^{\prime}|$ and $|d_{3}^{\prime}|$. For the initial condition $|d_{1}|<|d_{2}|<|d_{3}|$, the quantity $|d_{2}^{\prime}|$ has the intermediate value till it becomes less than $|d_{1}^{\prime}|$ and when $|d_{3}^{\prime}|$ becomes less than $|d_{1}^{\prime}|$, $|d_{3}^{\prime}|$ takes the intermediate value, so we observe the phenomena of double sudden change phenomenon (DSCP).
We observe the first sudden change when $|d_{2}^{\prime}| = |d_{1}^{\prime}|$ and second second change when $|d_{3}^{\prime}| = |d_{1}^{\prime}|$. The values of $p$ corresponding to DSCP are given by,
\begin{equation}
    p_{sc_{1}}^{T} = 1 - \left(\frac{|d_{1}|}{|d_{2}|} \right )^{\frac{1}{n}}
\quad \text{and} \quad
    p_{sc_{2}}^{T} = 1 - \left(\frac{|d_{1}|}{|d_{3}|} \right )^{\frac{1}{n}}
\end{equation}

We observe that the dynamics of TDD is preserved between $p_{sc_{1}}^{T}$ and $p_{sc_{2}}^{T}$ and has the value $|d_{1}|$, which is constant for the bit-flip channel. The parameter $p$ is given by, $p = 1 - exp(-\gamma t) $, where $\gamma$ is the decoherence rate. The time interval between which the TDD is frozen or preserved is given by,

\begin{equation}
    \Delta t_{frozen} = \frac{1}{n\gamma}ln\left(\frac{|d_{3}|}{|d_{2}|}\right)
\end{equation}

Interestingly, the time duration during which the TDD is frozen is inversely proportional to the number of times the bit-flip channel is acted upon the first subsystem, given the other parameters are constant during the evolution.


\subsection{Phase Flip Quantum Channel}

The phase flip channel, which is equivalent to the $\sigma_{z}$ (Pauli Z) operator, describes relative $\pi$-phase errors in the computational basis. The action of phase flip channel is shown in following example.
\begin{equation}
\ket{0} \rightarrow \ket{0} \quad \text{and} \quad \ket{1} \rightarrow - \ket{1}
\end{equation}
It is important to note that, the phase flip operation in one basis appears as a bit flip operation in another basis. This channel has no classical analog. In the phase flip channel, the information loss is not accompanied by the energy dissipation. This channel is is also known as dephasing channel. When a quantum system is isolated, the phase of the system is well defined. When a system interacts with its surroundings, the phase becomes uncertain; this process is referred to as dephasing. When quantum information is encoded in the phase, the dephasing processes add noise to the stored information. Dephasing processes are a subset of the unital channel family of decohering processes. Unital channels are linked to interactions that do not involve energy transfer from the state to its surroundings \cite{shaham2019implementation}.
The Kraus operators for phase flip quantum channel are as follows,
\begin{equation}
K_{0}=\sqrt{1-p/2}\begin{pmatrix}
1 & 0 \\
0 & 1
\end{pmatrix} \quad \text{and} \quad K_{1}=\sqrt{p/2} \begin{pmatrix}
1 &  0 \\
0 & -1
\end{pmatrix}
\end{equation}
 where, $p$ is the decoherence probability.\\
The coefficients of the evolved Bell-diagonal state after the first subsystem undergoes through this channel for n times are as follows,
\begin{equation}
    d_{1}^{\prime} = d_1(1-p)^{n} \quad d_2^{\prime} = d_{2}(1-p)^{n} \quad d_{3}^{\prime} = d_{3}
\end{equation}

\begin{figure}[hbt!]
\begin{subfigure}{.5\textwidth}
  \centering
  \includegraphics[width=0.8\linewidth]{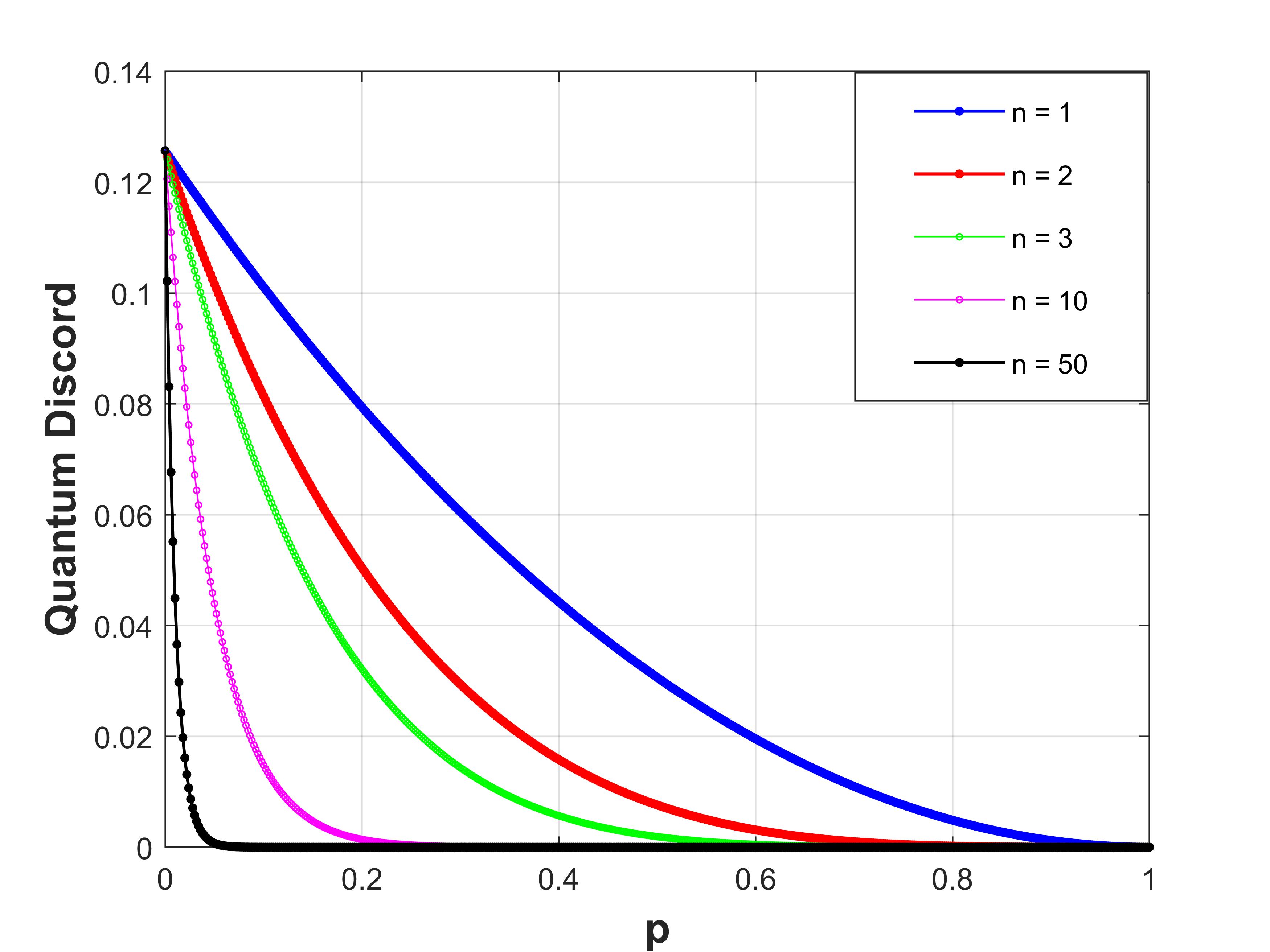}
  \label{fig0001}
\end{subfigure}%
\begin{subfigure}{.5\textwidth}
  \centering
  \includegraphics[width=.8\linewidth]{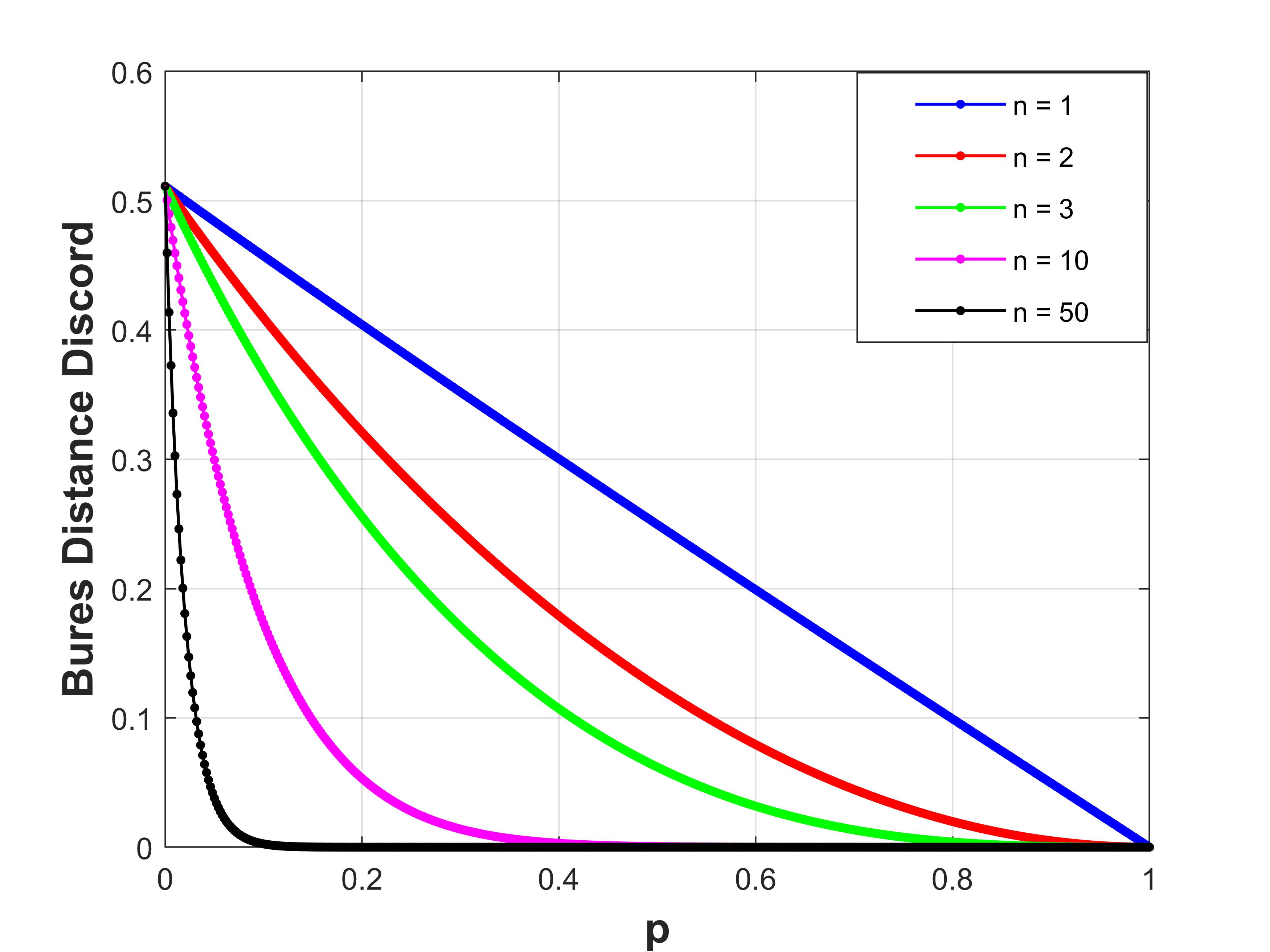}
  \label{fig0002}
\end{subfigure}
  \centering
\begin{subfigure}{.5\textwidth}
  \includegraphics[width=.8\linewidth]{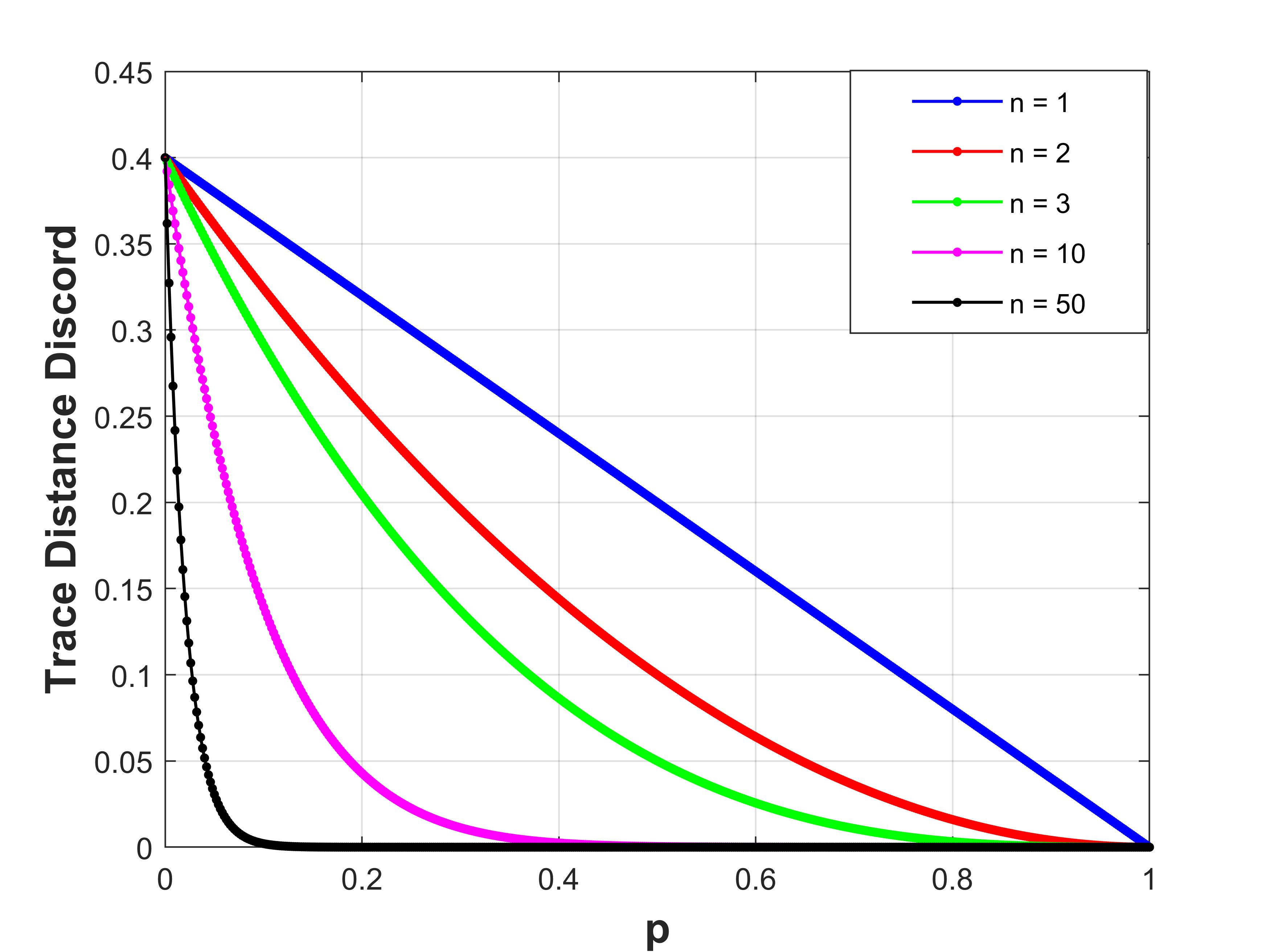}
  \label{fig0003}
\end{subfigure}
\caption{When the first subsystem undergoes through the Phase-Flip quantum channel $n$ times for Bell diagonal state ($d_{1} = 0.3$, $d_{2} = −0.4$ and $d_{3} = 0.56$) where colors Blue ($n = 1$), Red ($n = 2$), Green ($n = 3$), Magenta ($n = 10$), Black ($n = 50$)) (a) Quantum Discord (b) Bures Distance Discord (c) Trace Distance Discord.}
\label{fig300}
\end{figure}

The evolution of different entropic and geometric measures of quantum discord under phase-flip channel are shown in Fig.\ref{fig300}.

The discord measures considered here show no sudden change or freezing behaviour for the set of initial condition $|d_{1}|<|d_{2}|<|d_{3}|$. The presence or absence of sudden change behavior depends on the initial conditions. However for the above initial condition, the values of $|d_{1}^{\prime}|,|d_{2}^{\prime}|$ and $|d_{3}^{\prime}|$ preserve the magnitude order $|d_{1}^{\prime}|<|d_{2}^{\prime}|<|d_{3}^{\prime}|$ during the entire evolution for QD. In case of BDD and TDD, we do not observe any SCP. From the numerical expression of $A_{ijk}$, the term $A_{312}$ is greater than both the terms $A_{231}$ and $A_{123}$ for the entire range of $p$. In case of TDD, $|d^{\prime}_{2}|$ always remains the intermediate value as the decay factors of $|d^{\prime}_{1}|$ and $|d^{\prime}_{2}|$ are the same and $|d_{3}^{\prime}|$ has the maximum value. The correlation function $|d_{2}^{\prime}|$ remains the intermediate value during the entire evolution under this channel and hence, we do not observe any SCP. We observe that QD decays smoothly, whereas BDD and TDD decay sharply for $n = 1$. For higher values of $n$, all the three discord measures have a smooth decline.



\newpage
\subsection{Bit Phase Flip Quantum Channel}
The combination of bit flip and phase flip channel is known as a bit flip quantum channel. The action of this channel is equivalent to operating $\sigma_{y}$ (Pauli Y) on any single qubit state $\ket{\psi}$. 
On the individual bits, the bit-phase flip channel can be described as follows: \begin{equation}
\ket{0} \rightarrow +i \ket{1} \quad \text{and} \quad \ket{1} \rightarrow -i \ket{0}
\end{equation}
The Kraus operators for bit flip quantum channel are, 
\begin{equation}
K_{0}=\sqrt{1-p/2} \begin{pmatrix}
1 & 0 \\
0 & 1
\end{pmatrix} \text{and} \quad K_{1}=\sqrt{p/2}  \begin{pmatrix}
0 &  -i \\
i &  0
\end{pmatrix}
\end{equation}
where, $p$ is the decoherence probability.
After the first subsystem undergoes through this channel $n$ times, the coefficients of the evolved Bell-diagonal state obtained are, 
\begin{equation}
    d_{1}^{\prime} = d_1(1-p)^{n}, \quad d_2^{\prime} = d_{2}, \quad d_{3}^{\prime} = d_{3}(1-p)^{n}
\end{equation}

\begin{figure}[hbt!]
\begin{subfigure}{.5\textwidth}
  \centering
  \includegraphics[width=0.8\linewidth]{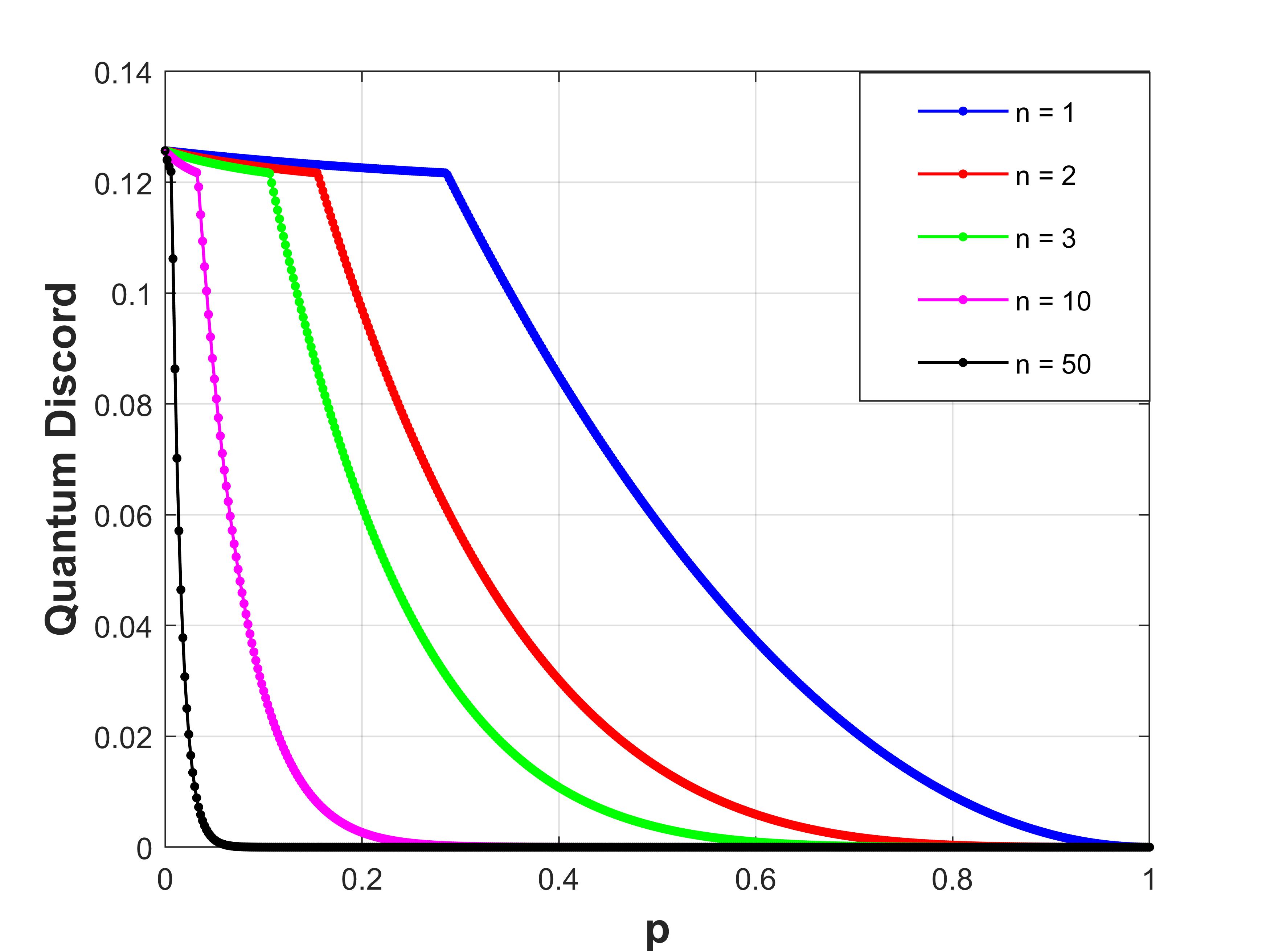}
  \label{fig00001}
\end{subfigure}%
\begin{subfigure}{.5\textwidth}
  \centering
  \includegraphics[width=.8\linewidth]{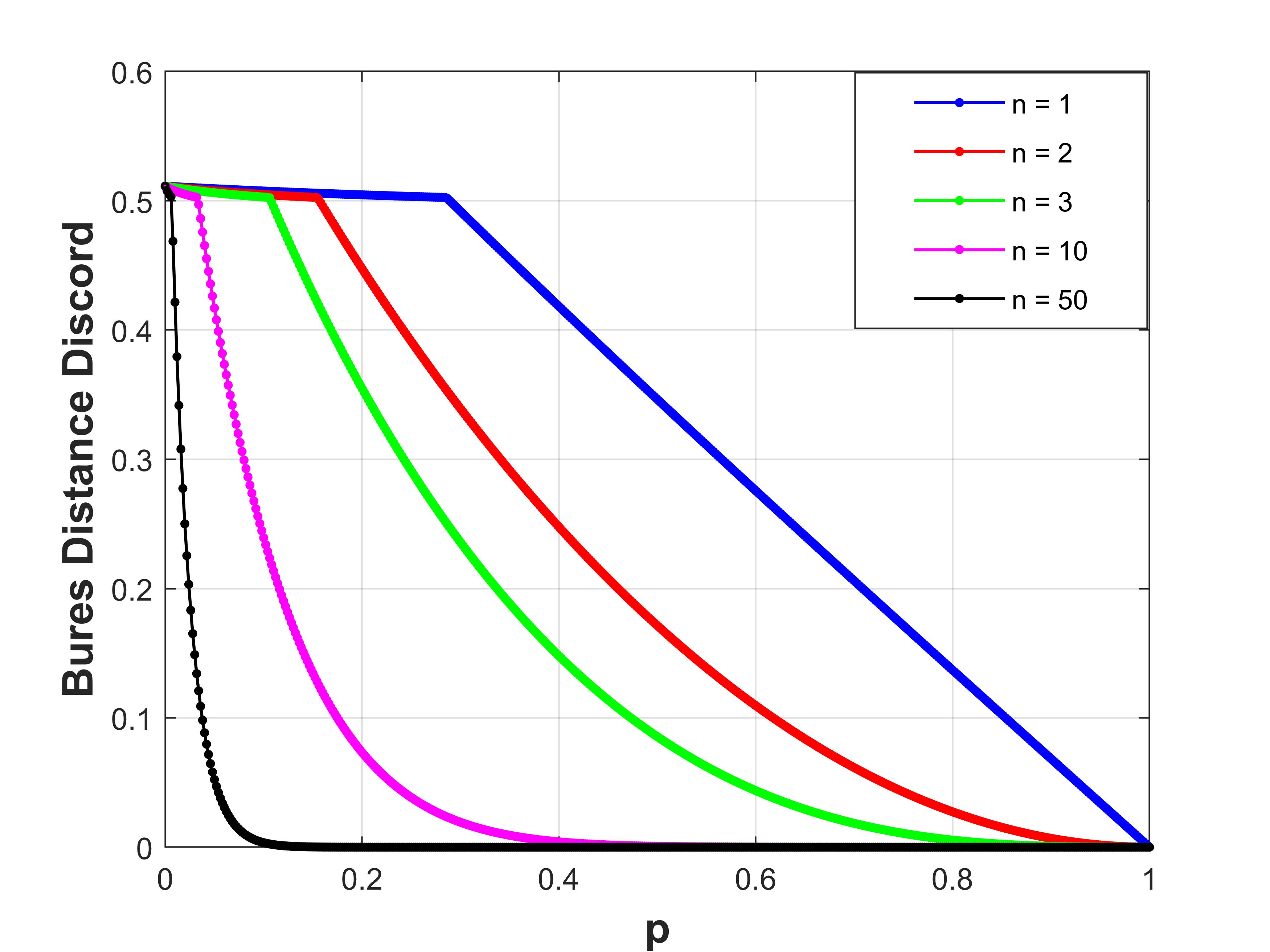}
  \label{fig00002}
\end{subfigure}
  \centering
\begin{subfigure}{.5\textwidth}
  \includegraphics[width=.8\linewidth]{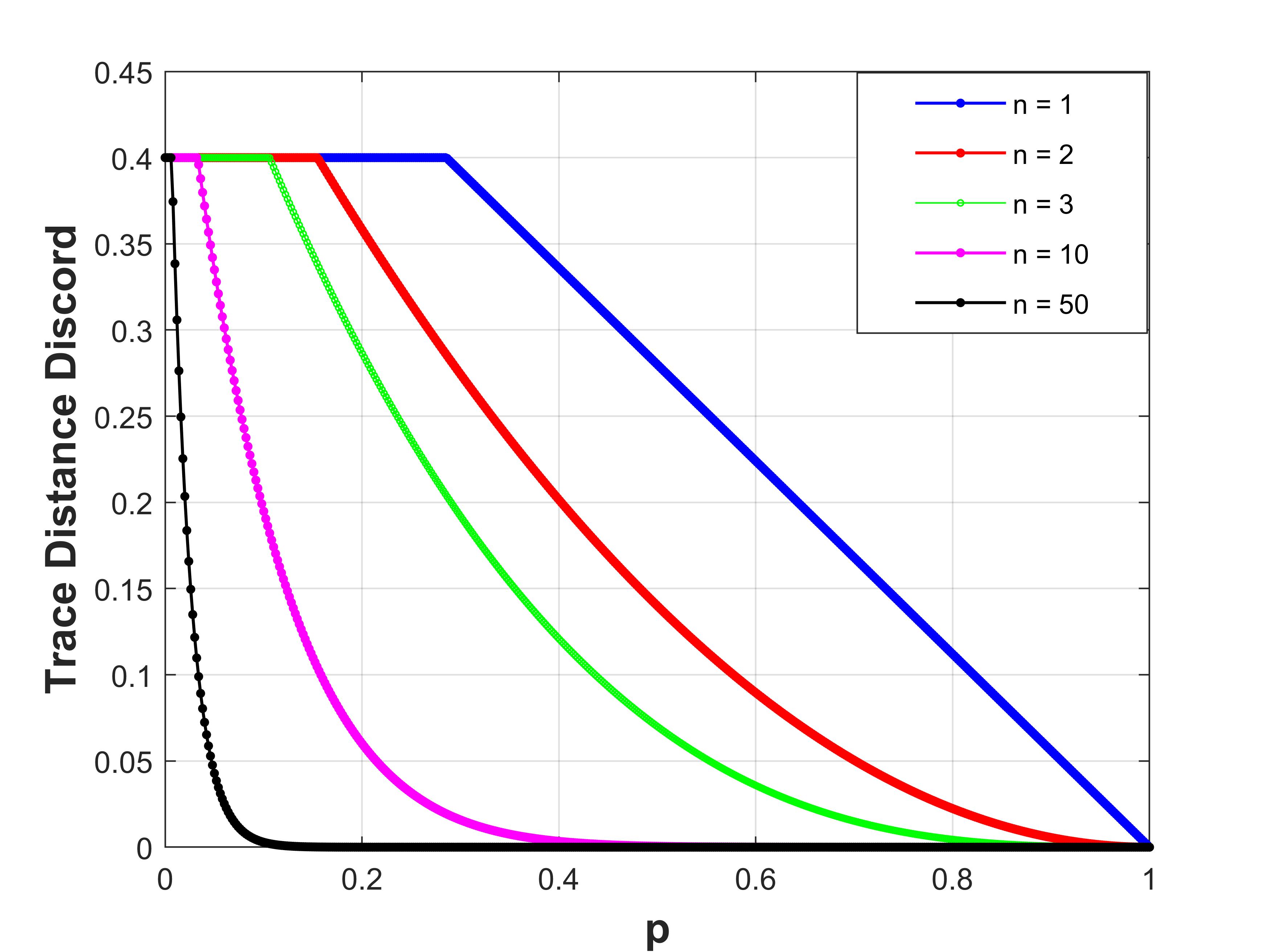}
  \label{fig00003}
\end{subfigure}
\caption{When the first subsystem undergoes through the Bit Phase Flip quantum channel $n$ times for Bell diagonal state ($d_{1} = 0.3$, $d_{2} = −0.4$ and $d_{3} = 0.56$). The evolution (a) Quantum Discord (b) Bures Distance Discord (c) Trace Distance Discord show sudden change phenomena at $p_{sc}^{1} = 0.28$ (Blue, $n =1$) , $p_{sc}^{2} = 0.15$ (Red, $n =2$), $p_{sc}^{3} = 0.10$ (Green, $n =3$), $p_{sc}^{10} = 0.03$ (Magenta, $n =10$) and $p_{sc}^{50} = 0.006$ (Black, $n = 50$).}
\label{fig3000}
\end{figure}

In this channel, we observe that all the discord measures show sudden change behavior only once. For the initial condition $|d_{1}|<|d_{2}|<|d_{3}|$, we observe that the correlation function $|d_{3}^{\prime}|$ decays with increase in $p$ and crosses the constant value of $|d_2^{\prime}|$ once.  In the case of quantum discord, the maximum value from the set \{$|d_{1}^{\prime}|,|d_{2}^{\prime}|,|d_{3}^{\prime}|$\} is $|d_{3}^{\prime}|$ till it becomes less than $|d_{2}^{\prime}|$. When $|d_{3}^{\prime}| = |d_{2}^{\prime}|$, we observe the SCP at the point ($p_{sc}^{Q}$),
\begin{equation}
p_{sc}^{Q} = 1 - \left(\frac{|d_{2}|}{|d_{3}|}\right)^{\frac{1}{n}}
\end{equation}
In the case of BDD, we observe the SCP when $A_{312}$ becomes equal to $A_{231}$. This leads to the condition $d_{2}^{\prime} = \pm d_{3}^{\prime}$ which is same as that of QD. However, we have considered the initial value of $d_{2}$ to be negative and $d_{3}$ to be positive, so we observe the SCP at $d_{2}^{\prime} = - d_{3}^{\prime}$. \\
In the case of TDD, $|d_{2}^{\prime}|$ is constant under this channel.
We observe SCP when the decaying correlation function $|d_{3}^{\prime}|$ becomes less than $|d_{2}^{\prime}|$ and $|d_{3}^{\prime}|$ takes the intermediate value for further evolution. We observe the sudden change at $|d_{2}^{\prime}|=|d_{3}^{\prime}|$. The value of decoherence parameter $p$ at which SCP occurs is,
\begin{equation}
    p_{sc}^{T} = 1 - \left(\frac{|d_{2}|}{|d_{3}|}\right)^\frac{1}{n}
\end{equation}
and duration upto which TDD is frozen from time $t = 0$ is,
\begin{equation}
     t_{frozen} = \frac{1}{n\gamma}ln\left(\frac{|d_{3}|}{|d_{2}|}\right)
\end{equation}


\subsection{Depolarising Quantum Channel}

The depolarising channel replaces the initial state of the qubit by a maximally mixed state $I/2$. It encompasses all Pauli noises such as bit flip, phase flip, and bit-phase flip which makes it the most detrimental noise \cite{Aolita}. 
If the depolarising error occurs, then the initial quantum state $\ket{\psi}$ evolves as the linear combination of $\sigma_{x}\ket{\psi}$, $\sigma_{y}\ket{\psi}$, $\sigma_{z}\ket{\psi}$ each, with equal probability \cite{preskill}.
It is worth mentioning that the depolarising channel reduces the volume of the Bloch sphere, implying that the original state is replaced with a maximally mixed state \cite{10.5555/1972505}. In addition, the depolarising channel is primarily used to investigate how decoherence impacts the qubit's symmetry.  
This channel randomises the qubit and changes the spin polarisation of the particles in physical circumstances \cite{preskill}.
The Kraus operators for depolarising channel are,

\begin{equation}
K_{0}=\sqrt{1-p}\begin{pmatrix}
1 & 0 \\
0 & 1
\end{pmatrix} \quad \text{and} \quad K_{1}=\sqrt{\frac{p}{3}}\begin{pmatrix}
0 &  1 \\
1 &  0
\end{pmatrix}
\end{equation}
\begin{equation}
K_{2}=\sqrt{\frac{p}{3}}\begin{pmatrix}
0 & -i \\
i &  0
\end{pmatrix} \quad \text{and} \quad K_{3}=\sqrt{\frac{p}{3}}\begin{pmatrix}
1 &  0 \\
0 & -1
\end{pmatrix}
\end{equation}
where, $p$ is the probability that the system gets affected by each of the depolarising errors i.e., bit flip, phase flip, and bit-phase flip, provided each error is equally likely. 


The coefficients of the Bell-diagonal state after the first subsystem undergoes through this channel $n$ times are,
\begin{equation}
    d_{1}^{\prime} = d_{1}\left(1-\frac{4p}{3}\right)^{n} \quad d_2^{\prime} = d_2\left(1-\frac{4p}{3}\right)^{n} \quad d_{3}^{\prime} = d_{3}\left(1-\frac{4p}{3}\right)^{n}
\end{equation}


\begin{figure}[hbt!]
\begin{subfigure}{.5\textwidth}
  \centering
  \includegraphics[width=0.8\linewidth]{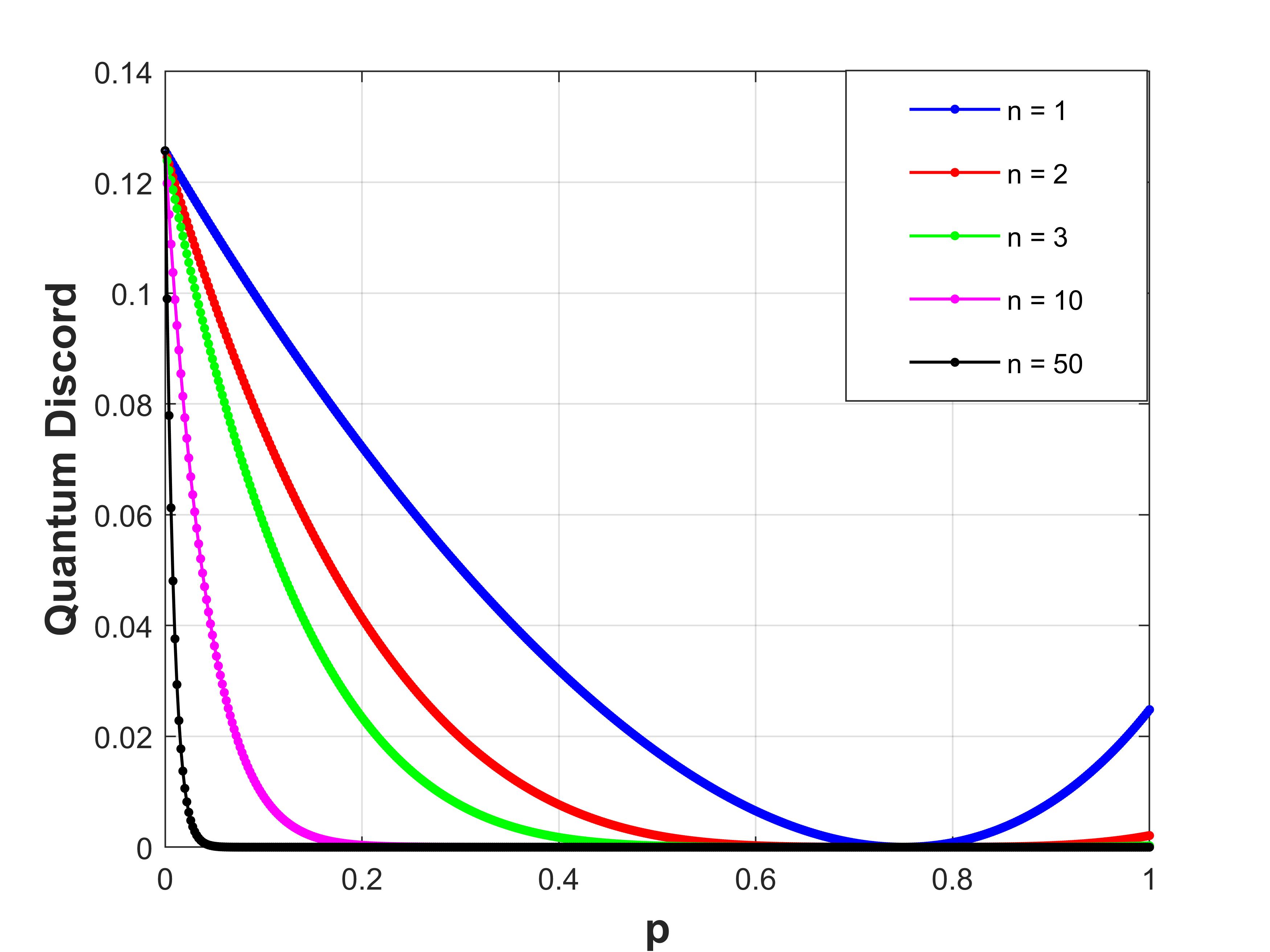}
  \label{fig000001}
\end{subfigure}%
\begin{subfigure}{.5\textwidth}
  \centering
  \includegraphics[width=.8\linewidth]{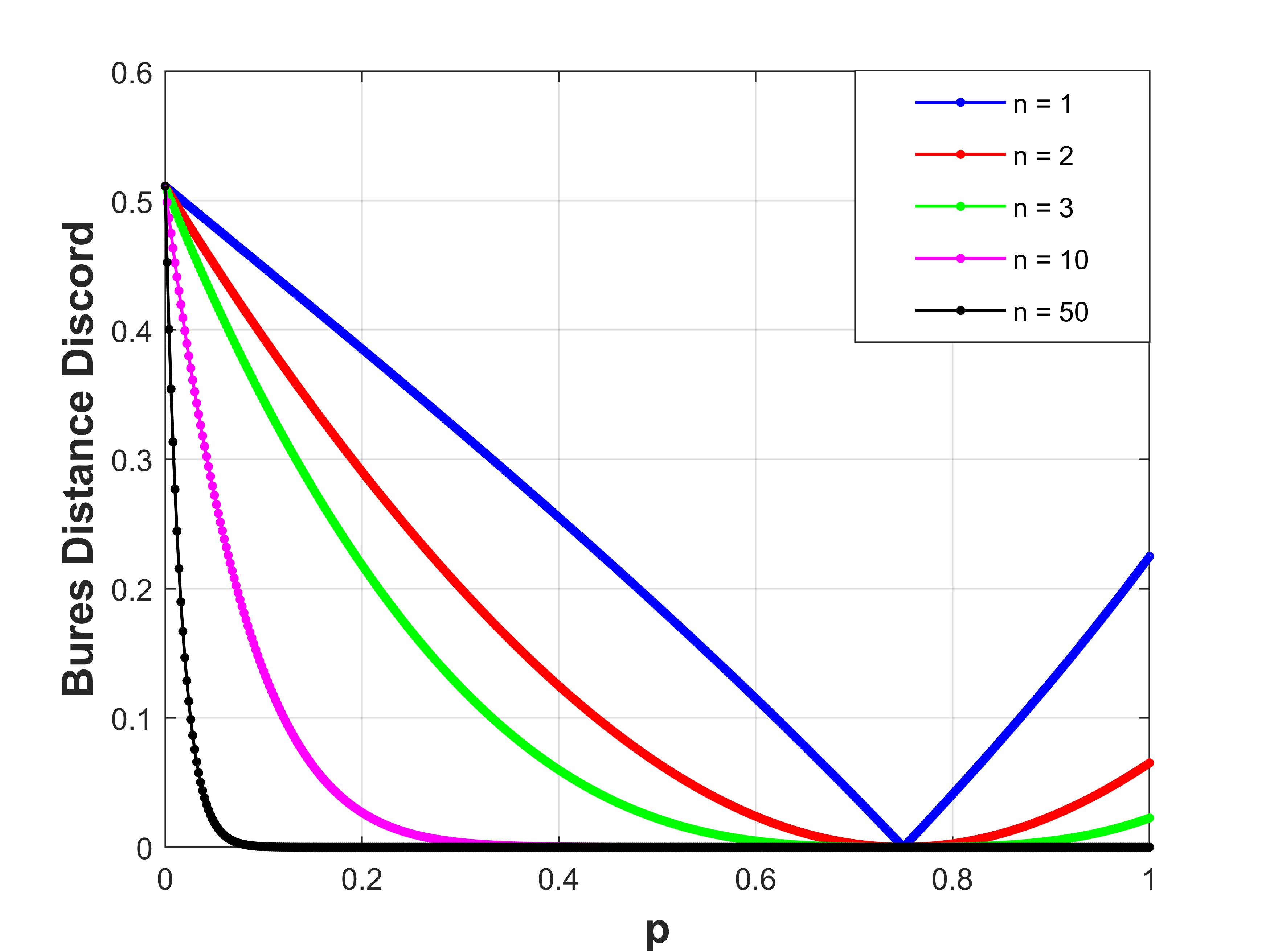}
  \label{fig000002}
\end{subfigure}
  \centering
\begin{subfigure}{.5\textwidth}
  \includegraphics[width=.8\linewidth]{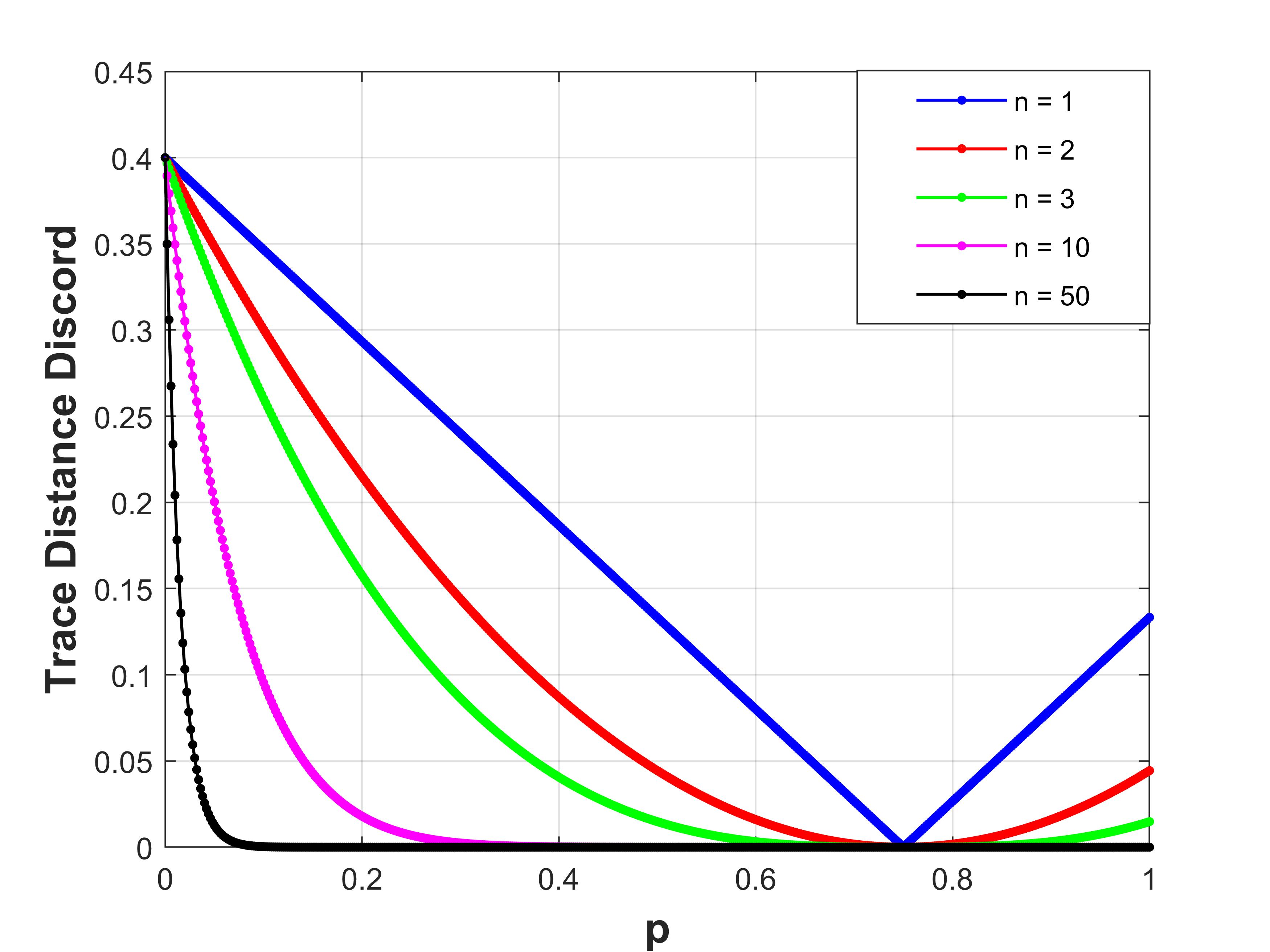}
  \label{fig000003}
\end{subfigure}
\caption{When the first subsystem undergoes through the Depolarising Channel $n$ times for Bell diagonal state ($d_{1} = 0.3$, $d_{2} = −0.4$ and $d_{3} = 0.56$). The colors, Blue ($n = 1$), Red ($n = 2$), Green ($n = 3$), Magenta ($n = 10$), Black ($n = 50$)) The evolution of (a) Quantum Discord (b) Bures Distance Discord (c) Trace Distance Discord show revival after $p =\frac{3}{4}$.}
\label{fig30000}
\end{figure}
From the Fig. \ref{fig30000}, we observe that the discord measures revive after a particular value of the decoherence probability ($p=3/4$). At this point, the system loses all its quantum correlations implying the state has become completely classical. 
We do not encounter SCP for any of the discord measures. The symmetrical nature of decay factor of the coefficients and the particular initial condition $|d_{1}|<|d_{2}|<|d_{3}|$ retain the order of the evolved coefficients as $|d_{1}^{\prime}|<|d_{2}^{\prime}|<|d_{3}^{\prime}|$. In the case of QD, $|d_{3}^{\prime}|$ and in the case of BDD, $A_{321}$ has the maximum value during the evolution, so we do not see any crossing of the determining factors/coefficients responsible for SCP. For TDD, $|d_{2}^{\prime}|$ takes the intermediate value during the entire evolution process.\\
A comparison of the evolution of the discord measures for $n= 1$ reveals that BDD and TDD decay sharply as compared to QD. Both geometric measures indicate a sharp recovery, but QD shows a gradual decrease and a gradual recovery. However, the decay of geometric discord measures becomes smoother for higher $n$ values.


\subsection{Generalized Amplitude Damping Quantum Channel (GADC)}
GADC is a two-parameter, qubit to qubit channel which includes both damping and environmental noise. It is commonly referred to as a rank four-channel since it has four Kraus operators. It depicts the dynamics of a two-level system in contact with a temperature bath of finite temperature. Although at zero temperature, the GADC reduces to amplitude damping channel. It is important to note that the GADC channel can be viewed as a qubit analog of the bosonic thermal channel \cite{Phy}. The Kraus operators for generalized amplitude damping quantum channel are, 
\begin{equation}
K_{0}=\sqrt{p} \begin{pmatrix}
1 & 0 \\
0 & \sqrt{1-\gamma}
\end{pmatrix} \quad \text{and} \quad K_{1}=\sqrt{1-p} \begin{pmatrix}
\sqrt{1-\gamma} &  0 \\
0 &  1
\end{pmatrix}
\end{equation}
\begin{equation}
K_{2}=\sqrt{p} \begin{pmatrix}
0 & \sqrt{\gamma} \\
0 & 0
\end{pmatrix} \quad \text{and} \quad K_{3}=\sqrt{1-p} \begin{pmatrix}
0 &  0 \\
\sqrt{\gamma} &  0
\end{pmatrix}
\end{equation}
where, $p$ is the decoherence probability and the parameter $\gamma$ signifies decay rate. 
The coefficients of the Bell-diagonal form after the first subsystem undergoes through this channel $n$ times, 
\begin{equation}
    d_{1}^{\prime} = d_{1}(1-p)^{\frac{n}{2}}, \quad d_2^{\prime} = d_2(1-p)^{\frac{n}{2}}, \quad d_{3}^{\prime} = d_{3}(1-p)^{n}
\end{equation}
In our analysis, we have fixed $p = \frac{1}{2}$ and replaced $\gamma$ by $p$. This combination preserves the Bell-diagonal form of the initial state for which discord measures have a closed-form expression.
\begin{figure}[hbt!]
\begin{subfigure}{.5\textwidth}
  \centering
  \includegraphics[width=0.8\linewidth]{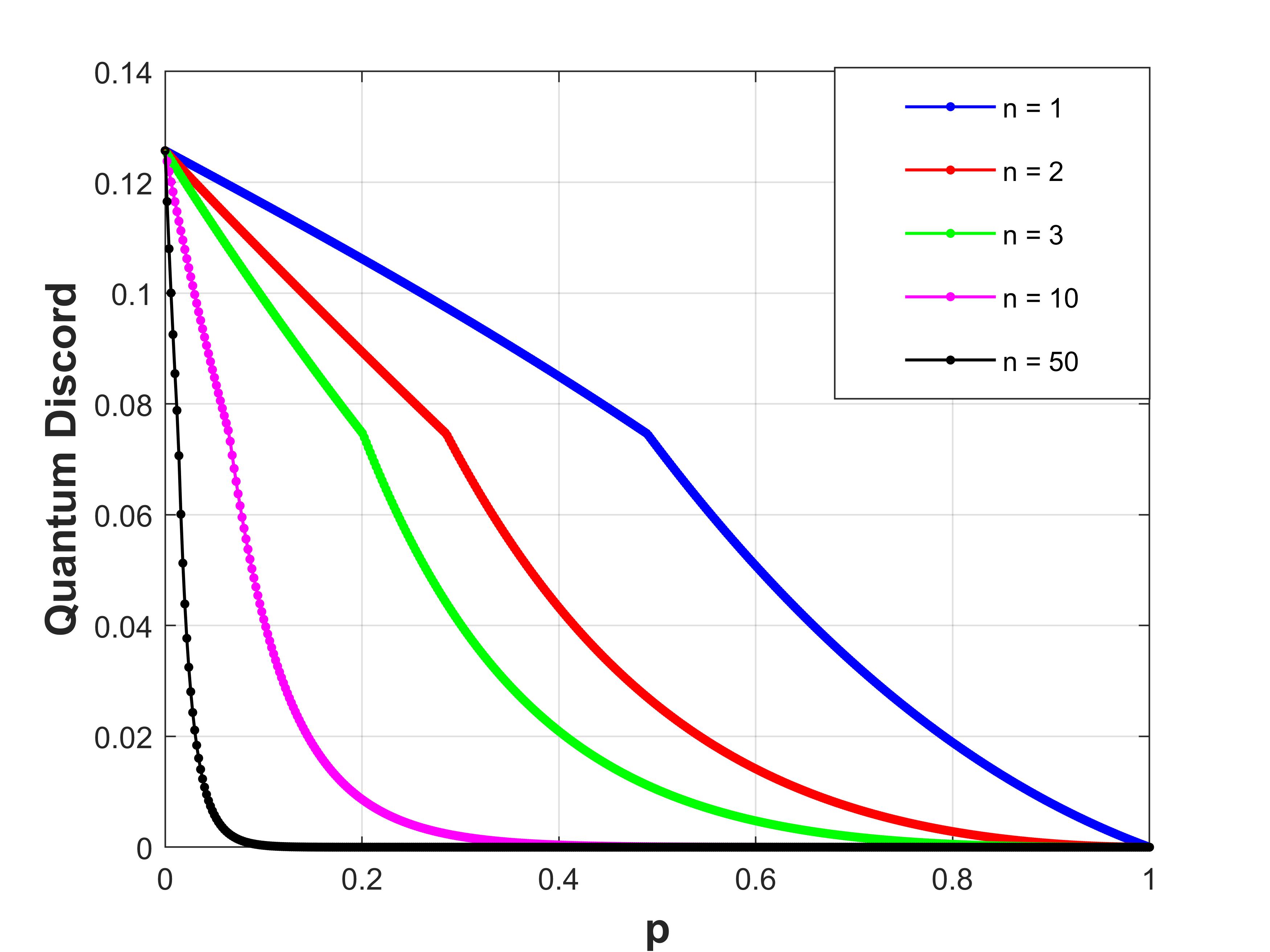}
  \label{fig041}
\end{subfigure}%
\begin{subfigure}{.5\textwidth}
  \centering
  \includegraphics[width=.8\linewidth]{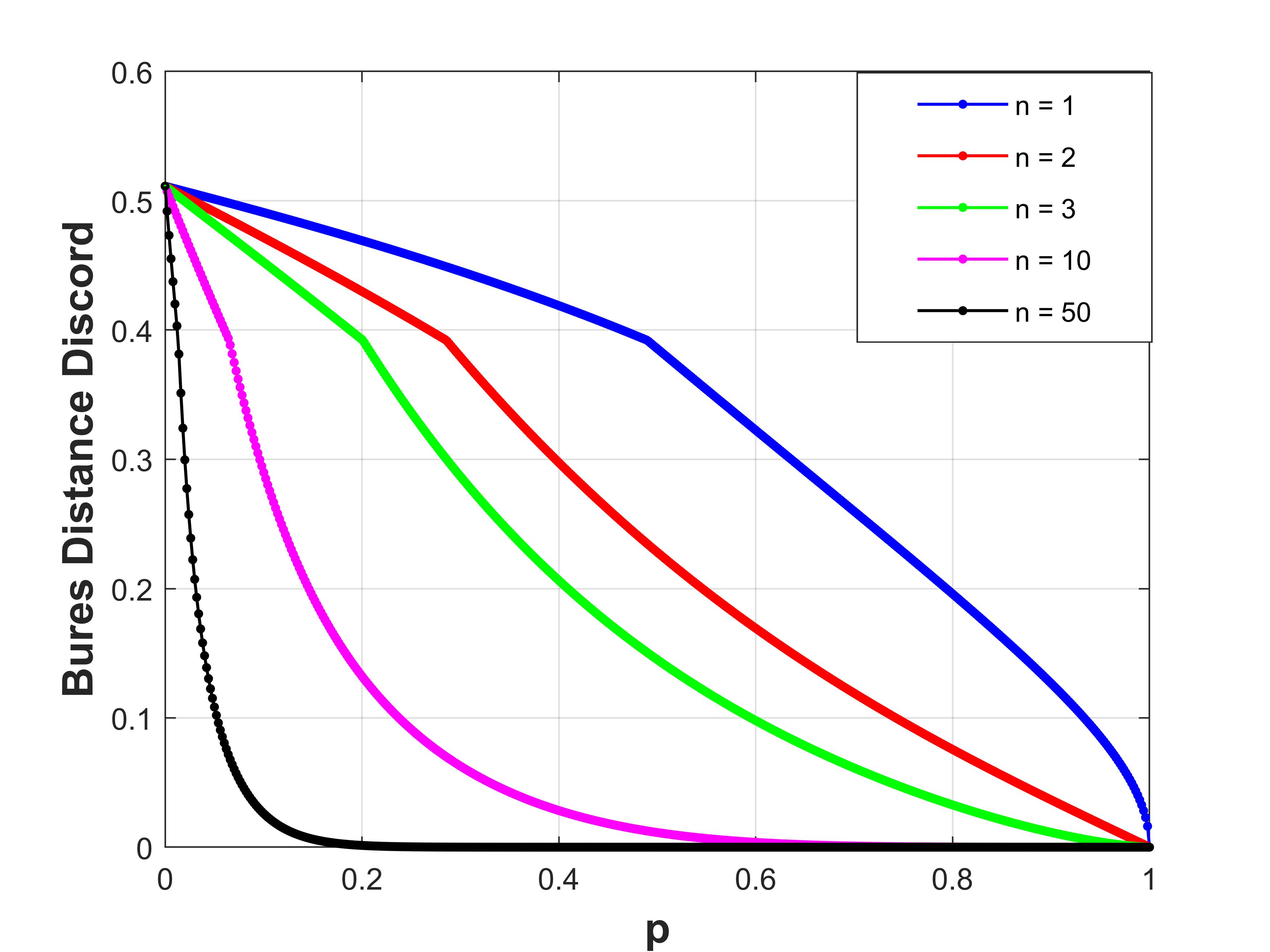}
  \label{fig05002}
\end{subfigure}
  \centering
\begin{subfigure}{.5\textwidth}
  \includegraphics[width=.8\linewidth]{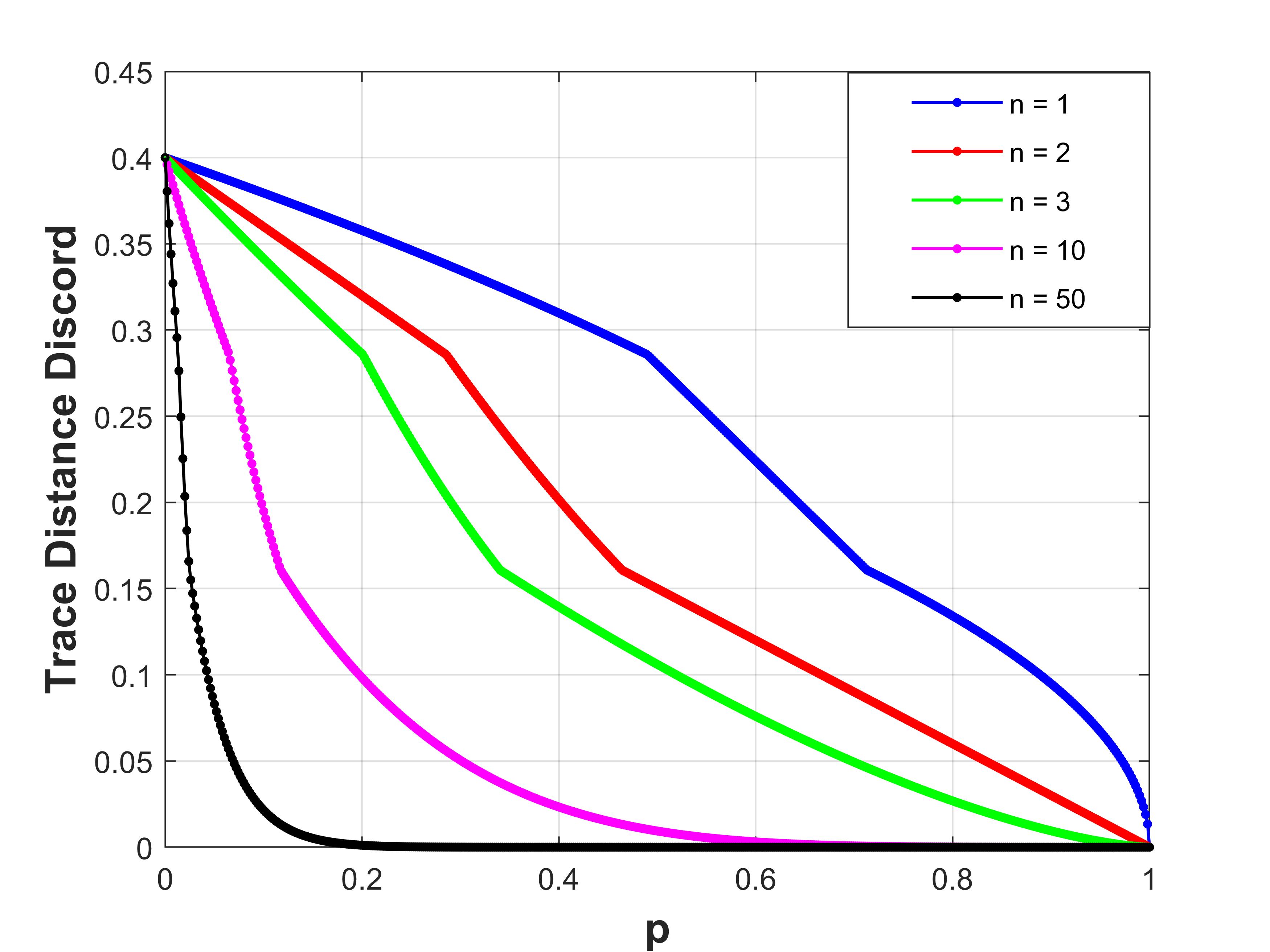}
  \label{fig00503}
\end{subfigure}
\caption{When first subsystem undergoes through the Generalized Amplitude Damping Quantum Channel $n$ times for Bell diagonal state ($d_{1} = 0.3$, $d_{2} = −0.4$ and $d_{3} = 0.56$)  (a) Quantum Discord (b) Bures Distance Discord show SCP at $p_{sc}^{1} = 0.49$ (Blue, $n = 1$), $p_{sc}^{2} = 0.28$ (Red, $n = 2$), $p_{sc}^{3} = 0.20$ (Green, $n = 3$), $p_{sc}^{10} = 0.06$ (Magenta, $n = 10$) and $p_{sc}^{50} = 0.01$ (Black, $n = 50$) (c) Trace Distance Discord shows two SCP, the first SCP matches with that of QD and BDD, the second SCP occurs at $p_{sc_{2}}^{1} = 0.71$, $p_{sc_{2}}^{2} = 0.46$, $p_{sc_{2}}^{3} = 0.33$, $p_{sc_{2}}^{10} = 0.12$ and $p_{sc_{2}}^{50} = 0.02$.}
\label{fig3400}
\end{figure}
Similar to previous cases, for the set of initial conditions considered in our work, we observe a single point of sudden change when $|d_{3}^{\prime}|$ becomes equal to $|d_{2}^{\prime}|$ for QD. This occurs at,
\begin{equation}
    p_{sc}^{Q} = 1 - \left(\frac{|d_{2}|}{|d_{3}|}\right)^\frac{2}{n}
\end{equation}
Similarly, we observe the SCP for BDD when $A_{231}$ crosses $A_{312}$ and becomes equal to $A_{312}$. This yields the condition $|d_{2}^{\prime}| = |d_{3}^{\prime}|$ which is same as that of quantum discord.\\
In the case of TDD, we observe two sudden changes when $|d_{3}^{\prime}| = |d_{2}^{\prime}|$ and $|d_{3}^{\prime}| = |d_{1}^{\prime}|$. The corresponding $p$ values for DSCP are,
\begin{equation}
    p_{sc_{1}}^{T} = 1 - \left(\frac{|d_{2}|}{|d_{3}|}\right)^\frac{2}{n} \quad \text{and} \quad p_{sc_{2}}^{T} = 1 - \left(\frac{|d_{1}|}{|d_{3}|}\right)^\frac{2}{n}
\end{equation}
The time interval between points of sudden is given by,
\begin{equation}
    \Delta t_{sc} = \frac{2}{n\gamma}ln\left(\frac{|d_{2}|}{|d_{1}|}\right)
\end{equation}
It is interesting to note that we do not observe freezing of TDD in this channel as none of the correlation functions are constant during the evolution.

\section{Dynamics of discord measures under Bi-side Markovian Channels}

We now investigate the scenarios when both the subsystems are subjected to locally evolving channels of same and different types for $n$ times that retain the Bell-diagonal form of the initial state. We systematically examine the following two cases,
\begin{enumerate}[label=\Roman*]
\item When both the subsytems $A$ and $B$ evolve under the same decoherence channels with different decoherence rates.
    
\item When both the subsystems $A$ and $B$ evolve under different decoherence channels with different decoherence rates.
\end{enumerate}

\subsection{Discord dynamics of Bell-Diagonal States under Bi-Side Markovian Channels of the same types}

Here, we consider the dynamics of discord measures of a two qubit Bell diagonal state undergoing through two independent local Markovian channels of the same type.
The three cases which preserve the Bell-diagonal form are : two independent local bit flip channels (BF-BF), local phase flip channels (PF-PF), local bit phase flip channels (BPF-BPF) acting on the subsystems of the joint Bell diagonal state.

\subsubsection{Local Bit flip channels ($BF^{A}-BF^{B}$)}
We consider the case when both the qubits are independently subjected to the bit flip channel for multiple times. The Kraus operators for this channel are,

\begin{equation}
K_{0}^{(A)}=\sqrt{1-\frac{p}{2}} I^{A} \otimes I^{B}, \quad K_{1}^{(A)}=\sqrt{\frac{p}{2}} \sigma_{x}^{A} \otimes I^{B}
\end{equation}

\begin{equation}
    K_{0}^{(B)}=I^{A} \otimes \sqrt{1-\frac{q}{2}} I^{B}, \quad K_{1}^{(B)}=I^{A} \otimes \sqrt{\frac{q}{2}} \sigma_{x}^{B}
\end{equation}
where, $p = 1 - exp(-\gamma_{1}t)$ and $q = 1 - exp(-\gamma_{2}t)$ are the decoherence probabilities and $\gamma_{1}, \gamma_{2}$ are decoherence rates of the channels acting on qubit $A$ and qubit $B$ respectively. The operators, $K_{0}^{(A)}$ and $K_{1}^{(A)}$ are the Kraus operators for the first subsystem and $K_{0}^{(B)}$, $K_{1}^{(B)}$ are operators for the second subsystem of the composite two qubit state. The corresponding coefficients of Bell diagonal states after undergoing through these operations for $n$ times are,
\begin{equation}
d_{1}^{\prime} = d_{1}, \quad d_{2}^{\prime} = (1-p)^{n}(1-q)^{n}d_{2}, \quad d_{3}^{\prime} = (1-p)^{n}(1-q)^{n} d_{3}
\end{equation}

\begin{figure}[hbt!]
  \subfloat[]{
	\begin{minipage}[c][1\width]{
	   0.3\textwidth}
	   \centering
	   \includegraphics[width=1.0\textwidth]{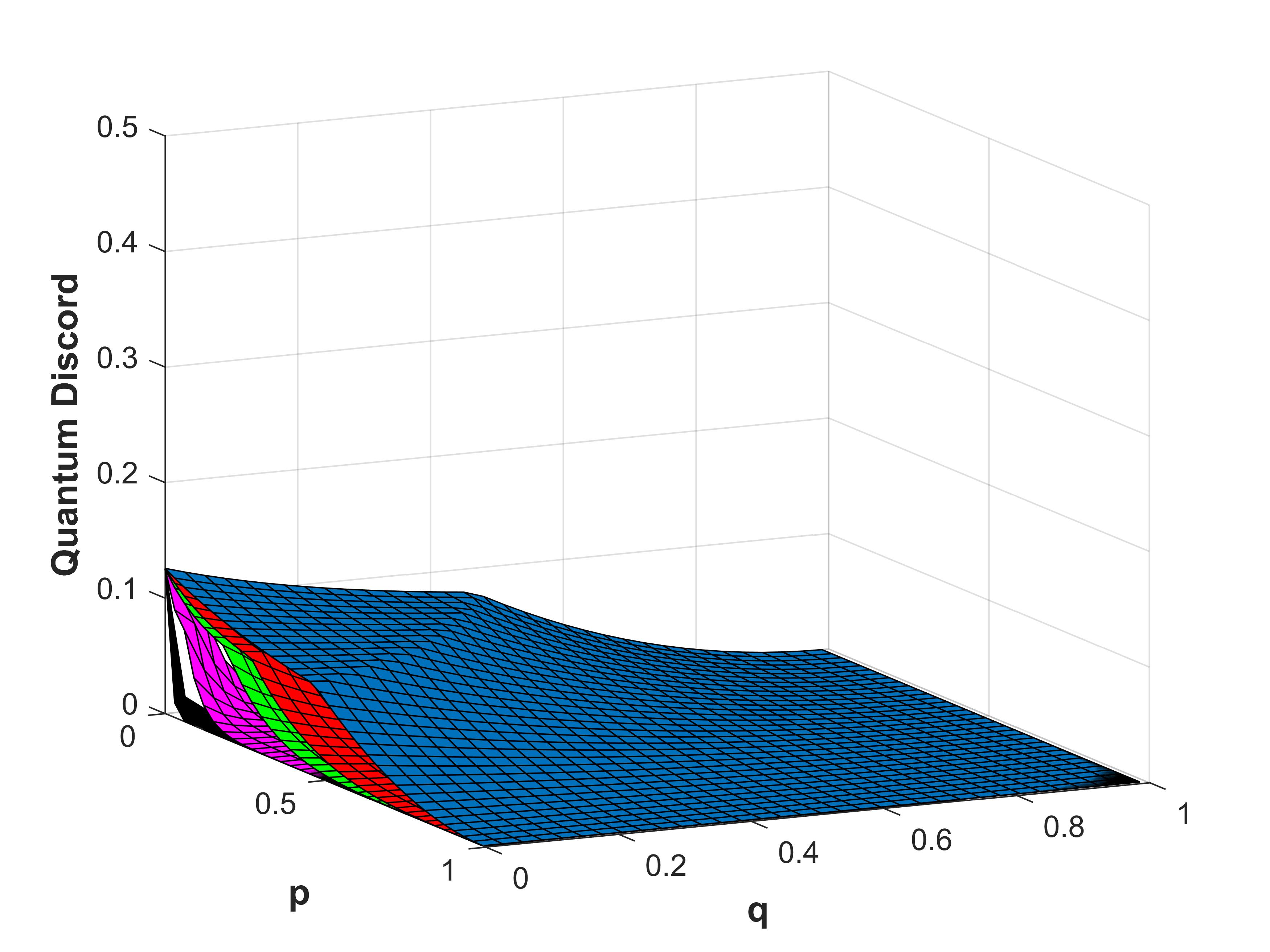}
	\end{minipage}}
 \hfill 	
  \subfloat[]{
	\begin{minipage}[c][1\width]{
	   0.3\textwidth}
	   \centering
	   \includegraphics[width=1.0\textwidth]{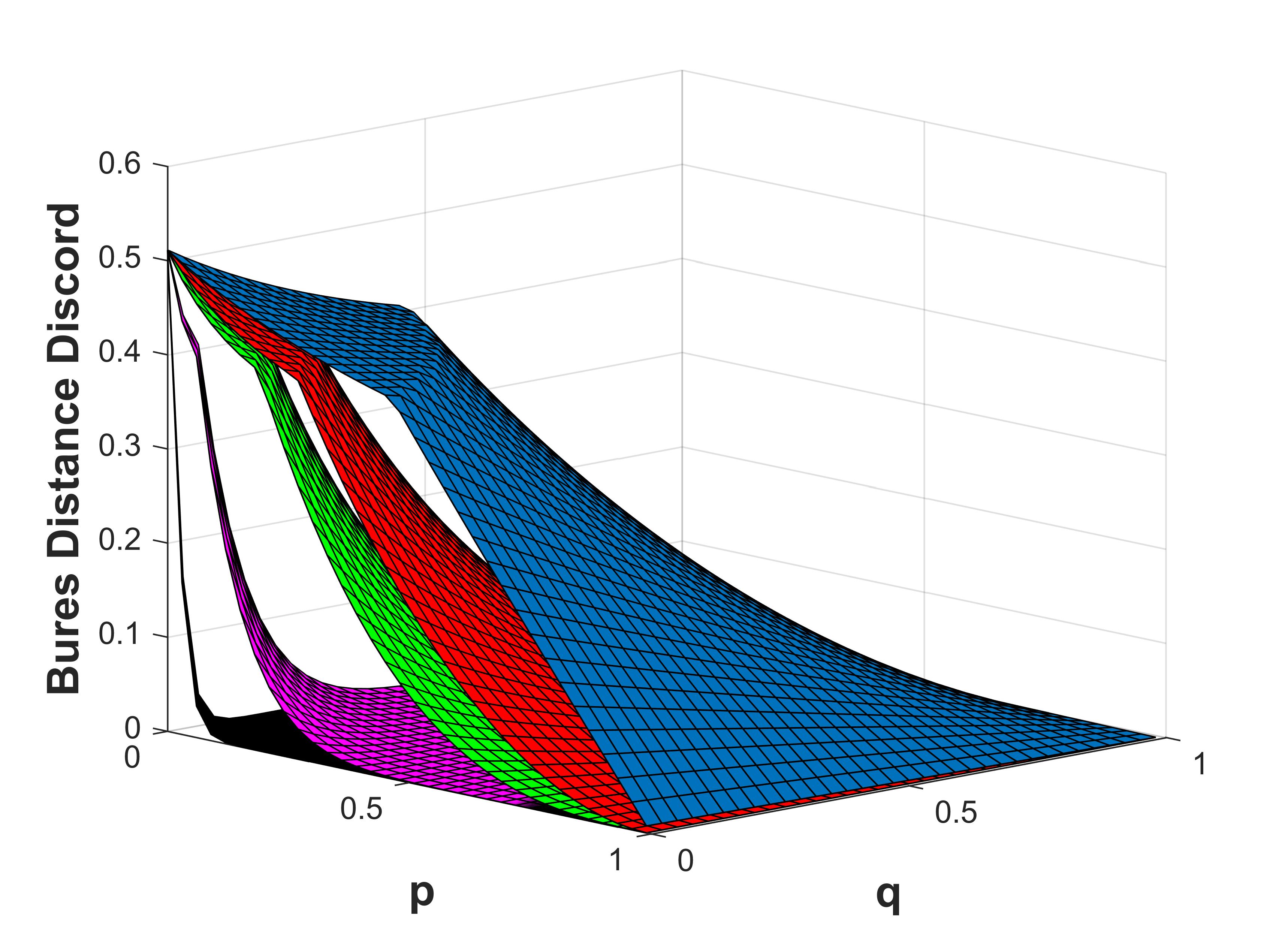}
	\end{minipage}}
 \hfill	
  \subfloat[]{
	\begin{minipage}[c][1\width]{
	   0.3\textwidth}
	   \centering
	   \includegraphics[width=1.0\textwidth]{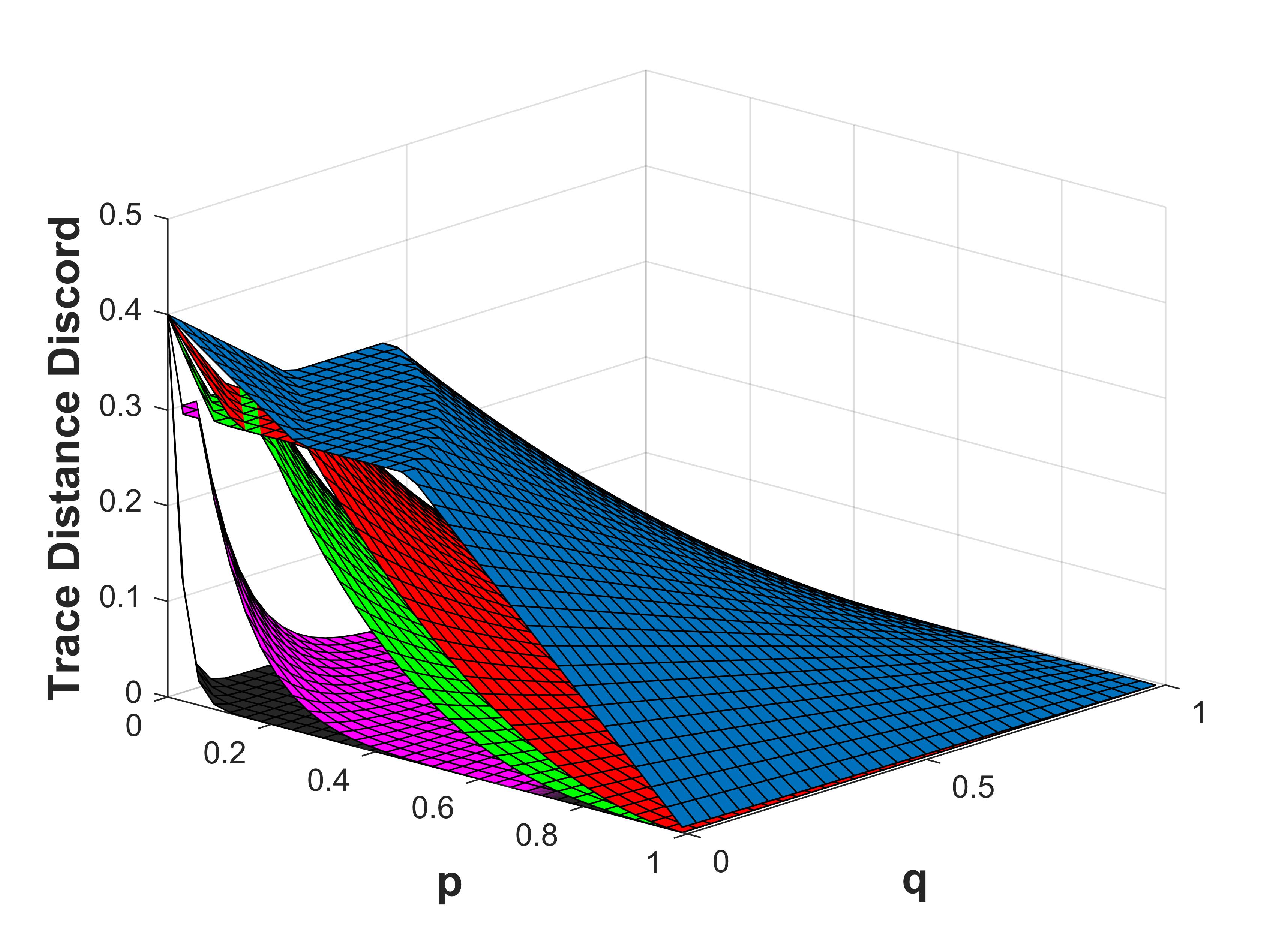}
	\end{minipage}}
\caption{When both the subsystems undergo through the same quantum channel $n$ times for Bell diagonal state ($d_{1} = 0.3$, $d_{2} = −0.4$ and $d_{3} = 0.56$) : Bit Flip Quantum Channel where colors Blue ($n = 1$), Red ($n = 2$), Green ($n = 3$), Magenta ($n = 10$), Black ($n = 50$)) (a) Quantum Discord (b) Bures Distance Discord (c) Trace Distance Discord}.
\label{fig150}
\end{figure}

\begin{figure}[ht]
  \subfloat[]{
	\begin{minipage}[c][1\width]{
	   0.45\textwidth}
	   \centering
	   \includegraphics[width=1.0\textwidth]{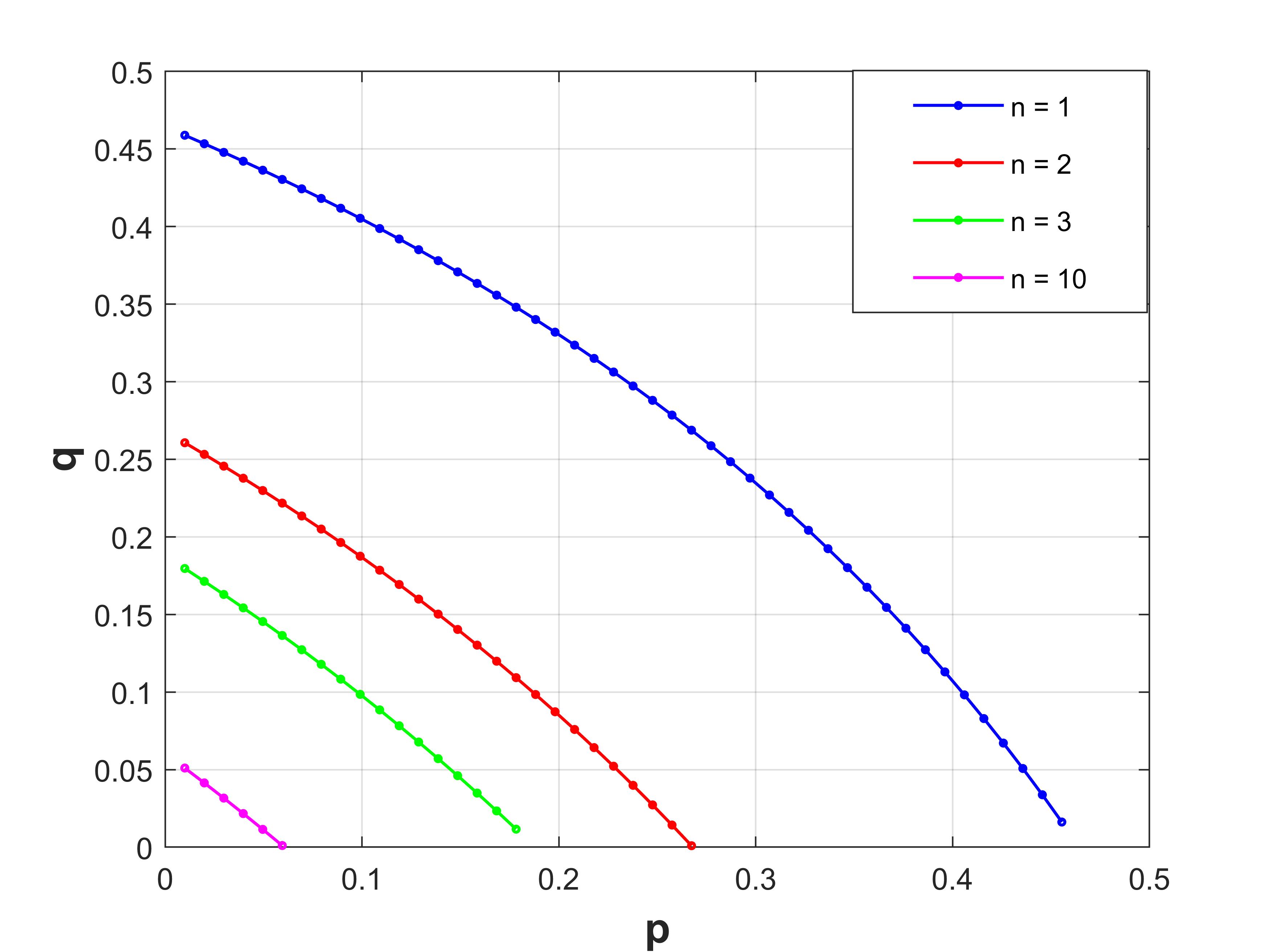}
	\end{minipage}}
 \hfill 	
  \subfloat[]{
	\begin{minipage}[c][1\width]{
	   0.45\textwidth}
	   \centering
	   \includegraphics[width=1.0\textwidth]{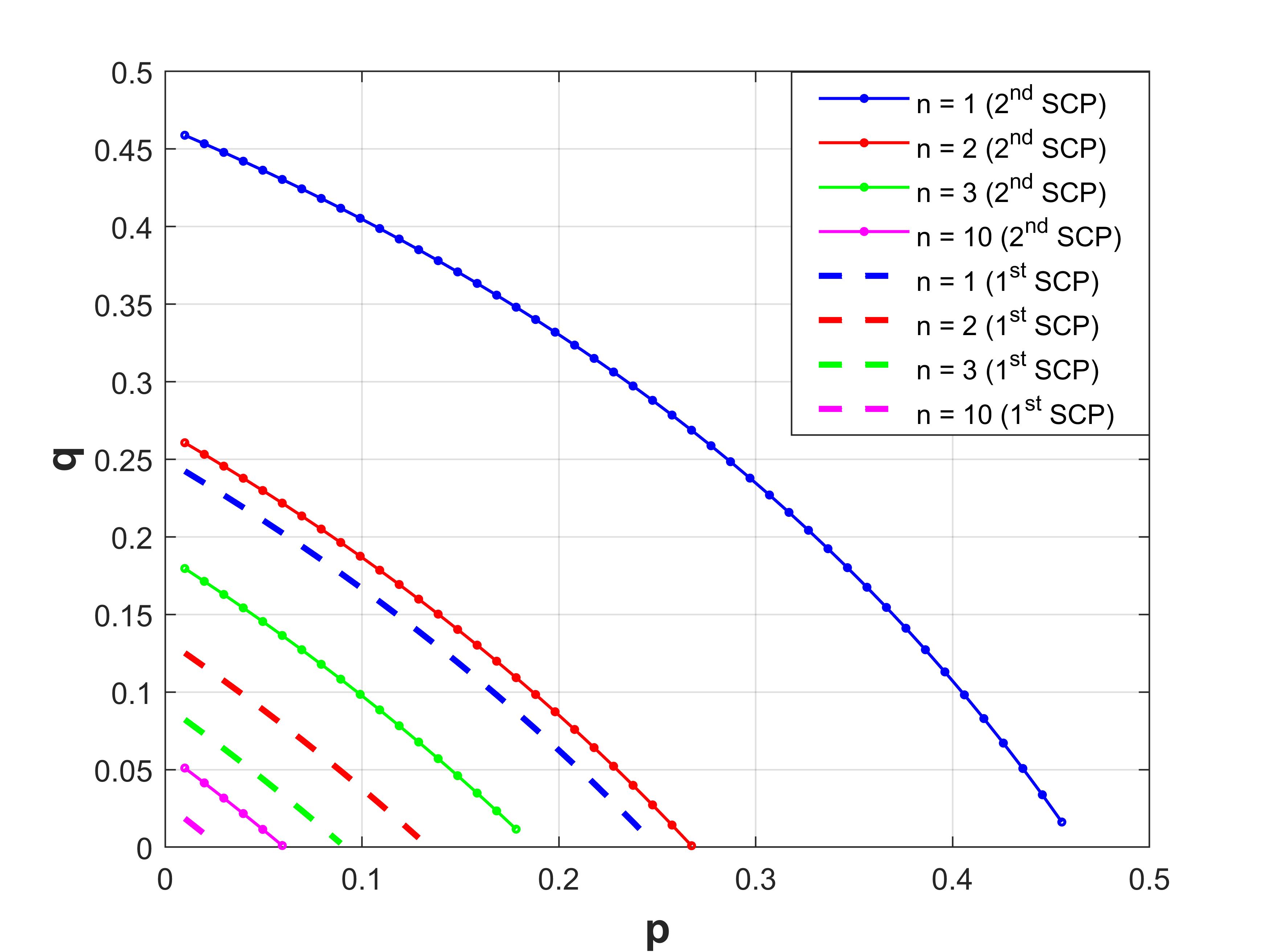}
	\end{minipage}}
\caption{ Constraint Relation(s) corresponding to SCP on $p-q$ plane for (a) Quantum Discord and Bures Distance Discord (b) Trace Distance Discord.}
\label{200}
\end{figure}

From the plots in Fig. \ref{fig150}, we observe that QD decays slowly in comparison to BDD and TDD. The plots of the discord measures with the individual decoherence probabilities ($p$ and $q$) of subsystem A and B in the bi-Markovian scenario show that the SCP occurs over a curve on the $p-q$ plane. The curves for SCP on the $p-q$ plane is shown in Fig. \ref{200}. Interestingly, the TDD shows freezing behavior between the curves of sudden change, this behavior is similar to the case of dynamics under single bit flip decoherence channel. The curve of sudden change on the $p-q$ plane is obtained as a constraint relation between $p$ and $q$. 
 

We can explain the sudden change phenomenon using similar reasons that we have used to describe the dynamics when only the first subsystem was subjected to different quantum channels. We get SCP due to different decay factors of the correlation functions, which leads to crossing of the correlation functions in the maximization step.  We observe the SCP of QD and BDD when $|d_{3}^{\prime}| = |d_{1}^{\prime}|$. This leads to the following constraint on $p$ and $q$, 

\begin{equation}
    (1-p)(1-q) = \left(\frac{|d_{1}|}{|d_{3}|}\right)^\frac{1}{n}
\end{equation}

In the case of TDD, we need to find the intermediate values among the absolute values of the correlation functions. Whenever the intermediate value changes during the evolution, we observe the SCP. We encounter the first SCP when $|d_{1}^{\prime}|=|d_{2}^{\prime}|$ and the second SCP when $|d_{1}^{\prime}|=|d_{3}^{\prime}|$. This leads to the following two constraints on $p$ and $q$ for observing first and second SCP,
\begin{equation}
     (1-p)(1-q) = \left(\frac{|d_{1}|}{|d_{2}|}\right)^\frac{1}{n} \quad \text{and} \quad (1-p)(1-q) = \left(\frac{|d_{1}|}{|d_{3}|}\right)^\frac{1}{n}
\end{equation}

\subsubsection{Local phase flip channels ($PF^{A}-PF^{B}$)}
The Kraus operators when both the subsystems are subjected to the locally independent phase flip channels are,
\begin{equation}
    K_{0}^{(A)}=\sqrt{1-\frac{p}{2}} I^{A} \otimes I^{B}, \quad K_{1}^{(A)}=\sqrt{\frac{p}{2}} \sigma_{z}^{A} \otimes I^{B}
\end{equation}

\begin{equation}
    K_{0}^{(B)}=I^{A} \otimes \sqrt{1-\frac{q}{2}} I^{B}, \quad K_{1}^{(B)}=I^{A} \otimes \sqrt{\frac{q}{2}} \sigma_{z}^{B}
\end{equation}


The coefficients of Bell diagonal state after undergoing through these operations for $n$ times are,

\begin{equation}
    d_{1}^{\prime}= (1-p)^{n}(1-q)^{n}d_{1}, \quad d_{2}^{\prime} = (1-p)^{n}(1-q)^{n}d_{2}, \quad d_{3}^{\prime} = d_{3}
\end{equation}

\begin{figure}[ht]
  \subfloat[]{
	\begin{minipage}[c][1\width]{
	   0.3\textwidth}
	   \centering
	   \includegraphics[width=1.0\textwidth]{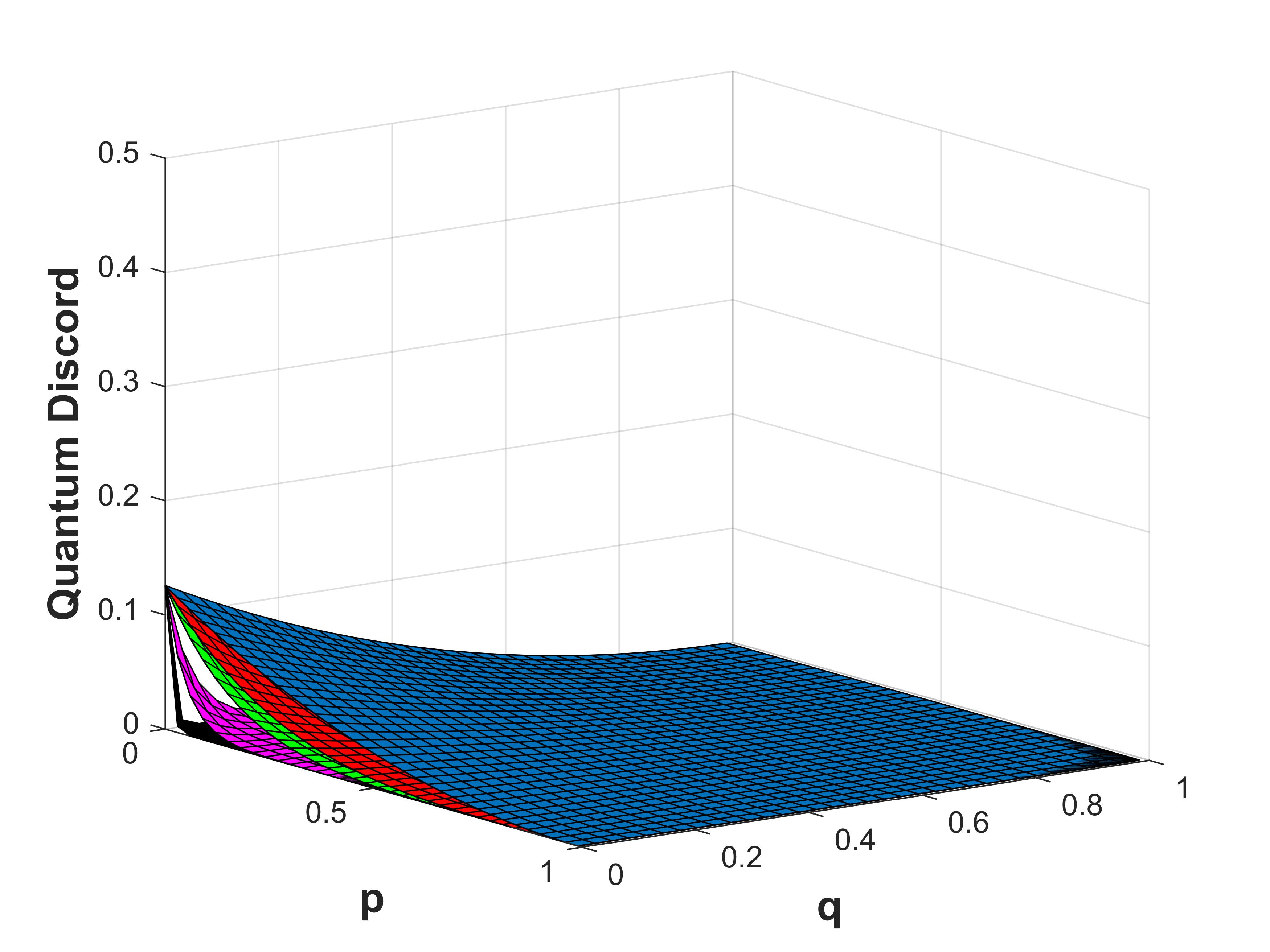}
	\end{minipage}}
 \hfill 	
  \subfloat[]{
	\begin{minipage}[c][1\width]{
	   0.3\textwidth}
	   \centering
	   \includegraphics[width=1.0\textwidth]{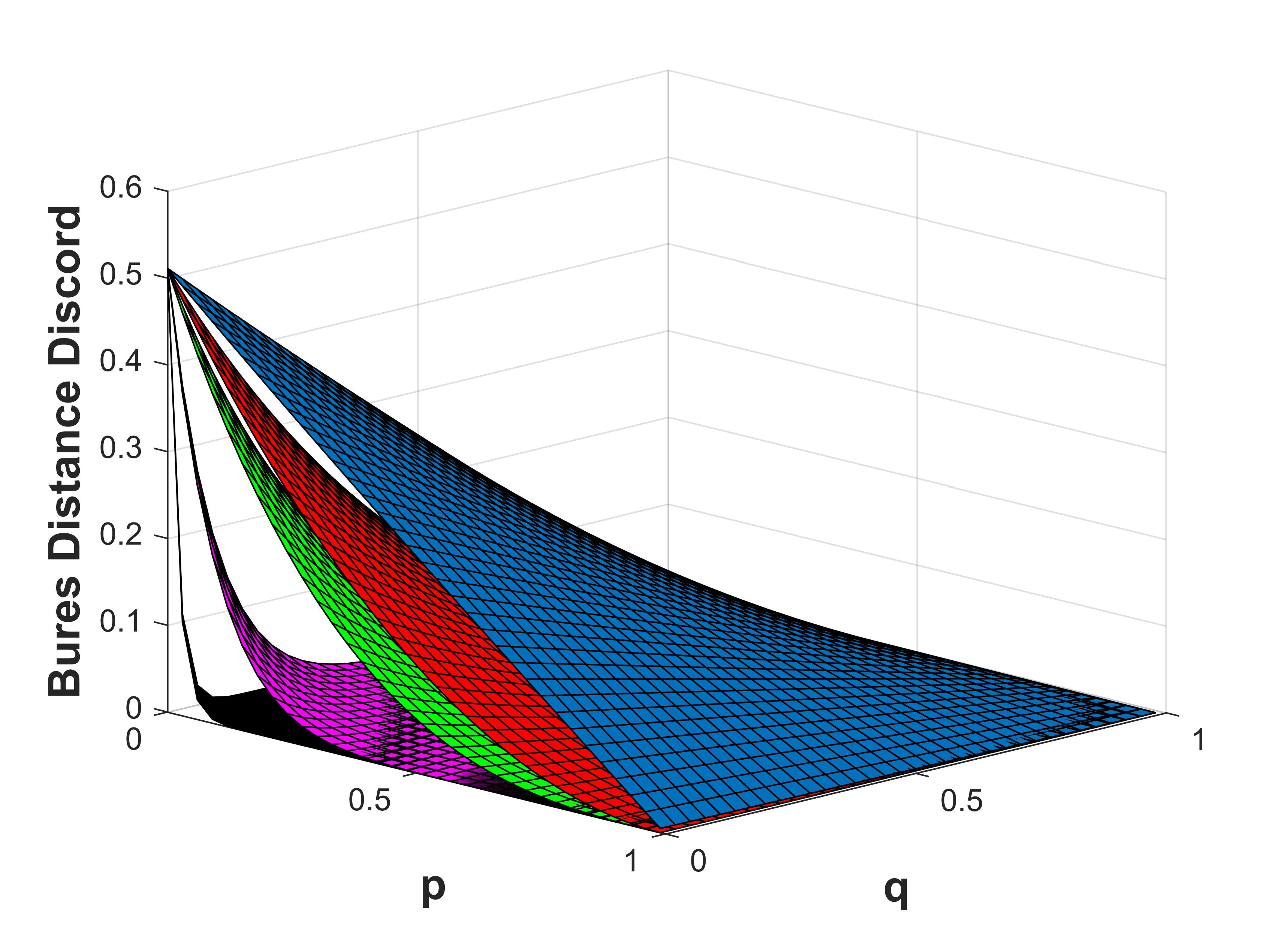}
	\end{minipage}}
 \hfill	
  \subfloat[]{
	\begin{minipage}[c][1\width]{
	   0.3\textwidth}
	   \centering
	   \includegraphics[width=1.0\textwidth]{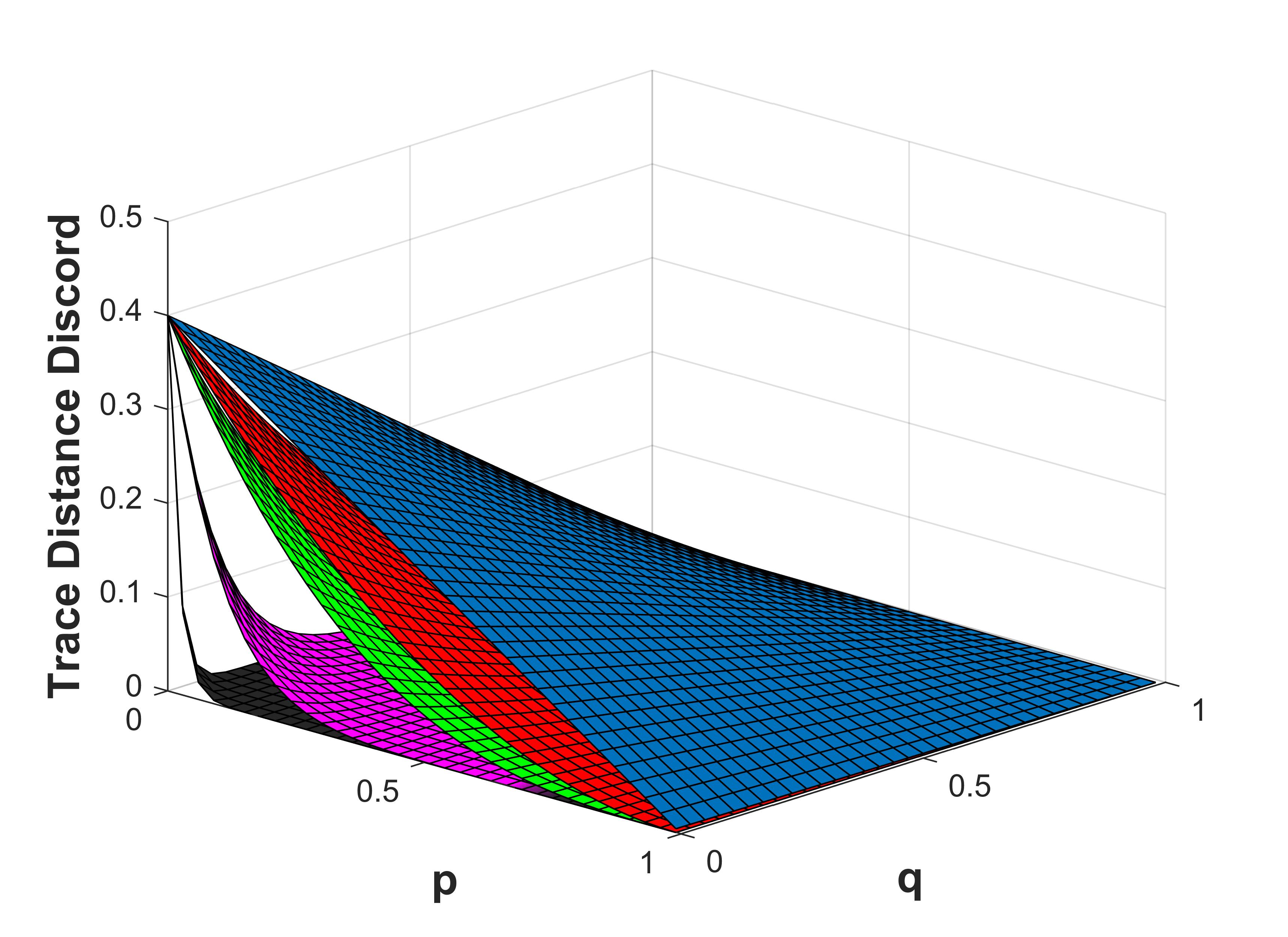}
	\end{minipage}}
\caption{When both the subsystems undergo through the same quantum channel $n$ times for Bell diagonal state ($d_{1} = 0.3$, $d_{2} = −0.4$ and $d_{3} = 0.56$) : Phase Flip Quantum Channel where colors Blue ($n = 1$), Red ($n = 2$), Green ($n = 3$), Magenta ($n = 10$), Black ($n = 50$)) (a) Quantum Discord (b) Bures Distance Discord (c) Trace Distance Discord}.
\label{fig500}
\end{figure}
In this case, we do not observe SCP in any of the correlation measures as the evolution of the set of initial conditions $|d_{1}|<|d_{2}|<|d_{3}|$ preserve the order i.e., $|d_{1}^{\prime}|<|d_{2}^{\prime}|<|d_{3}^{\prime}|$. From Fig. \ref{fig500}, we observe that QD decays slowly as compared to BDD and TDD.


\subsubsection{Local Bit Phase flip channels ($BPF^{A}-BPF^{B}$)}
 
 The Kraus operators of the locally independent bit-phase flip channels acting on both the subsystems are given by, 

\begin{equation}
K_{0}^{(A)}=\sqrt{1-\frac{p}{2}} I^{A} \otimes I^{B}, \quad K_{1}^{(A)}=\sqrt{\frac{p}{2}} \sigma_{y}^{A} \otimes I^{B} 
\end{equation}

\begin{equation}
K_{0}^{(B)}=I^{A} \otimes \sqrt{1-\frac{q}{2}} I^{B}, \quad K_{1}^{(B)}=I^{A} \otimes \sqrt{\frac{q}{2}} \sigma_{y}^{B}
\end{equation}
The coefficients of evolved Bell diagonal state are,
\begin{equation}
    d_{1}^{\prime} = (1-p)^{n}(1-q)^{n}d_{1}, \quad d_{2}^{\prime}=d_{2}, \quad d_{3}^{\prime} = (1-p)^{n}(1-q)^{n} d_{3}
\end{equation}

\begin{figure}[ht]
  \subfloat[]{
	\begin{minipage}[c][1\width]{
	   0.3\textwidth}
	   \centering
	   \includegraphics[width=1.0\textwidth]{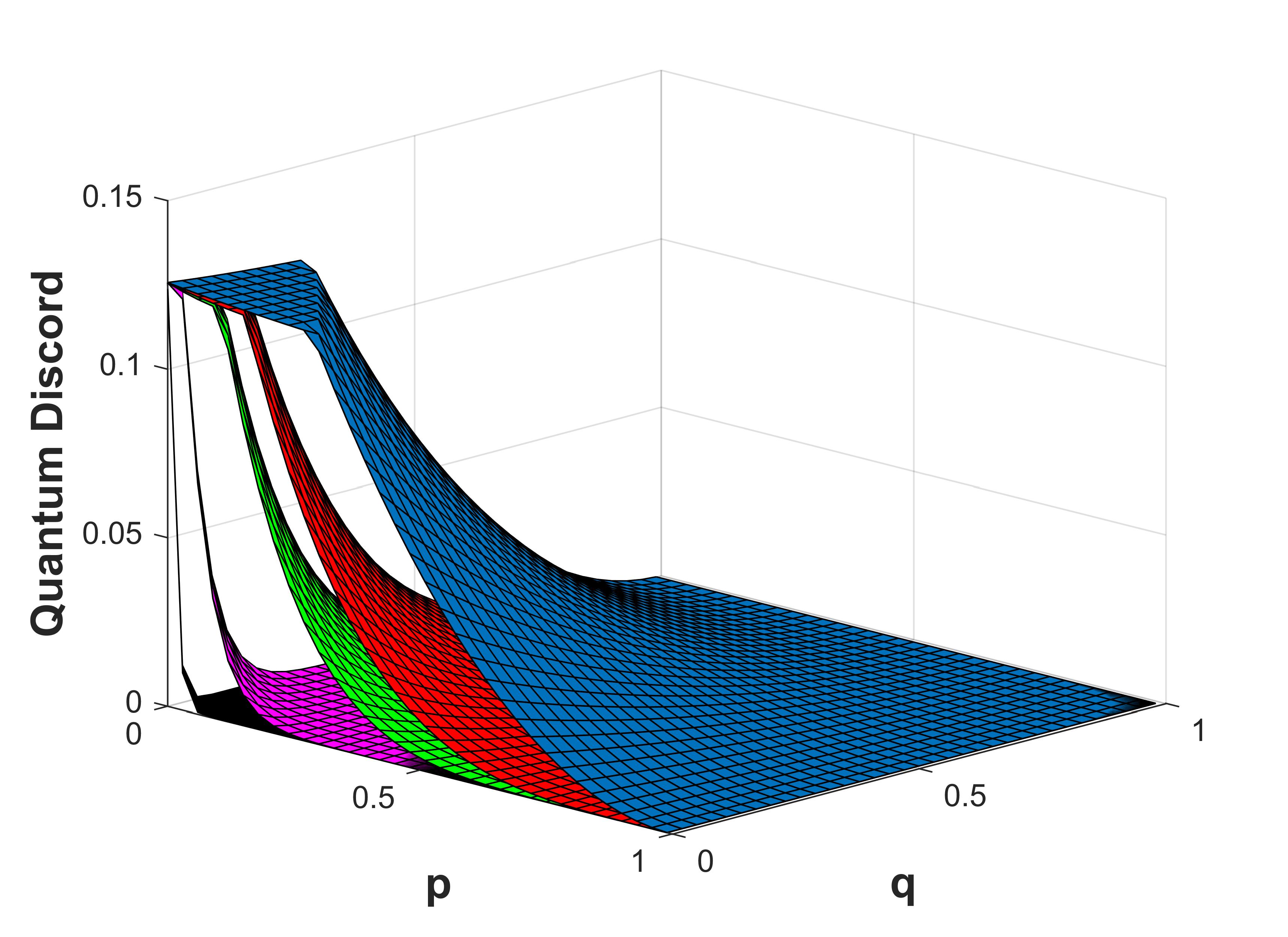}
	\end{minipage}}
 \hfill 	
  \subfloat[]{
	\begin{minipage}[c][1\width]{
	   0.3\textwidth}
	   \centering
	   \includegraphics[width=1.0\textwidth]{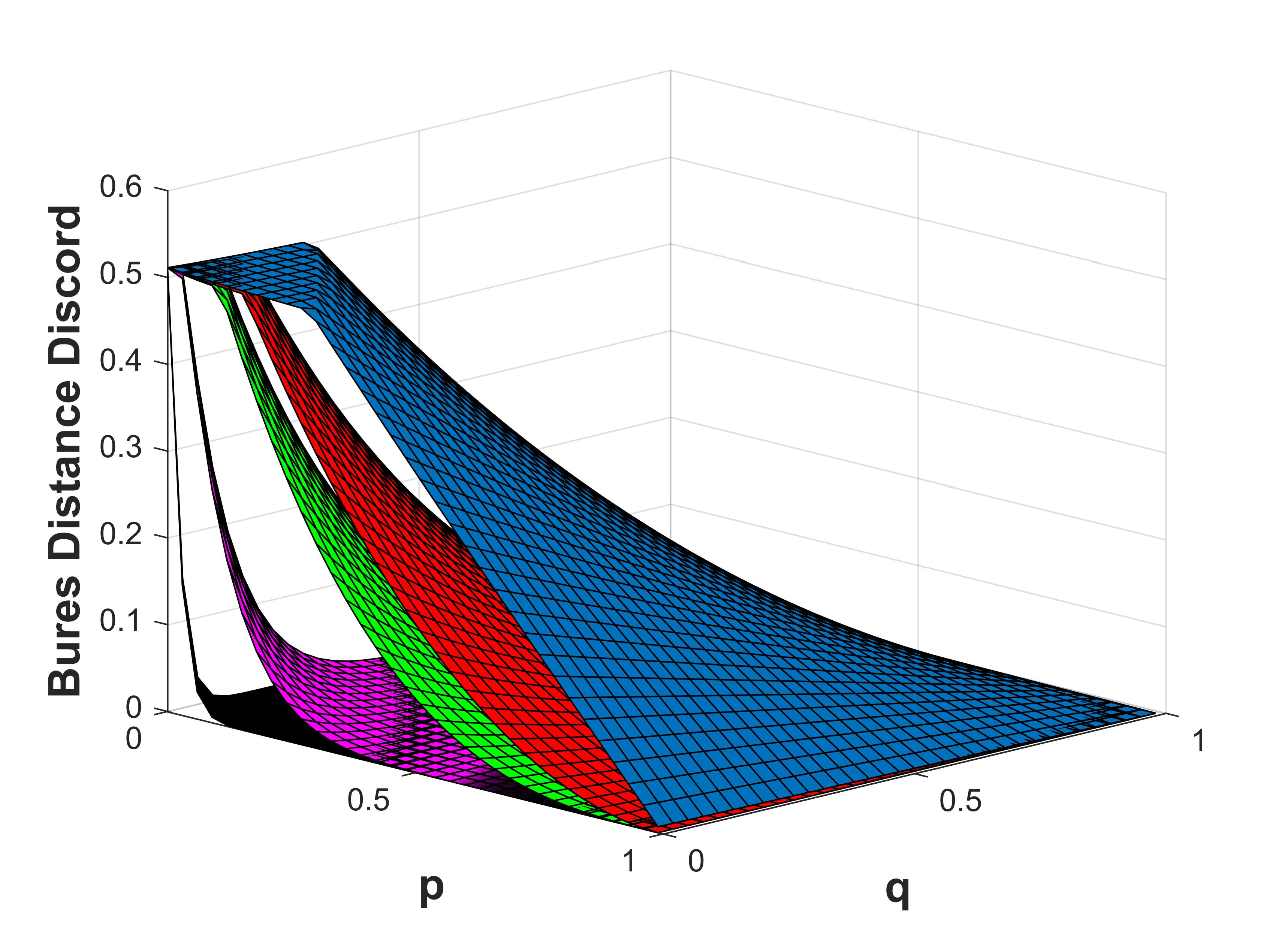}
	\end{minipage}}
 \hfill	
  \subfloat[]{
	\begin{minipage}[c][1\width]{
	   0.3\textwidth}
	   \centering
	   \includegraphics[width=1.0\textwidth]{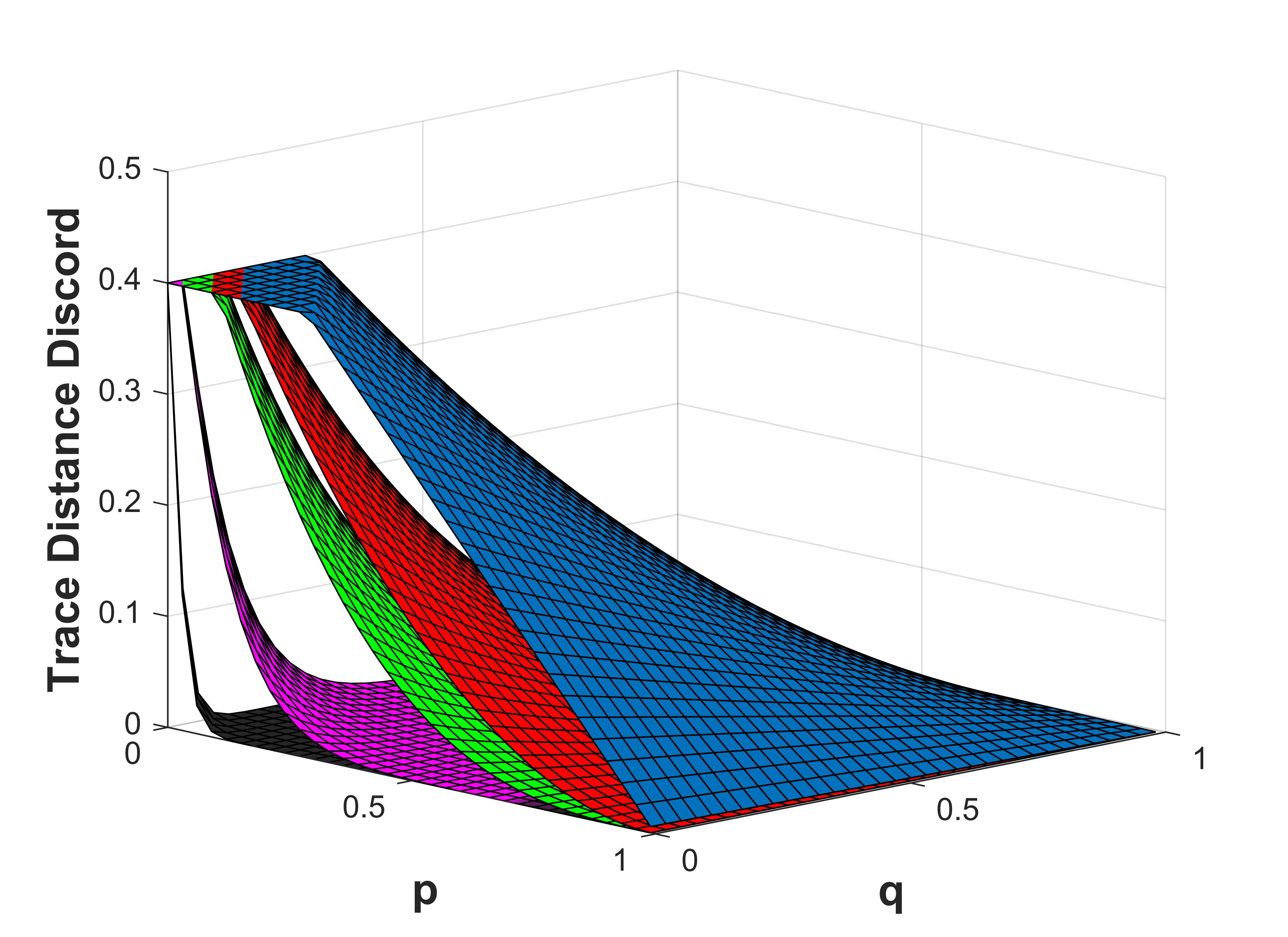}
	\end{minipage}}
\caption{When both the subsystems undergo through the same quantum channel $n$ times for Bell diagonal state ($d_{1} = 0.3$, $d_{2} = −0.4$ and $d_{3} = 0.56$) : Bit Phase Flip Quantum Channel where colors Blue ($n = 1$), Red ($n = 2$), Green ($n = 3$), Magenta ($n = 10$), Black ($n = 50$)) (a) Quantum Discord (b) Bures Distance Discord (c) Trace Distance Discord.}
\end{figure}
We observe the sudden change phenomenon of the discord measures when $|d_{3}^{\prime}|$=$|d_{2}^{\prime}|$. This leads to the following constraint on $p$ and $q$,
\begin{equation}
    (1-p)(1-q) = \left(\frac{|d_{2}|}{|d_{3}|}\right)^{\frac{1}{n}}
\end{equation}
We see that QD and BDD decay slowly until the region of sudden change, after which their decay becomes rapid. The TDD remains frozen until it reaches the SCP region, after which it begins to deteriorate quickly.


\newpage

\subsection{Dynamics of discord measures under Bi-side Markovian Channel of different types}
In this section, we investigate the dynamics of discord measures of a two qubit Bell diagonal state under two different local Markovian channels acted for $n$ times. We consider the cases wherein the form of the Bell-diagonal state is preserved such as, a bit-flip channel and a phase flip channel (BF-PF), a bit flip channel and a bit-phase-flip channel (BF-BPF), a phase flip channel and bit-phase-flip channel (PF-BPF).

\subsubsection{Local bit flip and phase flip channels $(BF^{A}-PF^{B})$}
Here, we consider the case when two locally independent quantum channels, a bit flip channel and a phase flip channel are acted on two subsystems A and B respectively for multiple times.

The corresponding Kraus operators are given by,

\begin{equation}
   K_{0}^{(A)}=\sqrt{1-\frac{p}{2}} I^{A} \otimes I^{B}, \quad K_{1}^{(A)}=\sqrt{\frac{p}{2}} \sigma_{x}^{A} \otimes I^{B}
\end{equation}

\begin{equation}
   K_{0}^{(B)}=I^{A} \otimes \sqrt{1-\frac{q}{2}} I^{B}, \quad K_{1}^{(B)}=I^{A} \otimes \sqrt{\frac{q}{2}}\sigma_{z}^{B}
\end{equation}

The coefficients of the evolved Bell diagonal state are as follows,

\begin{equation}
    d_{1}^{\prime}= (1-q)^{n}d_{1}, \quad d_{2}^{\prime} = (1-p)^{n}(1-q)^{n}d_{2} \quad d_{3}^{\prime} = (1-p)^{n}d_{3}
\end{equation}

\begin{figure}[ht]
  \subfloat[]{
	\begin{minipage}[c][1\width]{
	   0.3\textwidth}
	   \centering
	   \includegraphics[width=1.0\textwidth]{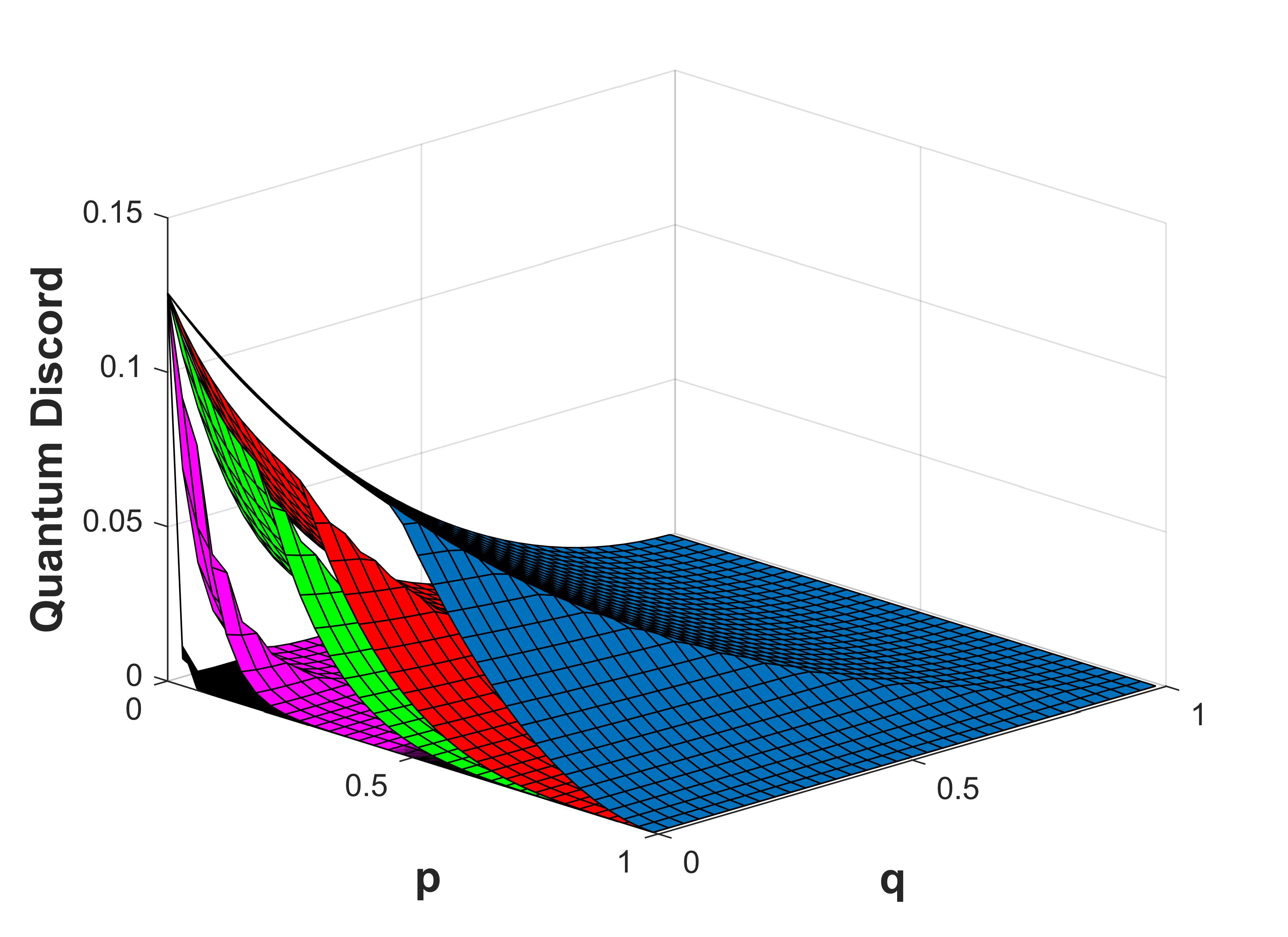}
	\end{minipage}}
 \hfill 	
  \subfloat[]{
	\begin{minipage}[c][1\width]{
	   0.3\textwidth}
	   \centering
	   \includegraphics[width=1.0\textwidth]{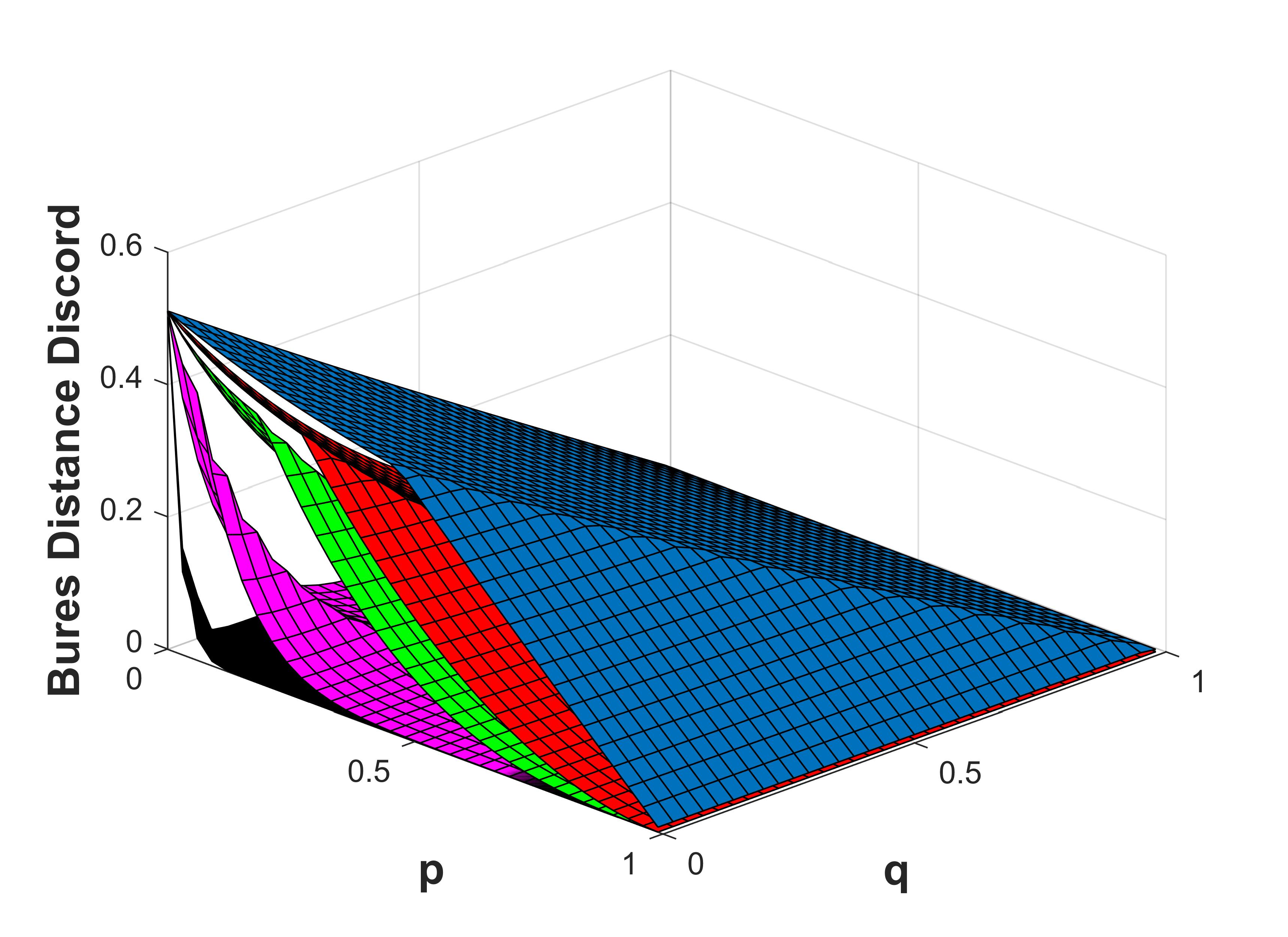}
	\end{minipage}}
 \hfill	
  \subfloat[]{
	\begin{minipage}[c][1\width]{
	   0.3\textwidth}
	   \centering
	   \includegraphics[width=1.0\textwidth]{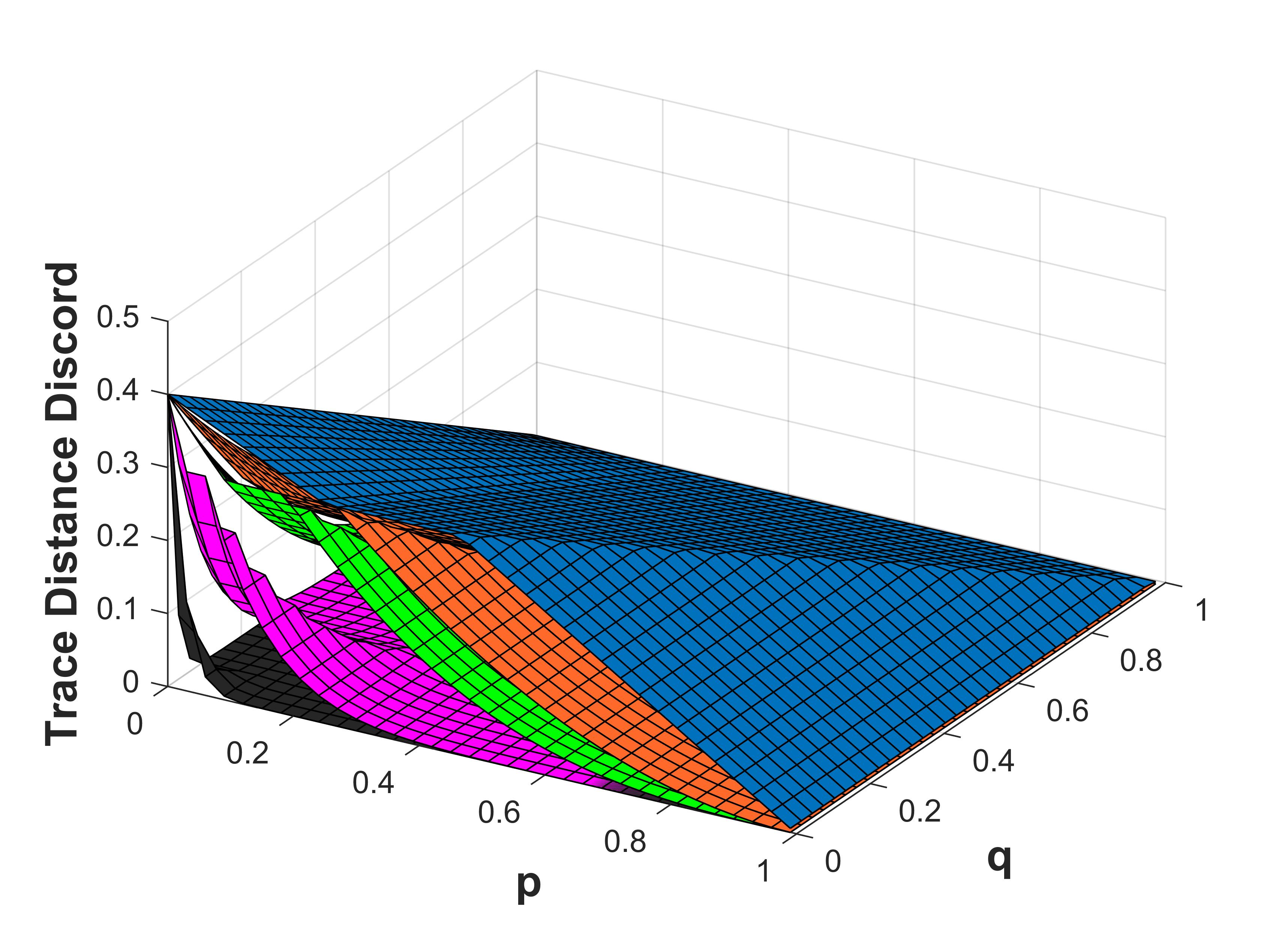}
	\end{minipage}}
\caption{When both the subsystems undergo through different quantum channels $n$ times for Bell diagonal state ($d_{1} = 0.3$, $d_{2} = −0.4$ and $d_{3} = 0.56$) : Bit Flip - Phase Flip Quantum Channel where colors Blue ($n = 1$), Red ($n = 2$), Green ($n = 3$), Magenta ($n = 10$), Black ($n = 50$)) (a) Quantum Discord (b) Bures Distance Discord (c) Trace Distance Discord.}
\end{figure}

In case of QD and BDD, we observe the region of sudden change phenomenon when $|d_{3}^{\prime}|=|d_{1}^{\prime}|$, this leads to the constraint,
\begin{equation}
    \frac{(1-p)}{(1-q)} = \left(\frac{|d_{1}|}{|d_{3}|}\right)^{\frac{1}{n}}
\end{equation}
For TDD, we observe two regions of sudden change phenomenon when $|d_{2}^{\prime}|= |d_{1}^{\prime}|$ and $|d_{3}^{\prime}|= |d_{1}^{\prime}|$. The value of $p$ for observing first SCP is,
\begin{equation}
    p = 1 - \left(\frac{|d_{1}|}{|d_{2}|}\right)^{\frac{1}{n}}
 \end{equation}
 where $q$ takes the value $0< q< 1$.
 The constraint relation in terms of $p$ and $q$ for second SCP is,
 \begin{equation}
    \frac{(1-p)}{(1-q)} = \left(\frac{|d_{1}|}{|d_{3}|}\right)^{\frac{1}{n}}
\end{equation}

\begin{figure}[htp]
	   \centering
	   \includegraphics[width=0.4\textwidth]{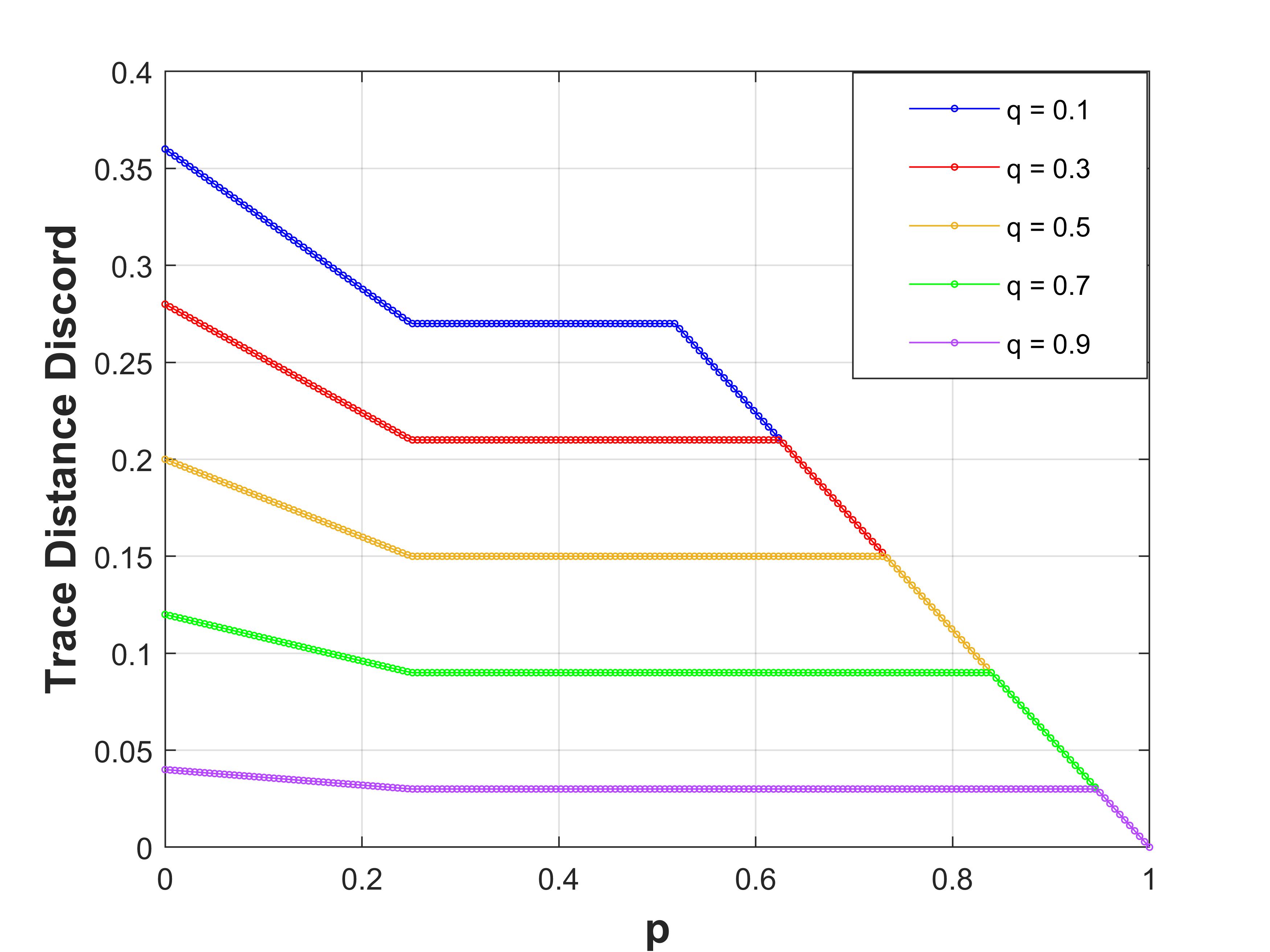}
\caption{Trace distance discord vs $p$ for five different values of $q$.}
\label{160}
\end{figure}

In Fig. $\ref{160}$, we plot the variation of TDD with a change in $p$ for different values of $q$. We observe that TDD remains constant between the two SCPs for fixed values of $q$. For this channel, we observe a freezing phenomenon for all values of $q$.

\subsubsection{Local Bit flip and Bit-phase flip channels $(BF^{A}-BPF^{B})$}
The Kraus operators for the case wherein the first subsystem is subjected to the bit flip channel and second subsystem to the bit phase flip channel are,
\begin{equation}
K_{0}^{(A)}=\sqrt{1-\frac{p}{2}} I^{A} \otimes I^{B}, \quad K_{1}^{(A)}=\sqrt{\frac{p}{2}} \sigma_{x}^{A} \otimes I^{B}
\end{equation}

\begin{equation}
    K_{0}^{(B)}=I^{A} \otimes \sqrt{1-\frac{q}{2}} I^{B}, \quad K_{1}^{(B)}=I^{A} \otimes \sqrt{\frac{q}{2}} \sigma_{y}^{B}
\end{equation}
The coefficients of the evolved Bell diagonal state after going through these channels for $n$ times are,

\begin{equation}
d_{1}^{\prime} = (1-q)^{n}d_{1}, \quad d_{2}^{\prime} = (1-p)^{n}d_{2}, \quad d_{3}^{\prime} = (1-p)^{n}(1-q)^{n} d_{3}
\end{equation}

\begin{figure}[ht]
  \subfloat[]{
	\begin{minipage}[c][1\width]{
	   0.3\textwidth}
	   \centering
	   \includegraphics[width=1.0\textwidth]{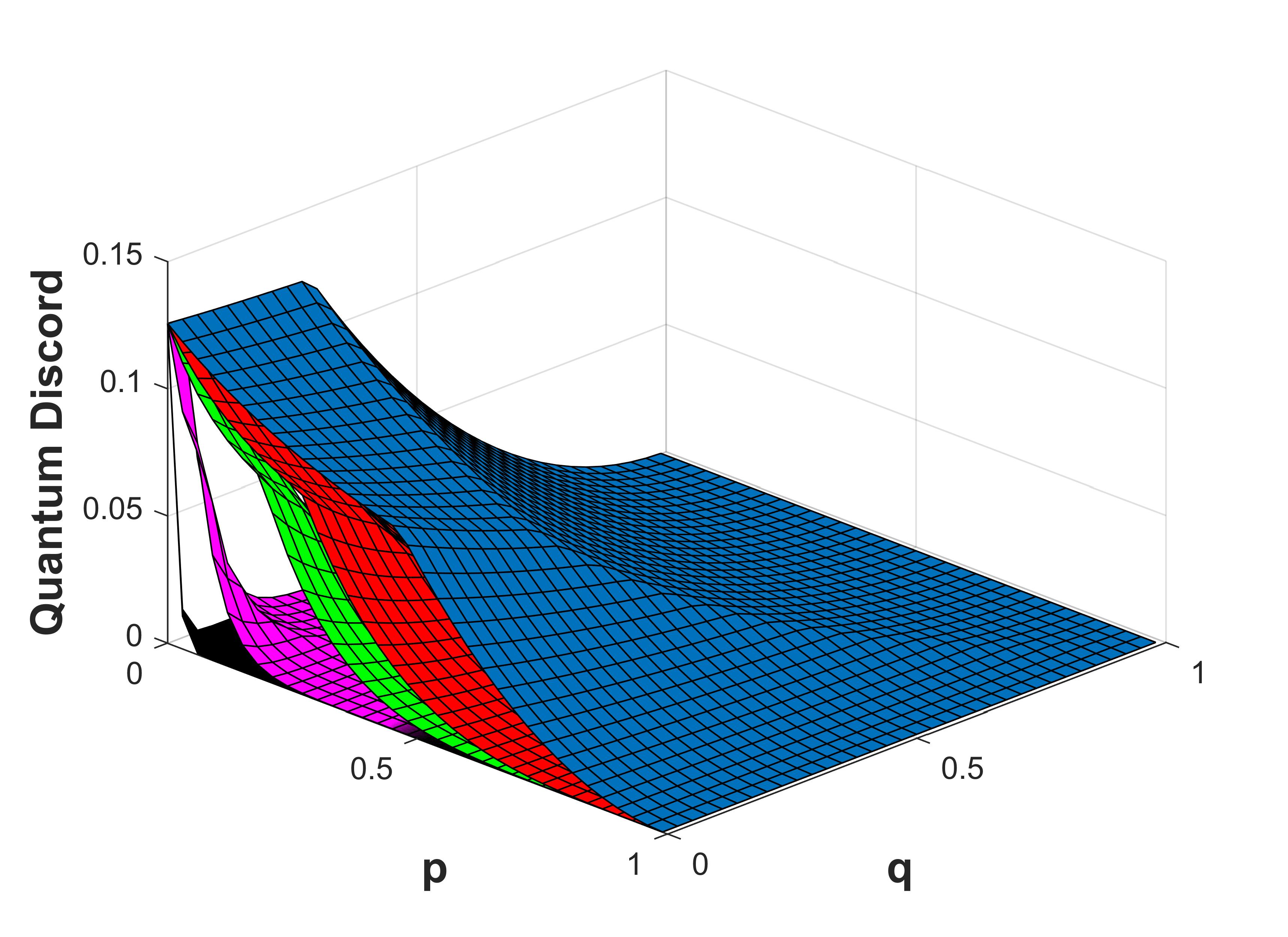}
	\end{minipage}}
 \hfill 	
  \subfloat[]{
	\begin{minipage}[c][1\width]{
	   0.3\textwidth}
	   \centering
	   \includegraphics[width=1.0\textwidth]{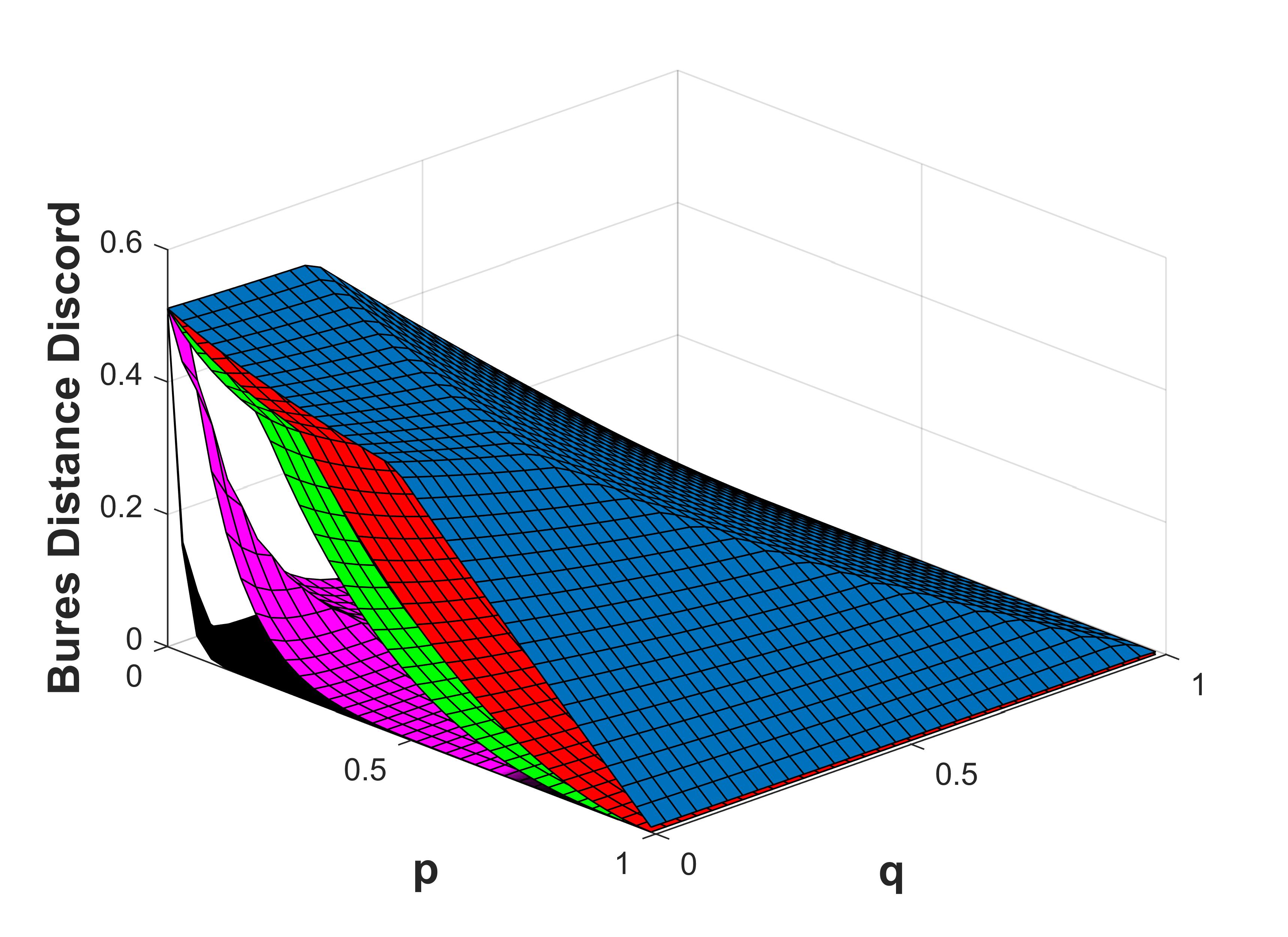}
	\end{minipage}}
 \hfill	
  \subfloat[]{
	\begin{minipage}[c][1\width]{
	   0.3\textwidth}
	   \centering
	   \includegraphics[width=1.0\textwidth]{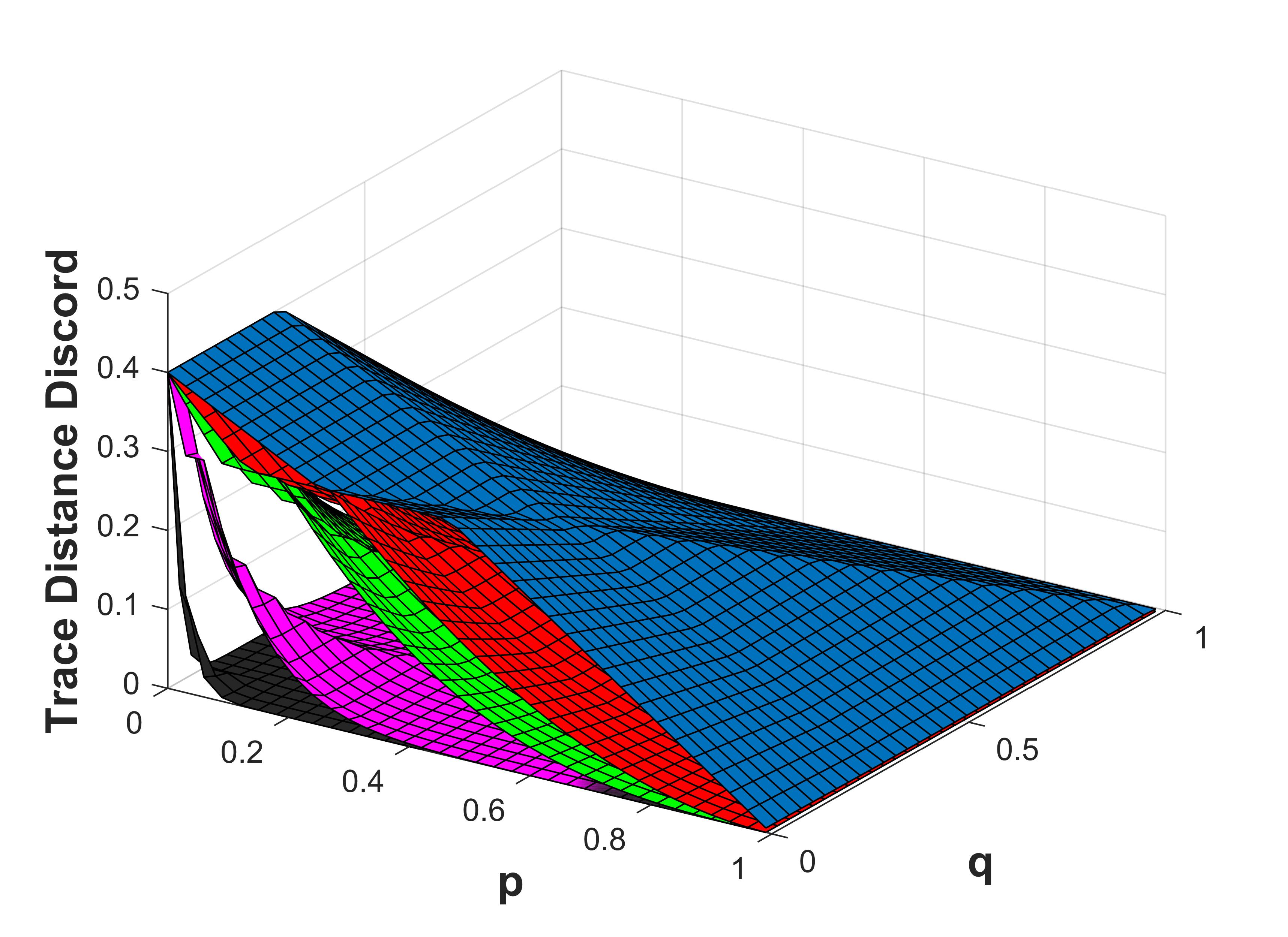}
	\end{minipage}}
\caption{When both the subsystems undergo through different quantum channels $n$ times for Bell diagonal state ($d_{1} = 0.3$, $d_{2} = −0.4$ and $d_{3} = 0.56$) : Bit Flip - Bit Phase Flip Quantum Channel where colors Blue ($n = 1$), Red ($n = 2$), Green ($n = 3$), Magenta ($n = 10$), Black ($n = 50$)) (a) Quantum Discord (b) Bures Distance Discord (c) Trace Distance Discord.}
\end{figure}

 \begin{figure}[ht]
  \subfloat[]{
	\begin{minipage}[c][1\width]{
	   0.45\textwidth}
	   \centering
	   \includegraphics[width=1.0\textwidth]{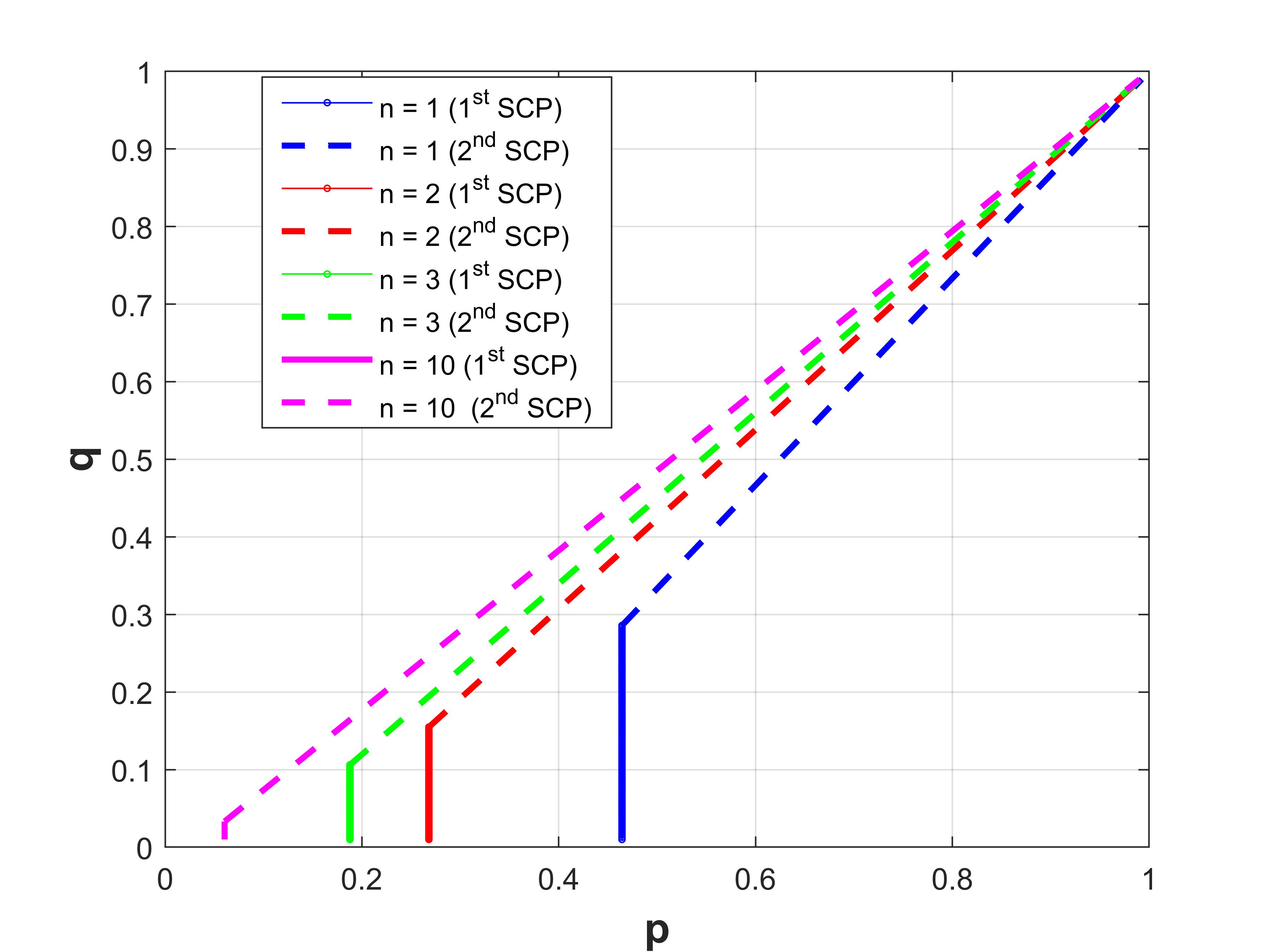}
	\end{minipage}}
 \hfill 	
  \subfloat[]{
	\begin{minipage}[c][1\width]{
	   0.45\textwidth}
	   \centering
	   \includegraphics[width=1.0\textwidth]{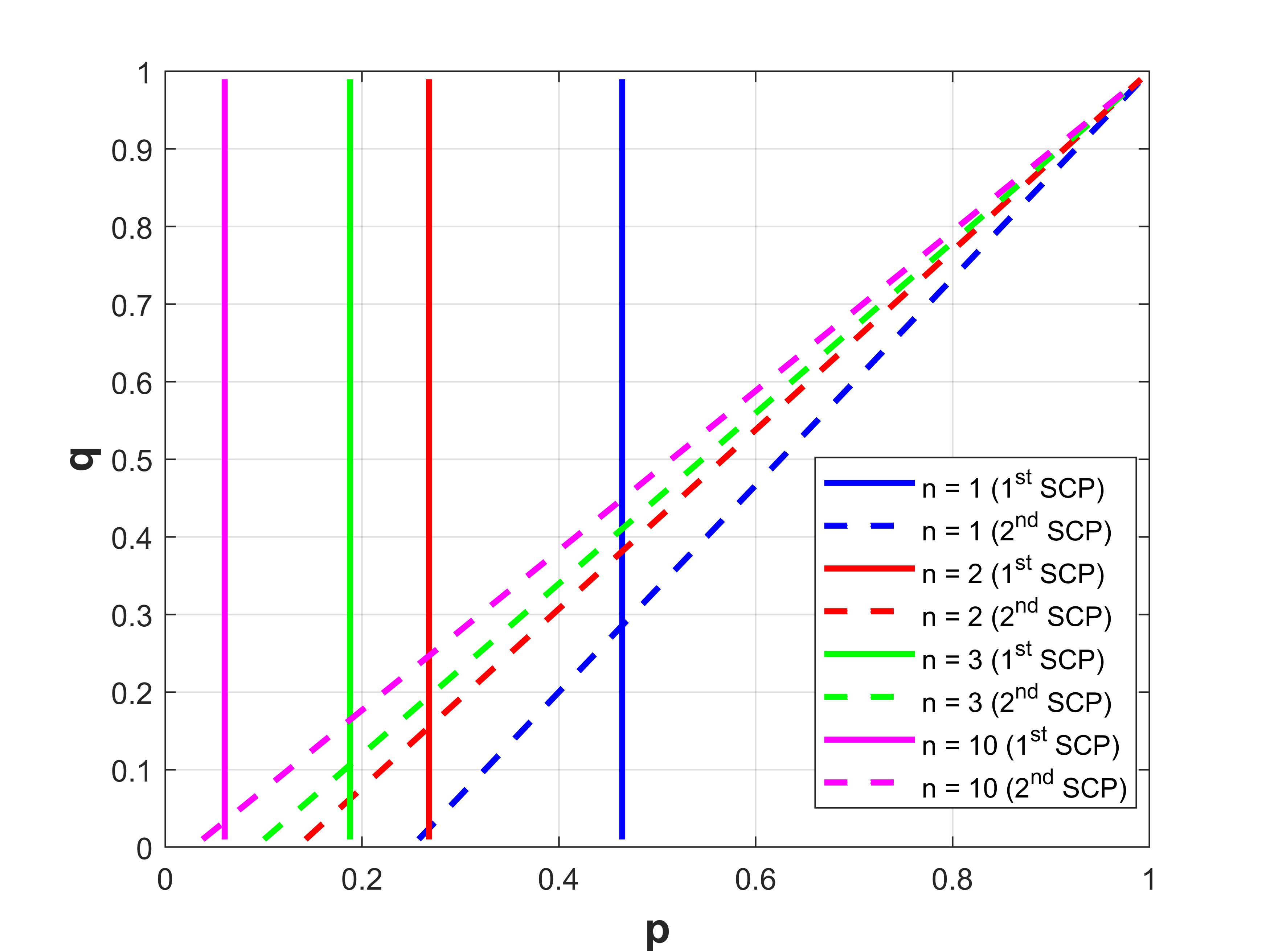}
	\end{minipage}}
\caption{ Constraint Relation corresponding to SCP on $p-q$ plane for (a) Quantum Discord and Bures Distance Discord (b) Trace Distance Discord.}
\label{560}
\end{figure}

We observe different set of constraints for SCP depending upon the value of $q$. If we consider the case when $|d_{3}^{\prime}|\geq |d_{2}^{\prime}|$, then $q$ varies from 0 to $q \leq q_{0} = 1 - \left(\frac{|d_{2}|}{|d_{3}|}\right)^{\frac{1}{n}}$. If we fix a certain $q$ which is less than $q_{0}$, from the decay factors of the evolved coefficients, we observe SCP when $|d_{3}^{\prime}| = |d_{1}^{\prime}|$. The value of $p$ corresponding to SCP for QD and BDD is then given by,
\begin{equation}
    p = 1 - \left(\frac{|d_{1}|}{|d_{3}|}\right)^{\frac{1}{n}}
\end{equation}

For the case when $|d_{2}^{\prime}| > |d_{3}^{\prime}|$, $q$ has to vary between $q_{0}$ and 1. If we fix a certain $q$ greater than $q_{0}$, we observe SCP when $|d_{2}^{\prime}| = |d_{1}^{\prime}|$. This leads to the constraint,
\begin{equation}
    \frac{(1-p)}{(1-q)} = \left(\frac{|d_{1}|}{|d_{2}|}\right)^{\frac{1}{n}}
\end{equation}

In case of TDD, we need to find the intermediate values among the absolute values of correlation functions. When $q$ varies between $0$ and $q_{0}$, $|d_{2}^{\prime}|\leq|d_{3}^{\prime}|$ for the entire range of $p$. From the decay factors of the coefficients, we observe SCP when $|d_{2}^{\prime}|$ = $|d_{1}^{\prime}|$ and $|d_{3}^{\prime}|$ = $|d_{1}^{\prime}|$. This leads to the constraint equations,
\begin{equation}
   \frac{(1-p)}{(1-q)} = \left(\frac{|d_{1}|}{|d_{2}|}\right)^{\frac{1}{n}} \quad \text{and} \quad p = 1 - \left(\frac{|d_{1}|}{|d_{3}|}\right)^{\frac{1}{n}}
\end{equation}

for $0<q\leq q_{0}$.

When $q > q_{0}$, $|d_{2}^{\prime}|>|d_{3}^{\prime}|$ for the entire range of $p$. From the decay factors of the coefficients, we observe SCP when $|d_{3}^{\prime}|$ = $|d_{1}^{\prime}|$ and $|d_{2}^{\prime}|$ = $|d_{1}^{\prime}|$, this results in the same constraint equations given by eq. (71). The curves for SCP on the $p-q$ plane is shown in Fig. \ref{560} for different values of $n$.



%




\subsubsection{Local Phase flip and Bit-phase flip channels $(PF^{A}-BPF^{B})$}
The Kraus operators for the case wherein the first subsystem is subjected to the phase flip channel and second subsystem to the bit phase flip channel are,


\begin{equation}
K_{0}^{(A)}=\sqrt{1-\frac{p}{2}} I^{A} \otimes I^{B}, \quad K_{1}^{(A)}=\sqrt{\frac{p}{2}} \sigma_{z}^{A} \otimes I^{B} 
\end{equation}

\begin{equation}
K_{0}^{(B)}=I^{A} \otimes \sqrt{1-\frac{p}{2}} I^{B}, \quad K_{1}^{(B)}=  I^{A} \otimes \sqrt{\frac{q}{2}} \sigma_{y}^{B}
\end{equation}
The coefficients of Bell diagonal states after undergoing through these operations $n$ times,
\begin{equation}
    d_{1}^{\prime} = (1-p)^{n}(1-q)^{n}d_{1}, \quad d_{2}^{\prime}= (1-p)^{n}d_{2}, \quad d_{3}^{\prime} = (1-q)^{n} d_{3}
\end{equation}

\begin{figure}[ht]
  \subfloat[]{
	\begin{minipage}[c][1\width]{
	   0.3\textwidth}
	   \centering
	   \includegraphics[width=1.0\textwidth]{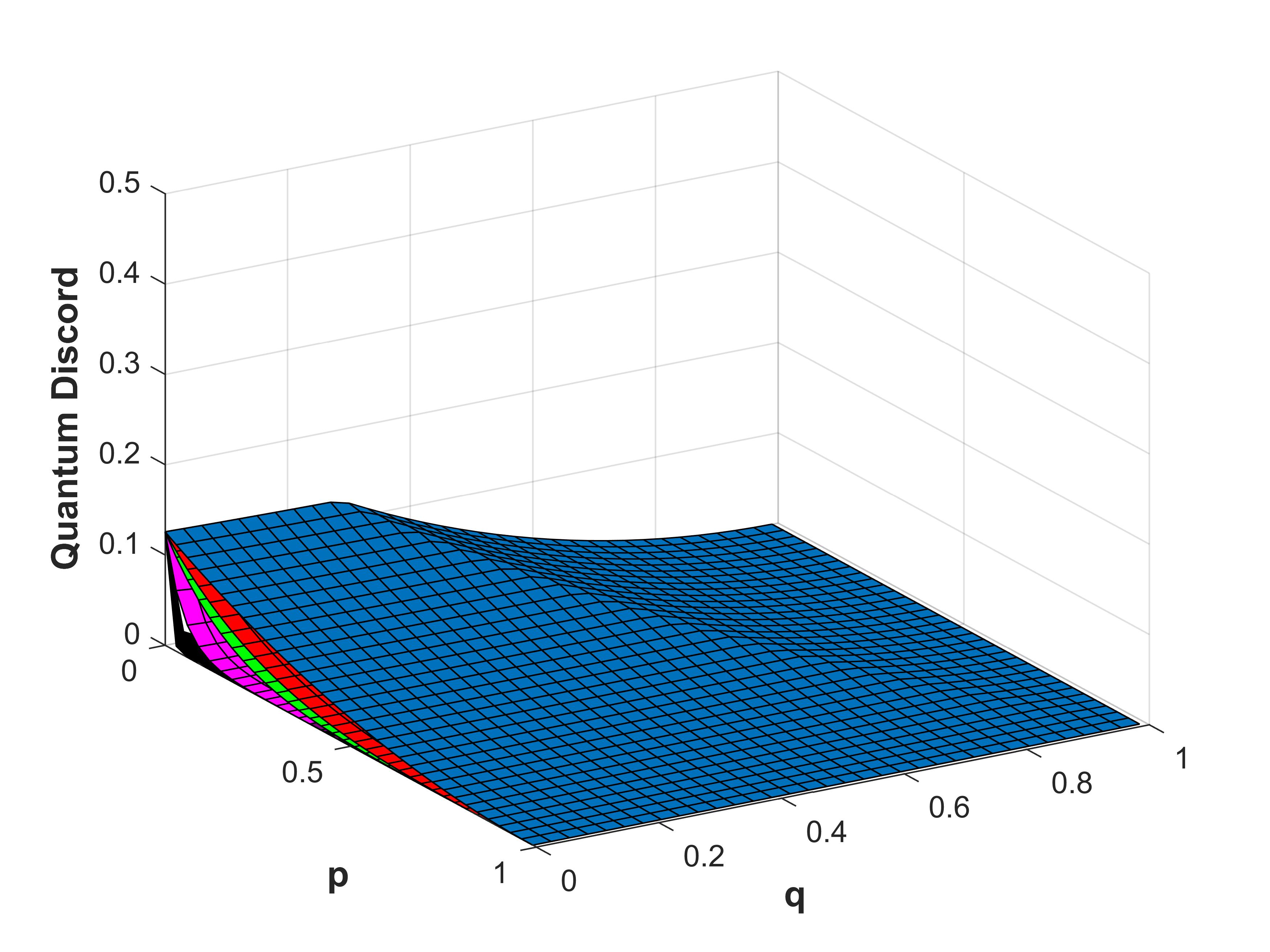}
	\end{minipage}}
 \hfill 	
  \subfloat[]{
	\begin{minipage}[c][1\width]{
	   0.3\textwidth}
	   \centering
	   \includegraphics[width=1.0\textwidth]{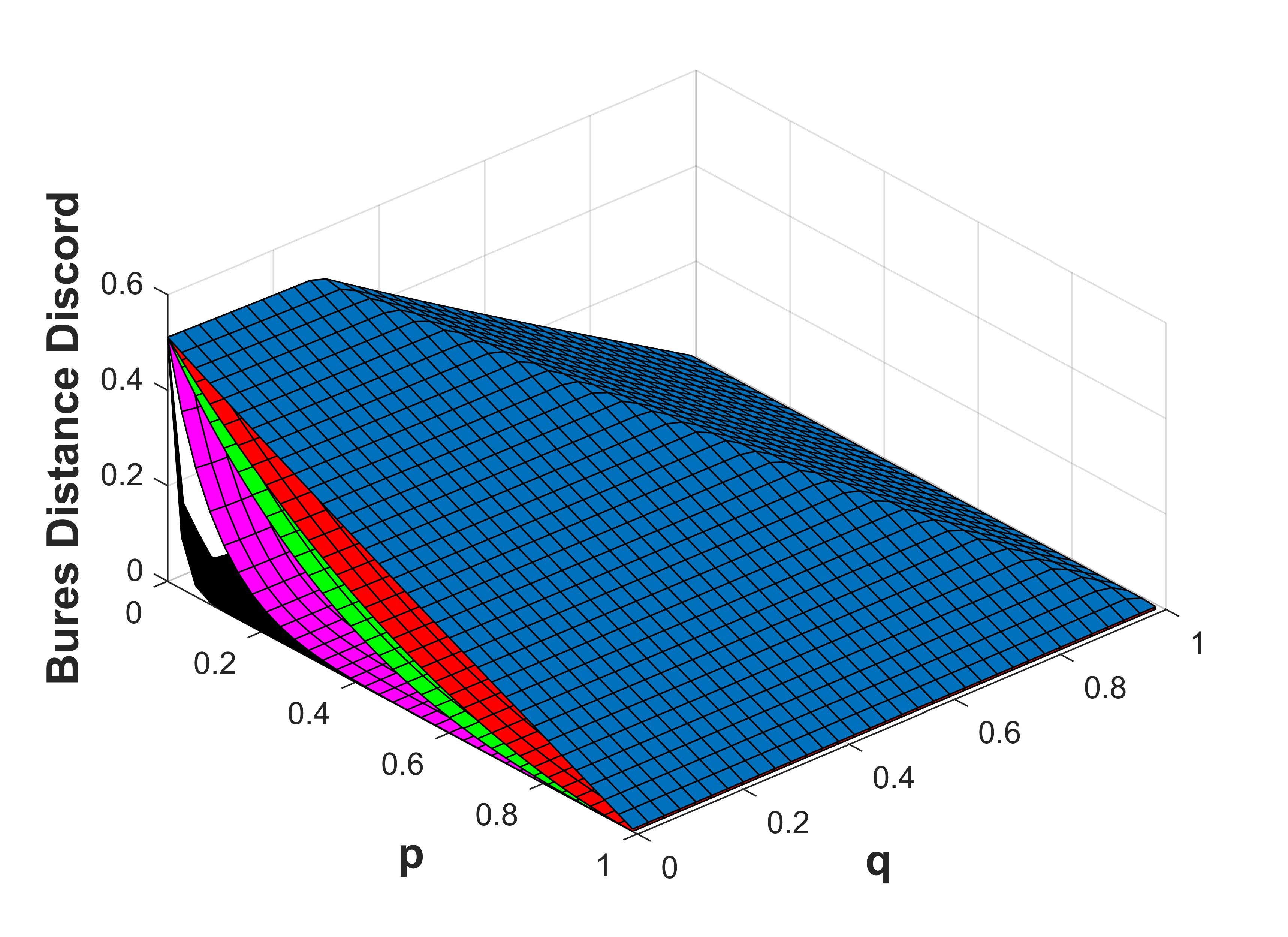}
	\end{minipage}}
 \hfill	
  \subfloat[]{
	\begin{minipage}[c][1\width]{
	   0.3\textwidth}
	   \centering
	   \includegraphics[width=1.0\textwidth]{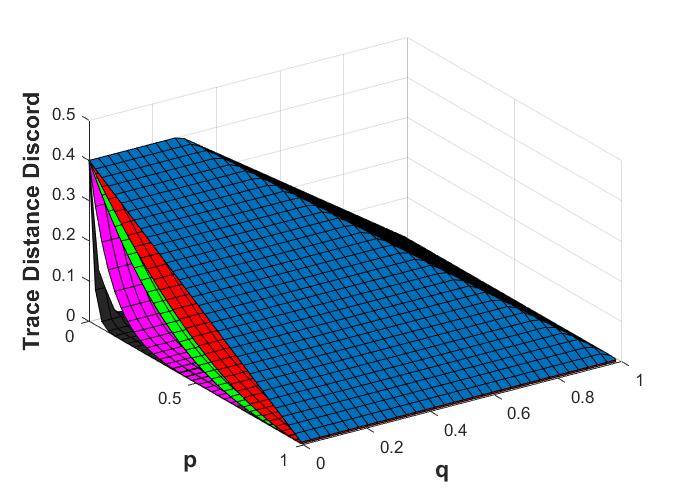}
	\end{minipage}}
\caption{When both the subsystems undergo through different quantum channels $n$ times for Bell diagonal state ($d_{1} = 0.3$, $d_{2} = −0.4$ and $d_{3} = 0.56$) : Bit Flip - Bit Phase Flip Quantum Channel where colors Blue ($n = 1$), Red ($n = 2$), Green ($n = 3$), Magenta ($n = 10$), Black ($n = 50$)) (a) Quantum Discord (b) Bures Distance Discord (c) Trace Distance Discord.}
\label{501}
\end{figure}

In this case, we encounter a single region of sudden change in the discord measures. By examining the decay factors, we get SCP for the discord measures when $|d_{3}^{\prime}|$ becomes equal to $|d_{2}^{\prime}|$. The constraint on $p$ and $q$ is obtained as,
\begin{equation}
    \frac{(1-p)}{(1-q)} = \left(\frac{|d_{3}|}{|d_{2}|}\right)^{\frac{1}{n}}
\end{equation}

From Fig.\ref{501}, we observe that QD is more robust against noise as it decays slowly than BDD and TDD.

\section{Conclusions}

In conclusion, we have investigated the dynamics of entropic quantum discord and geometric discord measures, namely, Bures distance discord and trace distance discord, when exposed to various decoherence channels. We have considered two cases, in the first case, the first qubit is subjected to Markovian channels multiple times, and in the second case, both the qubits of a two qubit $X$ state with maximally mixed marginals are subjected to two locally independent Markovian channels of the same, and different types for several times. The value of decoherence probabilities at which sudden change phenomena occur in discord measures are explicitly derived. Under the bit flip and bit phase flip channel, we find the preservation time during which trace distance discord remains unaffected under these noises and show its inverse proportionality to the number of times the channel acts upon the first subsystem. We also derive the expression for the time interval for trace distance discord between the two sudden changes in the generalized amplitude damping channel and find that its first point of sudden change is the same as that of quantum discord and Bures distance discord. In the case of the depolarizing channel, we demonstrated that revival occurs for all three discord measures after a particular value of decoherence probability which is independent of the number of times this channel is operated on the subsystem.
Further, we analyze the scenarios when bi-sided Markovian noises of the same and different types are acted independently on both the qubits for several times. We provide the constraint relations on decoherence probabilities for obtaining the sudden change and freezing of discord measures.

\section*{Acknowledgement}
The authors acknowledge the financial support from DST, India through Grant No.: DST/ICPS/QuST/Theme-1/2019/2020-21/01.

\printbibliography 

@misc{preskill,
  title={Lecture notes for ph219/cs219: Quantum information and computation},
  author={Preskill, John},
  year={2001}}

@book{10.5555/1972505,
author = {Nielsen, Michael A. and Chuang, Isaac L.},
title = {Quantum Computation and Quantum Information: 10th Anniversary Edition},
year = {2011},
isbn = {1107002176},
publisher = {Cambridge University Press},
address = {USA},
edition = {10th}
}

@article{LI20181,
title = {Concepts of quantum non-Markovianity: A hierarchy},
journal = {Physics Reports},
volume = {759},
year = {2018},
issn = {0370-1573},
doi = {https://doi.org/10.1016/j.physrep.2018.07.001},
url = {https://www.sciencedirect.com/science/article/pii/S0370157318301601},
author = {Li Li and Michael J.W. Hall and Howard M. Wiseman},
}

@article{Phy,
  title = {Information-theoretic aspects of the generalized amplitude-damping channel},
  author = {Khatri, Sumeet and Sharma, Kunal and Wilde, Mark M.},
  journal = {Phys. Rev. A},
  volume = {102},
  numpages = {31},
  year = {2020},
  publisher = {American Physical Society},
  doi = {10.1103/PhysRevA.102.012401},
  url = {https://link.aps.org/doi/10.1103/PhysRevA.102.012401}
}

@article{Werlang,
  title = {Quantum Correlations in Spin Chains at Finite Temperatures and Quantum Phase Transitions},
  author = {Werlang, T. and Trippe, C. and Ribeiro, G. A. P. and Rigolin, Gustavo},
  journal = {Phys. Rev. Lett.},
  volume = {105},
  numpages = {4},
  year = {2010},
  publisher = {American Physical Society},
  doi = {10.1103/PhysRevLett.105.095702},
  url = {https://link.aps.org/doi/10.1103/PhysRevLett.105.095702}
}

@ARTICLE{Pal,
       author = {{Pal}, A.~K. and {Bose}, I.},
        title = "{Quantum discord in the ground and thermal states of spin clusters}",
      journal = {Journal of Physics B Atomic Molecular Physics},
         year = 2011,
       volume = {44},
       number = {4},
          doi = {10.1088/0953-4075/44/4/045101},
archivePrefix = {arXiv},
       eprint = {1012.0650},
 primaryClass = {quant-ph},
       adsurl = {https://ui.adsabs.harvard.edu/abs/2011JPhB...44d5101P}
}

@ARTICLE{Luigi,
       author = {{Sarandy}, Marcelo S. and {de Oliveira}, Thiago R. and {Amico}, Luigi},
        title = "{Quantum Discord in the Ground State of Spin Chains}",
      journal = {International Journal of Modern Physics B},
         year = 2013,
       volume = {27},
          doi = {10.1142/S0217979213450306},
archivePrefix = {arXiv},
       eprint = {1208.4817},
 primaryClass = {quant-ph},
       adsurl = {https://ui.adsabs.harvard.edu/abs/2013IJMPB..2745030S}
}

@article{Ollivier,
  title = {Quantum Discord: A Measure of the Quantumness of Correlations},
  author = {Ollivier, Harold and Zurek, Wojciech H.},
  journal = {Phys. Rev. Lett.},
  volume = {88},
  numpages = {4},
  year = {2001},
  publisher = {American Physical Society},
  doi = {10.1103/PhysRevLett.88.017901},
  url = {https://link.aps.org/doi/10.1103/PhysRevLett.88.017901}
}

@article{henderson,
  title={Classical, quantum and total correlations},
  author={Henderson, Leah and Vedral, Vlatko},
  journal={Journal of physics A: mathematical and general},
  volume={34},
  number={35},
  year={2001},
  publisher={IOP Publishing}
}

@article{spehner,
  title={Geometric quantum discord with Bures distance},
  author={Spehner, Dominique and Orszag, Miguel},
  journal={New Journal of Physics},
  volume={15},
  number={10},
  year={2013},
  publisher={IOP Publishing}
}

@article{Bures,
 ISSN = {00029947},
 URL = {http://www.jstor.org/stable/1995012},
 author = {Donald Bures},
 journal = {Transactions of the American Mathematical Society},
 publisher = {American Mathematical Society},
 title = {An Extension of Kakutani's Theorem on Infinite Product Measures to the Tensor Product of Semifinite w*-Algebras},
 volume = {135},
 year = {1969}
}

@ARTICLE{Shi1,
       author = {{Shi}, Jia-Dong and {Wang}, Dong and {Ye}, Liu},
        title = "{Comparative explorations of the revival and robustness for quantum dynamics under decoherence channels}",
      journal = {Quantum Information Processing},
         year = 2016,
       volume = {15},
       number = {4},
          doi = {10.1007/s11128-015-1233-4},
       adsurl = {https://ui.adsabs.harvard.edu/abs/2016QuIP...15.1649S}
}

@article{Piani,
  title = {Negativity of quantumness and its interpretations},
  author = {Nakano, Takafumi and Piani, Marco and Adesso, Gerardo},
  journal = {Phys. Rev. A},
  volume = {88},
  numpages = {18},
  year = {2013},
  publisher = {American Physical Society},
  doi = {10.1103/PhysRevA.88.012117},
  url = {https://link.aps.org/doi/10.1103/PhysRevA.88.012117}
}

@article{Paula,
  title = {Geometric quantum discord through the Schatten 1-norm},
  author = {Paula, F. M. and de Oliveira, Thiago R. and Sarandy, M. S.},
  journal = {Phys. Rev. A},
  volume = {87},
  numpages = {4},
  year = {2013},
  publisher = {American Physical Society},
  doi = {10.1103/PhysRevA.87.064101},
  url = {https://link.aps.org/doi/10.1103/PhysRevA.87.064101}
}

@article{PianiM,
  title = {Problem with geometric discord},
  author = {Piani, M.},
  journal = {Phys. Rev. A},
  volume = {86},
  numpages = {3},
  year = {2012},
  publisher = {American Physical Society},
  doi = {10.1103/PhysRevA.86.034101},
  url = {https://link.aps.org/doi/10.1103/PhysRevA.86.034101}
}

@article{Modi,
author = {Modi, Kavan},
title = {A Pedagogical Overview of Quantum Discord},
journal = {Open Systems \& Information Dynamics},
volume = {21},
number = {01n02},
year = {2014},
doi = {10.1142/S123016121440006X},

URL = { 
        https://doi.org/10.1142/S123016121440006X
    
},
eprint = { 
        https://doi.org/10.1142/S123016121440006X
    
}
}

@article{Animesh,
  title = {Quantum Discord and the Power of One Qubit},
  author = {Datta, Animesh and Shaji, Anil and Caves, Carlton M.},
  journal = {Phys. Rev. Lett.},
  volume = {100},
  numpages = {4},
  year = {2008},
  publisher = {American Physical Society},
  doi = {10.1103/PhysRevLett.100.050502},
  url = {https://link.aps.org/doi/10.1103/PhysRevLett.100.050502}
}

@article{knill,
  title = {Power of One Bit of Quantum Information},
  author = {Knill, E. and Laflamme, R.},
  journal = {Phys. Rev. Lett.},
  volume = {81},
  numpages = {0},
  year = {1998},
  publisher = {American Physical Society},
  doi = {10.1103/PhysRevLett.81.5672},
  url = {https://link.aps.org/doi/10.1103/PhysRevLett.81.5672}
}

@article{1935, 
title={Discussion of Probability Relations between Separated Systems},
 volume={31}, DOI={10.1017/S0305004100013554}, 
number={4},
 journal={Mathematical Proceedings of the Cambridge Philosophical Society},
 publisher={Cambridge University Press}, 
author={Schrödinger, E.}, 
year={1935},
}

@article{Horodecki,
  title = {Quantum entanglement},
  author = {Horodecki, Ryszard and Horodecki, Pawe\l{} and Horodecki, Micha\l{} and Horodecki, Karol},
  journal = {Rev. Mod. Phys.},
  volume = {81},
  numpages = {0},
  year = {2009},
  publisher = {American Physical Society},
  doi = {10.1103/RevModPhys.81.865},
  url = {https://link.aps.org/doi/10.1103/RevModPhys.81.865}
}

@article{Usen,
  title = {Local versus nonlocal information in quantum-information theory: Formalism and phenomena},
  author = {Horodecki, Micha\l{} and Horodecki, Pawe\l{} and Horodecki, Ryszard and Oppenheim, Jonathan and Sen(De), Aditi and Sen, Ujjwal and Synak-Radtke, Barbara},
  journal = {Phys. Rev. A},
  volume = {71},
  numpages = {25},
  year = {2005},
  publisher = {American Physical Society},
  doi = {10.1103/PhysRevA.71.062307},
  url = {https://link.aps.org/doi/10.1103/PhysRevA.71.062307}
}

@article{Bell1964,
  title = {On the Einstein Podolsky Rosen paradox},
  author = {Bell, J. S.},
  journal = {Physics Physique Fizika},
  volume = {1},
  numpages = {6},
  year = {1964},
  publisher = {American Physical Society},
  doi = {10.1103/PhysicsPhysiqueFizika.1.195},
  url = {https://link.aps.org/doi/10.1103/PhysicsPhysiqueFizika.1.195}
}

@article{Blaylock,
author = {Blaylock,Guy },
title = {The EPR paradox, Bell’s inequality, and the question of locality},
journal = {American Journal of Physics},
volume = {78},
number = {1},
year = {2010},
doi = {10.1119/1.3243279},
URL = {https://doi.org/10.1119/1.3243279},
eprint = {https://doi.org/10.1119/1.3243279 }}

@article{PhysRevLett.105.190502,
  title = {Necessary and Sufficient Condition for Nonzero Quantum Discord},
  author = {Daki\ifmmode \acute{c}\else \'{c}\fi{}, Borivoje and Vedral, Vlatko and Brukner, \ifmmode \check{C}\else \v{C}\fi{}aslav},
  journal = {Phys. Rev. Lett.},
  volume = {105},
  year = {2010},
  publisher = {American Physical Society},
  doi = {10.1103/PhysRevLett.105.190502},
  url = {https://link.aps.org/doi/10.1103/PhysRevLett.105.190502}
}

@article{dakic,
  title={Quantum discord as resource for remote state preparation},
  author={Daki{\'c}, Borivoje and Lipp, Yannick Ole and Ma, Xiaosong and Ringbauer, Martin and Kropatschek, Sebastian and Barz, Stefanie and Paterek, Tomasz and Vedral, Vlatko and Zeilinger, Anton and Brukner, {\v{C}}aslav and others},
  journal={Nature Physics},
  volume={8},
  number={9},
  year={2012},
  publisher={Nature Publishing Group}
}

@article{PhysRevA.86.034101,
  title = {Problem with geometric discord},
  author = {Piani, M.},
  journal = {Phys. Rev. A},
  volume = {86},
  numpages = {3},
  year = {2012},
  publisher = {American Physical Society},
  doi = {10.1103/PhysRevA.86.034101},
  url = {https://link.aps.org/doi/10.1103/PhysRevA.86.034101}
}

@article{Roga_2016,
	doi = {10.1088/1751-8113/49/23/235301},
	url = {https://doi.org/10.1088/1751-8113/49/23/235301},
	year = 2016,
	publisher = {{IOP} Publishing},
	volume = {49},
	number = {23},
	author = {W Roga and D Spehner and F Illuminati},
	title = {Geometric measures of quantum correlations: characterization, quantification, and comparison by distances and operations},
	journal = {Journal of Physics A: Mathematical and Theoretical}
}

@article{PhysRevA.80.024103,
  title = {Robustness of quantum discord to sudden death},
  author = {Werlang, T. and Souza, S. and Fanchini, F. F. and Villas Boas, C. J.},
  journal = {Phys. Rev. A},
  volume = {80},
  numpages = {4},
  year = {2009},
  publisher = {American Physical Society},
  doi = {10.1103/PhysRevA.80.024103},
  url = {https://link.aps.org/doi/10.1103/PhysRevA.80.024103}
}

@article{PhysRevA.81.014101,
  title = {Non-Markovian effect on the quantum discord},
  author = {Wang, Bo and Xu, Zhen-Yu and Chen, Ze-Qian and Feng, Mang},
  journal = {Phys. Rev. A},
  volume = {81},
  numpages = {4},
  year = {2010},
  publisher = {American Physical Society},
  doi = {10.1103/PhysRevA.81.014101},
  url = {https://link.aps.org/doi/10.1103/PhysRevA.81.014101}
}

@article{Mazzola,
  title = {Sudden Transition between Classical and Quantum Decoherence},
  author = {Mazzola, L. and Piilo, J. and Maniscalco, S.},
  journal = {Phys. Rev. Lett.},
  volume = {104},
  numpages = {4},
  year = {2010},
  publisher = {American Physical Society},
  doi = {10.1103/PhysRevLett.104.200401},
  url = {https://link.aps.org/doi/10.1103/PhysRevLett.104.200401}
}

@Article{Xu2010,
author={Xu, Jin-Shi
and Xu, Xiao-Ye
and Li, Chuan-Feng
and Zhang, Cheng-Jie
and Zou, Xu-Bo
and Guo, Guang-Can},
title={Experimental investigation of classical and quantum correlations under decoherence},
journal={Nature Communications},
year={2010},
day={12},
volume={1},
number={1},
issn={2041-1723},
doi={10.1038/ncomms1005},
url={https://doi.org/10.1038/ncomms1005}
}

@article{PhysRevLett.93.140404,
  title = {Finite-Time Disentanglement Via Spontaneous Emission},
  author = {Yu, Ting and Eberly, J. H.},
  journal = {Phys. Rev. Lett.},
  volume = {93},
  numpages = {4},
  year = {2004},
  publisher = {American Physical Society},
  doi = {10.1103/PhysRevLett.93.140404},
  url = {https://link.aps.org/doi/10.1103/PhysRevLett.93.140404}
}

@article{Animesh2,
author = { Madhok, Vaibhav and  Datta, Animesh},
title = { Quantum discord as a resource in quantum communication},
journal = {International Journal of Modern Physics B},
volume = {27},
number = {01n03},
year = {2013},
doi = {10.1142/S0217979213450410},

URL = { 
        https://doi.org/10.1142/S0217979213450410
    
},
eprint = { 
        https://doi.org/10.1142/S0217979213450410
    
}
}

@article{PhysRevA.87.042115,
  title = {One-norm geometric quantum discord under decoherence},
  author = {Montealegre, J. D. and Paula, F. M. and Saguia, A. and Sarandy, M. S.},
  journal = {Phys. Rev. A},
  volume = {87},
  numpages = {6},
  year = {2013},
  publisher = {American Physical Society},
  doi = {10.1103/PhysRevA.87.042115},
  url = {https://link.aps.org/doi/10.1103/PhysRevA.87.042115}
}

@ARTICLE{Pirandola,
       author = {{Pirandola}, Stefano},
        title = "{Quantum discord as a resource for quantum cryptography}",
      journal = {Scientific Reports},
         year = 2014,
       volume = {4},
          eid = {6956},
          doi = {10.1038/srep06956},
archivePrefix = {arXiv},
       eprint = {1309.2446},
 primaryClass = {quant-ph},
       adsurl = {https://ui.adsabs.harvard.edu/abs/2014NatSR...4E6956P}
}

@article{Madhok,
  title = {Interpreting quantum discord through quantum state merging},
  author = {Madhok, Vaibhav and Datta, Animesh},
  journal = {Phys. Rev. A},
  volume = {83},
  numpages = {4},
  year = {2011},
  publisher = {American Physical Society},
  doi = {10.1103/PhysRevA.83.032323},
  url = {https://link.aps.org/doi/10.1103/PhysRevA.83.032323}
}

@Article{Cianciaruso2015,
author={Cianciaruso, Marco
and Bromley, Thomas R.
and Roga, Wojciech
and Lo Franco, Rosario
and Adesso, Gerardo},
title={Universal freezing of quantum correlations within the geometric approach},
journal={Scientific Reports},
year={2015},
volume={5},
number={1},
issn={2045-2322},
doi={10.1038/srep10177},
url={https://doi.org/10.1038/srep10177}
}

@Article{Shi2016,
author={Shi, Jia-Dong
and Wang, Dong
and Ye, Liu},
title={Comparative explorations of the revival and robustness for quantum dynamics under decoherence channels},
journal={Quantum Information Processing},
year={2016},
volume={15},
number={4},
issn={1573-1332},
doi={10.1007/s11128-015-1233-4},
url={https://doi.org/10.1007/s11128-015-1233-4}
}

@article{SHI2016843,
title = {Revival and robustness of Bures distance discord under decoherence channels},
journal = {Physics Letters A},
volume = {380},
number = {7},
year = {2016},
issn = {0375-9601},
doi = {https://doi.org/10.1016/j.physleta.2015.11.039},
url = {https://www.sciencedirect.com/science/article/pii/S037596011501035X},
author = {Jia-dong Shi and Dong Wang and Yang-cheng Ma and Liu Ye}
}

@article{PhysRevA.86.062313,
  title = {Operational meaning of discord in terms of teleportation fidelity},
  author = {Adhikari, Satyabrata and Banerjee, Subhashish},
  journal = {Phys. Rev. A},
  volume = {86},
  numpages = {5},
  year = {2012},
  publisher = {American Physical Society},
  doi = {10.1103/PhysRevA.86.062313},
  url = {https://link.aps.org/doi/10.1103/PhysRevA.86.062313}
}

@article{Wang,
	doi = {10.1088/0253-6102/71/5/555},
	url = {https://doi.org/10.1088/0253-6102/71/5/555},
	year = 2019,
	publisher = {{IOP} Publishing},
	volume = {71},
	number = {5},
	author = {Yao-Kun Wang and Shao-Ming Fei and Zhi-Xi Wang},
	title = {Dynamics of Quantum Coherence in Bell-Diagonal States under Markovian Channels},
	journal = {Communications in Theoretical Physics}
}

@article{Zhao,
  title={Coherence evolution in two-qubit system going through amplitude damping channel},
  author={M. Zhao and Teng Ma and Yuquan Ma},
  journal={Science China Physics, Mechanics \& Astronomy},
  year={2017},
  volume={61}
}

@article{Bromley,
  title = {Frozen Quantum Coherence},
  author = {Bromley, Thomas R. and Cianciaruso, Marco and Adesso, Gerardo},
  journal = {Phys. Rev. Lett.},
  volume = {114},
  numpages = {6},
  year = {2015},
  publisher = {American Physical Society},
  doi = {10.1103/PhysRevLett.114.210401},
  url = {https://link.aps.org/doi/10.1103/PhysRevLett.114.210401}
}

@article{Luo,
  title = {Quantum discord for two-qubit systems},
  author = {Luo, Shunlong},
  journal = {Phys. Rev. A},
  volume = {77},
  numpages = {6},
  year = {2008},
  publisher = {American Physical Society},
  doi = {10.1103/PhysRevA.77.042303},
  url = {https://link.aps.org/doi/10.1103/PhysRevA.77.042303}
}

@article{Siciu,
author = {Suciu,Serban  and Isar,Aurelian },
title = {Gaussian geometric discord in terms of Hellinger distance},
journal = {AIP Conference Proceedings},
volume = {1694},
number = {1},
year = {2015},
doi = {10.1063/1.4937239},

URL = { 
        https://aip.scitation.org/doi/abs/10.1063/1.4937239
    
},
eprint = { 
        https://aip.scitation.org/doi/pdf/10.1063/1.4937239
    
}

}

@article{Werner,
  title = {Quantum states with Einstein-Podolsky-Rosen correlations admitting a hidden-variable model},
  author = {Werner, Reinhard F.},
  journal = {Phys. Rev. A},
  volume = {40},
  numpages = {0},
  year = {1989},
  publisher = {American Physical Society},
  doi = {10.1103/PhysRevA.40.4277},
  url = {https://link.aps.org/doi/10.1103/PhysRevA.40.4277}
}

@misc{nielsen2002quantum,
  title={Quantum computation and quantum information},
  author={Nielsen, Michael A and Chuang, Isaac},
  year={2002},
  publisher={American Association of Physics Teachers}
}

@article{Ciccarello_2014,
	doi = {10.1088/1367-2630/16/1/013038},
	url = {https://doi.org/10.1088/1367-2630/16/1/013038},
	year = 2014,
	publisher = {{IOP} Publishing},
	volume = {16},
	number = {1},
	author = {F Ciccarello and T Tufarelli and V Giovannetti},
	title = {Toward computability of trace distance discord},
	journal = {New Journal of Physics}
}

@ARTICLE{Aolita,
       author = {{Aolita}, Leandro and {de Melo}, Fernando and {Davidovich}, Luiz},
        title = "{Open-system dynamics of entanglement:a key issues review}",
      journal = {Reports on Progress in Physics},
         year = 2015,
       volume = {78},
       number = {4},
          doi = {10.1088/0034-4885/78/4/042001},
archivePrefix = {arXiv},
       eprint = {1402.3713},
 primaryClass = {quant-ph},
       adsurl = {https://ui.adsabs.harvard.edu/abs/2015RPPh...78d2001A}
}

@Inbook{Céleri2017,
author={C{\'e}leri, Lucas C.
and Maziero, Jonas},
title={The Sudden Change Phenomenon of Quantum Discord},
bookTitle={Lectures on General Quantum Correlations and their Applications},
year={2017},
publisher={Springer International Publishing},
address={Cham},
isbn={978-3-319-53412-1},
doi={10.1007/978-3-319-53412-1_15},
url={https://doi.org/10.1007/978-3-319-53412-1_15}
}

@book{bharucha1997elements,
  title={Elements of the Theory of Markov Processes and their Applications},
  author={Bharucha-Reid, Albert T},
  year={1997},
  publisher={Courier Corporation}
}

@article{PhysRevLett.103.210401,
  title = {Measure for the Degree of Non-Markovian Behavior of Quantum Processes in Open Systems},
  author = {Breuer, Heinz-Peter and Laine, Elsi-Mari and Piilo, Jyrki},
  journal = {Phys. Rev. Lett.},
  volume = {103},
  numpages = {4},
  year = {2009},
  publisher = {American Physical Society},
  doi = {10.1103/PhysRevLett.103.210401},
  url = {https://link.aps.org/doi/10.1103/PhysRevLett.103.210401}
}

@article{Vacchini_2011,
	doi = {10.1088/1367-2630/13/9/093004},
	url = {https://doi.org/10.1088/1367-2630/13/9/093004},
	year = 2011,
	publisher = {{IOP} Publishing},
	volume = {13},
	number = {9},
	author = {Bassano Vacchini and Andrea Smirne and Elsi-Mari Laine and Jyrki Piilo and Heinz-Peter Breuer},
	title = {Markovianity and non-Markovianity in quantum and classical systems},
	journal = {New Journal of Physics}
}

@article{Glick2020,
author={Glick, Jennifer R.
and Adami, Christoph},
title={Markovian and Non-Markovian Quantum Measurements},
journal={Foundations of Physics},
year={2020},
volume={50},
number={9},
issn={1572-9516},
doi={10.1007/s10701-020-00362-4},
url={https://doi.org/10.1007/s10701-020-00362-4}
}

@article{PhysRevLett.85.2014,
  title = {Sophisticated Quantum Search Without Entanglement},
  author = {Meyer, David A.},
  journal = {Phys. Rev. Lett.},
  volume = {85},
  numpages = {0},
  year = {2000},
  publisher = {American Physical Society},
  doi = {10.1103/PhysRevLett.85.2014},
  url = {https://link.aps.org/doi/10.1103/PhysRevLett.85.2014}
}

@article{PhysRevLett.83.1054,
  title = {Separability of Very Noisy Mixed States and Implications for NMR Quantum Computing},
  author = {Braunstein, S. L. and Caves, C. M. and Jozsa, R. and Linden, N. and Popescu, S. and Schack, R.},
  journal = {Phys. Rev. Lett.},
  volume = {83},
  numpages = {0},
  year = {1999},
  publisher = {American Physical Society},
  doi = {10.1103/PhysRevLett.83.1054},
  url = {https://link.aps.org/doi/10.1103/PhysRevLett.83.1054}
}

@article{PhysRevA.59.1070,
  title = {Quantum nonlocality without entanglement},
  author = {Bennett, Charles H. and DiVincenzo, David P. and Fuchs, Christopher A. and Mor, Tal and Rains, Eric and Shor, Peter W. and Smolin, John A. and Wootters, William K.},
  journal = {Phys. Rev. A},
  volume = {59},
  numpages = {0},
  year = {1999},
  publisher = {American Physical Society},
  doi = {10.1103/PhysRevA.59.1070},
  url = {https://link.aps.org/doi/10.1103/PhysRevA.59.1070}
}

@article{PhysRevA.81.042105,
  title = {Quantum discord for two-qubit $X$ states},
  author = {Ali, Mazhar and Rau, A. R. P. and Alber, G.},
  journal = {Phys. Rev. A},
  volume = {81},
  numpages = {7},
  year = {2010},
  publisher = {American Physical Society},
  doi = {10.1103/PhysRevA.81.042105},
  url = {https://link.aps.org/doi/10.1103/PhysRevA.81.042105}
}

@Article{Kenfack2017,
author={Kenfack, Lionel Tenemeza
and Tchoffo, Martin
and Fai, Lukong Cornelius},
title={Dynamics of tripartite quantum entanglement and discord under a classical dephasing random telegraph noise},
journal={The European Physical Journal Plus},
year={2017},
volume={132},
number={2},
doi={10.1140/epjp/i2017-11364-5},
url={https://doi.org/10.1140/epjp/i2017-11364-5}
}

@Article{Pal2012,
author={Pal, A. K.
and Bose, I.},
title={Markovian evolution of classical and quantum correlations in transverse-field XY model},
journal={The European Physical Journal B},
year={2012},
volume={85},
number={8},
doi={10.1140/epjb/e2012-30108-1},
url={https://doi.org/10.1140/epjb/e2012-30108-1}
}

@article{SCHLOSSHAUER20191,
title = {Quantum decoherence},
journal = {Physics Reports},
volume = {831},
year = {2019},
issn = {0370-1573},
doi = {https://doi.org/10.1016/j.physrep.2019.10.001},
url = {https://www.sciencedirect.com/science/article/pii/S0370157319303084},
author = {Maximilian Schlosshauer}
}

@book{kraus1983states,
  title={States, Effects, and Operations Fundamental Notions of Quantum Theory: Lectures in Mathematical Physics at the University of Texas at Austin},
  author={Kraus, Karl and B{\"o}hm, Arno and Dollard, John D and Wootters, WH},
  year={1983},
  publisher={Springer}
}

@article{helstrom1969quantum,
  title={Quantum detection and estimation theory},
  author={Helstrom, Carl W},
  journal={Journal of Statistical Physics},
  volume={1},
  number={2},
%  pages={231--252},
  year={1969},
  publisher={Springer}
}

@article{castro2018entanglement,
  title={Entanglement content of quasiparticle excitations},
  author={Castro-Alvaredo, Olalla A and De Fazio, Cecilia and Doyon, Benjamin and Sz{\'e}cs{\'e}nyi, Istv{\'a}n M},
  journal={Physical review letters},
  volume={121},
  number={17},
%  pages={170602},
  year={2018},
  publisher={APS}
}

@article{calabrese2012entanglement,
  title={Entanglement negativity in quantum field theory},
  author={Calabrese, Pasquale and Cardy, John and Tonni, Erik},
  journal={Physical review letters},
  volume={109},
  number={13},
%  pages={130502},
  year={2012},
  publisher={APS}
}

@article{calabrese2004entanglement,
  title={Entanglement entropy and quantum field theory},
  author={Calabrese, Pasquale and Cardy, John},
  journal={Journal of statistical mechanics: theory and experiment},
  volume={2004},
  number={06},
%  pages={P06002},
  year={2004},
  publisher={IOP Publishing}
}

@article{shaham2019implementation,
  title={Implementation of controllable universal unital optical channels},
  author={Shaham, A and Karni, T and Eisenberg, HS},
  journal={Optics Express},
  volume={27},
  number={17},
%  pages={23839--23848},
  year={2019},
  publisher={Optical Society of America}
}

@article{plenio2005logarithmic,
  title={Logarithmic negativity: a full entanglement monotone that is not convex},
  author={Plenio, Martin B},
  journal={Physical review letters},
  volume={95},
  number={9},
 % pages={090503},
  year={2005},
  publisher={APS}
}

\end{document}